\documentclass[useAMS,usenatbib]{mnras}
\usepackage{graphicx,epsfig,longtable}
\usepackage{epstopdf}
\usepackage{color}
\usepackage{xcolor}
\usepackage{amsmath}
\usepackage{amssymb}
\usepackage{tabularx} 
\usepackage{blindtext}

\newcommand{\Msun}{$M_{\sun}$}

\title[Nuclear uncertainties and r-process in NS mergers]{Impact of systematic nuclear uncertainties on composition and decay heat of dynamical and disk ejecta in compact binary mergers}

\author[I.~Kullmann et al.]
  {I. Kullmann$^1$, S.~Goriely$^1$, O.~Just$^{2,3}$, A. Bauswein$^{2,4}$,  H.-T. Janka$^5$ \\ 
  $^1$Institut d'Astronomie et d'Astrophysique, CP-226, Universit\'e Libre de Bruxelles, 1050 Brussels, Belgium\\
  $^2$GSI Helmholtzzentrum f\"ur Schwerionenforschung, Planckstrasse 1, 64291 Darmstadt, Germany\\
  $^3$Astrophysical Big Bang Laboratory, RIKEN Cluster for Pioneering Research, 2-1 Hirosawa, Wako, Saitama 351-0198, Japan\\
  $^4$Helmholtz Research Academy Hesse for FAIR (HFHF), GSI Helmholtz Center for Heavy Ion Research, Campus Darmstadt,  Germany\\
  $^5$Max-Planck-Institut f\"ur Astrophysik, Postfach 1317, 85741 Garching, Germany   
 }
\date{Released 2022 Xxxxx XX}

\pagerange{\pageref{firstpage}--\pageref{lastpage}} \pubyear{2021}

\def\ga{\,\,\raise0.14em\hbox{$>$}\kern-0.76em\lower0.28em\hbox
{$\sim$}\,\,}
\def\la{\,\,\raise0.14em\hbox{$<$}\kern-0.76em\lower0.28em\hbox
{$\sim$}\,\,}
\def\Msun{$M_{\odot}$}

\begin{document}

\label{firstpage}

\maketitle

\begin{abstract}
Theoretically predicted yields of elements created by the rapid neutron capture (r-) process carry potentially large uncertainties associated with incomplete knowledge of nuclear properties and approximative hydrodynamical modelling of the matter ejection processes.
We present an in-depth study of the nuclear uncertainties by varying theoretical nuclear input models that describe the experimentally unknown neutron-rich nuclei. This includes two frameworks for calculating the radiative neutron capture rates and 14 different models for nuclear masses, $\beta$-decay rates and fission properties.
Our r-process nuclear network calculations are based on detailed hydrodynamical simulations of dynamically ejected material from NS-NS or NS-BH binary mergers plus the secular ejecta from BH-torus systems.  
The impact of nuclear uncertainties on the r-process abundance distribution and the early radioactive heating rate is found to be modest (within a factor of $\sim20$ for individual $A>90$ abundances and a factor of 2 for the heating rate). However, the impact on the late-time heating rate is more significant and depends strongly on the contribution from fission.
We witness significantly higher sensitivity to the nuclear physics input if only a single trajectory is used compared to considering ensembles with a much larger number of trajectories (ranging between 150 and 300), and the quantitative effects of the nuclear uncertainties strongly depend on the adopted conditions for the individual trajectory. 
We use the predicted Th/U ratio to estimate the cosmochronometric age of six metal-poor stars and find the impact of the nuclear uncertainties to be up to 2 Gyr.
\end{abstract}

\begin{keywords}
Nuclear reactions, nucleosynthesis, abundances -- Neutron star mergers
\end{keywords}

\section{Introduction}
\label{sec_intro}

Through a series of neutron captures and $\beta$-decays on light nuclei, the rapid neutron capture process (or r-process) can explain the production of about 50\% of the stable (and some long-lived) neutron-rich nuclides heavier than iron, as initially proposed by \citet{burbidge1957,cameron1957}. 
Since then and until a few years ago, the astrophysical site(s) of the r-process remained essentially unknown. 
Recently, our understanding of the astrophysical site for the r-process and the nucleosynthesis has improved drastically both in terms of observational data but also due to computational advancements of site-specific simulations.
Current r-process calculations rely on advanced hydrodynamical simulations which provide the conditions of the astrophysical environment where the nucleosynthesis occurs and a full nuclear reaction network with regularly improved nuclear physics input is usually applied in a post-processing step. 

Until a decade ago, the neutrino-driven wind launched during a core-collapse supernova (CCSN) \citep[e.g.,][]{takahashi1994,qian1996,hoffman1997,Arcones2007} was a very popular candidate for the r-process site. 
However, the conditions required for a successful r-process have not been obtained in the most sophisticated existing models \citep[e.g.,][and references therein]{witti1994,takahashi1994,Hudepohl2010,roberts2010,Fischer2010,wanajo2011,
janka2012,Mirizzi2016,wanajo2018c}.
The focus shifted towards binary neutron star (NS) mergers after hydrodynamical models managed to demonstrate that a significant amount of material ($10^{-3}-10^{-2}$~\Msun) can become unbound in the dynamical phase of NS-NS mergers \citep[e.g.,][]
{Ruffert1997, Ruffert1999,Freiburghaus1999,rosswog1999, Ruffert2001, Janka2002, Oechslin2007,goriely2011, hotokezaka2013, bauswein2013,Just2015,radice2018b,foucart2016}.
The final observational confirmation that r-process material is synthesized in NS-NS mergers came in 2017 with the first gravitational wave detection from a NS-NS merger, GW170817 \citep[e.g.,][]{abbott2017c} accompanied by an electromagnetic signal, AT2017gfo \citep[][]{abbott2017d,kasen2017,Drout2017,Villar2017, Cowperthwaite2017,Kilpatrick2017}.
Such a signal in the optical electromagnetic spectrum in the aftermath of a NS merger event is often referred to as a ``kilonova'' \citep[e.g.,][]{li1998, roberts2011, metzger2010, goriely2011, barnes2013,Kulkarni2005,Tanaka2013} and is powered by the decay of freshly produced radioactive r-process elements.
Therefore, modelling kilonova light curves requires knowledge of the energy released by the newly synthesized radioactive species in the ejecta, i.e., the rate of heat released through $\beta$-decays, fission, and $\alpha$-decays. Moreover, the heat generated by the various decay components thermalizes in the material with different efficiencies, which again affects the light curve. 
Before they escape, kilonova photons will be absorbed and re-emitted by atomic transitions of the elements in the opaque inner region of the ejecta.
A major source of opacity is believed to be due to the presence of lanthanides ($Z = 57-71$) 
and actinides ($Z=90-100$) in the ejecta \citep[][]{Kasen2013,Tanaka2013}. 
Thus, the amount of lanthanide and actinide elements produced by the r-process in NS merger models can significantly affect the kilonova light curve. 
However, it is not observationally settled yet whether binary NS mergers can also produce the heaviest of the solar r-process elements and whether they are the only r-process site or if other sites contribute to the Galactic enrichment. For example, outflows associated with jets in core-collapse supernovae or from accretion tori around black holes forming in collapsing, rapidly rotating massive stars, so-called collapsars \citep{macfadyen1999}, were considered in numerous papers  \citep[e.g.,][]{Cameron2001,cameron2003,winteler2012,Nishimura2015,Mosta2018,siegel2019,Grimmett2020, Just2021,Reichert2021,cowan2021,Just2022,Reichert2022,Siegel2022}.
In particular, the question of the earliest possible onset of r-process nucleosynthesis from NS-NS mergers during the galactic chemical evolution comes with significant uncertainties. 
One problem may be that the time required to form the first NS binary systems plus their in-spiral phase before merging is too long to explain the presence of Eu in metal-poor halo stars, and therefore another r-process site might be needed \citep{Argast2004,Hotokezaka2018,cote2019}. This conclusion, however, has been weakened by several chemical evolution models \citep[e.g.,][see also \citet{Roederer2016,Ji2016} for the observation of r-process enhanced stars in the ultra-faint dwarf galaxy Reticulum 2 which points towards a pollution by a rare r-process source with a relatively large enrichment early in its history]{Shen2015,dvorkin2020,Voort2020,Voort2022}. More work in this regard is required to settle this debate.


\begin{figure}
\begin{center}
\includegraphics[width=\columnwidth]{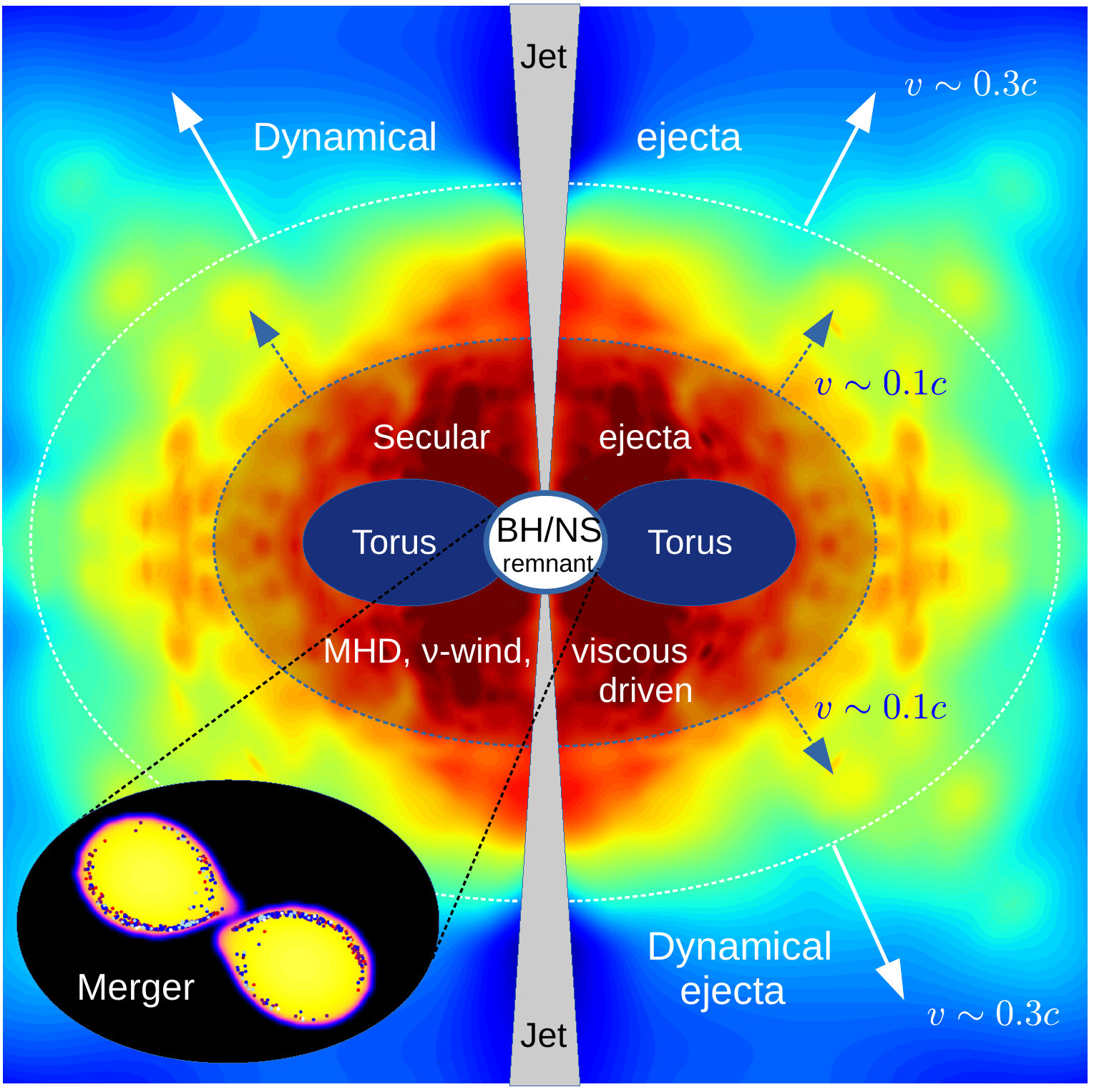}
\caption{(colour online)
A compact object (NS-NS or NS-BH) merger leads to either a NS or BH remnant producing ejecta via different mechanisms. During the first few ms after merging the dynamical ejecta (outermost sphere) become unbound due to tidal torques and shock heating from the contact interface (bottom left insert) with typical ejecta velocities of $\sim0.3$~c (where c is the speed of light). The material ejected on the dynamical time scale surrounds the remnant plus torus system that forms at later times and its corresponding secular ejecta with velocities of $\sim0.1$~c. A jet is also expected to form, drilling through the ejecta and producing a short $\gamma$-ray burst.
}
\label{fig_ejphases}
\end{center}
\end{figure}

During and shortly after the merging, the NS-NS system undergoes several mass ejection phases (see Fig.~\ref{fig_ejphases}), which depend on the binary system parameters and the properties of the nuclear equation of state (EoS) \citep[e.g.][]{Just2015,cowan2021,Perego17,Shibata19}. 
In the first phase, matter is ejected dynamically both from the tidal tails of the NSs after they interact as well as from the shock-heated material originating from the contact interface.
The ratio between the cold tidal and hot shocked ejecta mainly depends on the mass ratio between both stars $q=M_1/M_2$ (with $M_2\ge M_1$) and the nuclear EoS, where the relative amount of cold tidal ejecta typically grows with smaller $q$ and stiffer EoSs \citep{bauswein2013}.
In addition,  it was discussed already in early papers \citep{Ruffert1997,Ruffert2001} that the inclusion of weak nucleonic reactions of electron neutrinos ($\nu_e$) with free neutrons ($n$) and electron antineutrinos ($\bar{\nu}_e$) with free protons ($p$) as well as their inverse reactions
\begin{equation}
\begin{aligned}
 & \nu_e+ n \rightleftharpoons p + e^- , \label{eq:betareac}  \\
 & \bar{\nu}_e + p \rightleftharpoons n + e^+  .
\end{aligned}
\end{equation}
might significantly change the initial composition of the ejecta. Indeed, this was later shown by numerical modelling \citep[e.g.,][and references therein]{wanajo2014, goriely2015a, martin2018, radice2016, foucart2016,kullmann2021}. Since the r-process nucleosynthesis is very sensitive to the initial composition, in particular to the neutron richness of the material, it is critical to base r-process calculations on hydrodynamical simulations which include such neutrino reactions. 

Similar to NS-NS mergers, a significant amount of neutron-rich material can also be ejected from NS-black hole (BH) mergers \citep[e.g.,][]{lattimer1974,Janka1999,Etienne2008,Etienne2009,Kyutoku2013,Bauswein2014,Just2015,Kiuchi2015, Fernandez2017,Fernandez2020,Kruger2020} so that they are also viable candidates for the r-process site. The evolution from two orbiting, in-spiralling compact objects to a BH or a long-lived NS remnant depends on the initial conditions of the system. 
In NS-BH mergers with sufficiently small BH masses and NS-NS mergers with sufficiently high total masses and/or a sufficiently soft EoS, a BH-torus system is formed right after the merger. Alternatively, if the NS remnant can be stabilized against a prompt collapse in a NS-NS merger, it will evolve as a (hyper- or super-) massive NS and either undergo a delayed collapse after some time or remain indefinitely stable. In both kinds of remnants, BH-torus and massive NS systems, material can be expelled due to turbulent viscosity, neutrino heating, or magnetic-fields effects \citep[e.g.,][]{Metzger2014,Siegel2018,HosseinNouri2018,Christie2019,miller2019, Fernandez2019,Fujibayashi2018,Fujibayashi2020b,Fujibayashi2020, perego2014,Perego2021,Just2015,Just2022,Fahlman2022}. 

Great efforts have been put into hydrodynamical simulations of different NS-NS or NS-BH merger configurations in the last decade. 
An important goal for the community is to develop astrophysically consistent models which can estimate the total amount of ejecta, either by combining models consistently or, ideally, by covering the entire evolution of the merging system.  
Recently, improved models including fully relativistic approaches have managed to cover the dynamical phase and the late-time evolution of the remnant and secular ejecta \citep{Kawaguchi2021,Hayashi2021}. 
Note that the NS-NS simulations of \citet{Kawaguchi2021} involve a mapping between different codes; see also \citet{Just2015} for an early attempt to combine simulations of different phases of the merger in a consistent way, i.e., (approximately) preserving the global properties, such as BH mass and spin and disk mass.
However, due to computational constraints, most models only cover one phase of the evolution, i.e., either the dynamical or the secular phase.

In this work, we will base our r-process nucleosynthesis calculations on state-of-the-art hydrodynamical simulations that estimate the dynamical ejecta of two NS-NS and one NS-BH merger systems and, separately, two BH-torus simulations of the post-merger phase. By combining the appropriate hydrodynamical models for the dynamical and secular ejecta,  we will estimate the nucleosynthesis composition of the total ejecta. 

After high-resolution spectra became available for metal-poor stars, it has become possible to derive the abundances of long-lived radioactive elements such as thorium and uranium in stellar spectra of r-process-enhanced metal-poor stars  \citep{Sneden96,Cayrel01}. This has opened the door for an independent age-dating method of individual stars and, therefore, also sets lower limits on the age of our Galaxy. 
The method, called cosmochronometry \citep{Fowler60,Butcher1987,Goriely2001,Schatz2002}, relies on the known radioactive decay time (i.e., half-life) of long-lived radioactive nuclei and the theoretical prediction of the initial abundance (before decay) of the involved nuclei.
By combining the known radioactive decay time for $^{238}$U and $^{232}$Th of $\tau(^{238}\mathrm{U}) = 6.45$~Gyr and $\tau(^{232}\mathrm{Th}) = 20.27$~Gyr, respectively, an age estimate of the star ($t^*_{\mathrm{U},\mathrm{Th}} $) can be calculated from 
\begin{equation}
\log \Big( \frac{\mathrm{Th}}{\mathrm{U}} \Big)_{\mathrm{obs}} = \log \Big( \frac{\mathrm{Th}}{\mathrm{U}} \Big)_\mathrm{r} + \log \Big[ \exp \Big( \frac{1}{\tau(\mathrm{U})} - \frac{1}{\tau(\mathrm{Th})} \Big)  \Big] \times t^*_{\mathrm{U},\mathrm{Th}} 
\label{eq_cosmo}
\end{equation}
where $(\mathrm{Th}/\mathrm{U})_{\mathrm{obs}}$ and $(\mathrm{Th}/\mathrm{U})_{\mathrm{r}}$ are the observationally derived and model-dependent r-process abundance ratios, respectively, of $^{232}$Th and $^{238}$U (after decay of the shorter-lived actinides).
Here, we use the Th/U notation commonly used in astronomy for the abundance ratios, which is identical to the molar fraction\footnote{The mass fraction $X_i$ of a nucleus $i$ shown in the figures herein is defined as $X_i=Y_i \times A$ where $A$ is the mass number of the nucleus.} 
ratios $Y_{^{232}\mathrm{Th}}/Y_{^{238}\mathrm{U}}$.
Other cosmochronometers have been applied in the literature as, for example, the Th/Eu ratio since Eu is relatively easy to identify in the spectra of metal-poor stars. This is in contrast to U, which is extremely difficult to extract from stellar spectra and thus, only a handful of stars have U measurements.
However, Eu and Th are widely separated in atomic mass, which makes the theoretical estimates more uncertain compared to U and Th, which are neighbouring elements \citep{Goriely2001,Goriely2016d}.
The uncertainty in Th/Eu chronometry is also related to the long half-life of Th.
Considering the Th/U ratio obtained in our r-process calculations, we will make use of Eq.~\ref{eq_cosmo} to estimate the age  of six metal-poor r-process enriched stars with both Th and U lines in their spectra.

In addition to the astrophysical modelling uncertainties, r-process nucleosynthesis calculations rely on nuclear data for thousands of nuclei\footnote{There are typically about $5000$ nuclei with $Z\le 110$ lying between the valley of stability and the neutron drip line, where the latter is defined by the applied nuclear mass model (see Sec.~\ref{sec_network_input}).}. Although masses and $\beta$-decay rates are known for many neutron-rich nuclei, these need to be estimated from nuclear models for almost all nuclei involved during the neutron irradiation. Additionally, no reaction rates are known experimentally for unstable neutron-rich nuclei produced during the r-process irradiation.
Thus, theoretical models are crucial to predict fundamental nuclear properties such as nuclear masses, $\alpha$- and $\beta$-decay rates, radiative neutron capture rates and fission probabilities, all of which enter into the r-process reaction network as input.
Despite much progress, these nuclear models are still affected by a variety of uncertainties, in particular for the complex description of exotic neutron-rich nuclei.
In addition to the radiative neutron captures and the reverse photodisintegrations, all charged-particle fusion reactions on light and medium mass elements and $\beta$-delayed processes become important during the nucleosynthesis.  If the r-process reaches the fissile region, fission processes such as neutron-induced, spontaneous and $\beta$-delayed fission, together with the corresponding fission fragment distribution, have to be taken into account for all fissioning nuclei.  
Previous works have shown that the nuclear uncertainties can significantly affect the results for the r-process abundances \cite[e.g.,][]{Goriely2001,Schatz2002,surman2014,Caballero2014,Mendoza-Temis2015,Eichler2015,goriely2015,Martin2016,Liddick2016,Mumpower2016,Nishimura2016,Bliss2017, Denissenkov2018, Vassh2019,Nikas2020,sprouse2020a,McKay2020,Giuliani2020, Lund2022} and, therefore, for the radioactive heating rate that gives rise to the observable kilonova emission \cite[e.g.,][]{Rosswog2017,zhu2018,wu2019,Even2020,Zhu2021,Barnes2021}.
However, the conclusions of these studies may change with, for example, the inclusion of weak nucleonic interactions in the dynamical ejecta, the ability for the r-process to reach the fissile region or the nuclear physics model adopted, as discussed in \citet{Lemaitre2021,kullmann2021}. 

To estimate the sensitivity of the r-process nucleosynthesis yields to the nuclear input, many studies up to now have been using Monte Carlo (MC) type simulations \citep[e.g.,][]{Mumpower2016,Nikas2020}.
In this approach, reaction or decay rates are modified in a given range, independently of the changes of other reactions, and these modified sets of reactions are used to compute the nucleosynthesis. This approach assumes that the rates are uncorrelated; this may be the case in particular for experimentally determined cross-sections or temperature-dependent $\beta$-decay rates, for which the properties are intrinsic to each specific case and do not affect rates of neighbouring nuclei. However, in the case of theoretically derived rates, those are predicted by a given model that defines the correlations between all reactions, either locally (e.g., by modifying a given structure property of a specific nucleus) or globally (e.g., by changing the model adopted to describe the structure properties of interacting nuclei or the interaction with nucleons or photons). 
For example, if a rate is uncertain because of the still unknown mass of the target nucleus, this will affect the production as well as destruction rates of this nucleus through a modification of the corresponding $Q$-value. Consequently, none of the (destruction or production) rates can be changed independently from the other. Similarly, theoretical rates are correlated by the underlying reaction model and the many nuclear models adopted to describe the ingredients of the reaction model, such as nuclear masses, nuclear level densities, photon strength functions, and optical potentials. With this in mind, we will study the uncertainties related to theoretical nuclear models on the r-process nucleosynthesis by varying the input between global models that have been adjusted on the full set of available experimental data. 
This approach has, however, the drawback of considering only systematic model-correlated  and not statistical uncertainties. The propagation of statistical (or parameter) uncertainties, as done for example by \citet{sprouse2020a}, are not considered here because model uncertainties are expected to dominate over parameter uncertainties for exotic neutron-rich nuclei, as shown  by \citet{Goriely2014a} in the context of mass predictions. Systematic uncertainties are inherently correlated by the underlying model, while statistical ones are not and can therefore be applied for example through MC techniques. It remains extremely complex to build parameter uncertainties around each of the model uncertainties and consistently propagate them. Such an extensive propagation of both model and parameter uncertainties will need to be considered in the future.

This paper is organized as follows: in Sec.~\ref{sec_astro_mods} we introduce our NS-NS and NS-BH merger models and their basic properties. Sec.~\ref{sec_network_input} presents the r-process network and the nuclear models used as input and varied in this work. 
The results of our nucleosynthesis calculations and the impact of consistently propagating nuclear uncertainties into r-process calculations, including abundance distributions, heating rates and cosmochronometric age estimates, are discussed in Sec.~\ref{sec_nucuncert}.
In addition, a comparison to other works can be found in Sec.~\ref{sec_nucuncert}.
A summary and conclusions are given in Sec.~\ref{sec_concl}. 


\section{Astrophysical models} 
\label{sec_astro_mods}

Our nucleosynthesis calculations are based on advanced computational models of the binary NS or NS-BH merger systems. 
However, fully consistent models, which include weak nucleonic reactions and cover all evolution phases over a wide range of binary parameters are not available at the present time. 
Our selection of hydrodynamical models is not meant to be exhaustive concerning the possible astrophysical scenarios for the r-process. In order to keep the amount of calculations tractable, we only consider here a small number of hydrodynamical models, which, however, are found to be representative regarding the nucleosynthesis-relevant properties of the various ejecta components, namely the final composition of the ejecta and the time evolution of its decay heat.
We have applied an approximate model for the total NS-NS (NS-BH) merger ejecta by combining three models for the dynamical ejecta from \citet{ardevol-pulpillo2019,kullmann2021,Just2015} with two BH-torus models from \citet{Just2015}.

The NS-NS merger simulations that produce the dynamical ejecta are based on a relativistic smoothed-particle-hydrodynamics (SPH) code coupled to the so-called improved leakage-equilibration-absorption scheme (ILEAS) for neutrino transport \citep{ardevol-pulpillo2019}.
The ILEAS scheme goes beyond a conventional leakage scheme by accounting for re-absorption with a ray-tracing algorithm in the optically thin conditions, and adding an equilibration step in the optically thick regions in addition to an improved estimate of the neutrino diffusion time scales.
ILEAS was designed to be more computationally efficient than the most sophisticated neutrino transport calculations in the literature; still, it has been shown to reproduce local neutrino interaction rates as well as the global effects of neutrinos found by state-of-the-art neutrino-transport solutions at the level of 10-15\% \citep{ardevol-pulpillo2019}.

To model the contribution from the NS-NS dynamical ejecta, we have chosen two systems with a total mass of $M_1+M_2=2.7$~\Msun; one symmetric (1.35~\Msun - 1.35~\Msun) and one asymmetric (1.25~\Msun - 1.45~\Msun) merger with the SFHo EoS \citep{steiner2013}, from now on referred to as model SFHo-135-135 and SFHo-125-145, respectively. For the NS-NS systems, a delayed collapse of the HMNS remnant to a BH is expected to take place soon after the dynamical merger phase, i.e., some $\sim10$~ms after the stars touch.
Although the gravitational collapse did not yet occur at the end of the simulations at about $\sim10$~ms after merging, we do not consider ejecta launched from the HMNS on time-scales longer than $\sim10$~ms in this study. 
Even though the amount of matter ejected during the HMNS evolution can, in principle, be comparable to the other ejecta components, the ejecta masses and other properties still carry large uncertainties, mainly because the magnetohydrodynamic effects responsible for angular momentum transport are yet poorly understood \citep[see, e.g.,][for discussions of HMNS ejecta]{perego2014,Fujibayashi2018,Kiuchi2018,Ciolfi2020,Aguilera-Miret2022,Palenzuela2022,Shibata2021b}.
In our cases, the HMNS phase will be relatively short. Thus, neglecting the HMNS ejecta is an acceptable approximation and is assumed to introduce uncertainties at the same level as the uncertainties presently existing in hydrodynamical simulations. 
The NS-BH binary system consists of a $1.1$~\Msun\ NS and a $2.3$~\Msun\ BH\footnote{Note that, despite the low BH mass, the mass, dynamics, and neutron excess of the ejecta can be considered as representative of NS-BH mergers.}, also modelled with the SFHo EoS \citep[referred to as sfho\_1123 in][]{Just2015}, from now on named SFHo-11-23 (this model does not include neutrino effects which may be a decent approximation for the mostly tidally ejected material; ignoring neutrino emission and absorption may be a valid approximation because of the rapid expansion of mostly tidally ejected material in NS-BH mergers \citep[e.g., see][]{Roberts2017}). 

Given the challenge to obtain fully consistent models covering the entire merger evolution and the large uncertainties related to the viscosity of the disk models, we have chosen two BH-torus systems that approximately match the configurations obtained after the merger in models SFHo-135-135, SFHo-125-145, and SFHo-11-23. For the post-merger phase, we used a BH mass of 3~\Msun\ and torus masses of 0.1 and 0.3~\Msun, named models M3a8m1a5 and M3a8m3a5-v2 in \citet{Just2015}. The BH-torus simulations are performed in Newtonian hydrodynamics with a modified gravitational potential \citep{Artemova1996} to approximately model some relativistic effects like an innermost stable circular orbit. 
The neutrino transport is described by a truncated two-moment scheme \citep{Just2015b}, and two alpha-viscosity approaches called ``type 1'' and ``type 2'' are applied in model M3a8m1a5 and M3a8m3a5-v2, respectively, to include the effects of turbulent angular momentum transport.

 \begin{table*}
\centering
\caption{Parameters of the merger models: model name, mass ratio $q=M_1/M_2$ for the NS masses $M_1$ and $M_2$, mean electron fraction and mass fraction of free neutrons at the network initiation time, ejected mass, number of trajectories in the subset, total number of trajectories, fraction of outflow mass in the subset adopted, BH mass, BH spin and torus mass.
Both NS+NS merger systems have the same total mass of $M_1 + M_2 = 2.7$~\Msun, and the $Y_e$ is estimated at $\rho_\mathrm{net}$ for the merger models and when the temperature drops below 10~GK for the BH-torus models. Both BH-torus models use $\alpha_{vis}=0.05$  for the viscosity and model M3A8m3a5-v2 uses a ``type 2'' viscosity, as described in \citet{Just2015}.}
 \begin{tabular}{lcccccccccc}
 \hline  \hline 
 Model name & $q$ & $\langle Y_e \rangle$ & $\langle X_n^0 \rangle$ & $M_{\mathrm{ej}}$ & $N_\mathrm{sub}$ & $N_\mathrm{tot}$ & $\frac{M_{\mathrm{sub}}}{M_{\mathrm{ej}}}$ &  $M_{\mathrm{BH}}$ & $A_{\mathrm{BH}}$ & $M_{\mathrm{torus}}$  \\ 
                       &  &  &  & [$10^{-3}$~\Msun] & & & & [\Msun] & & [$10^{-3}$~\Msun] \\ 
 \hline 
 SFHo-125-145 & 0.86 & 0.24 & 0.56 & 8.7 & 266 & 4398 & 0.15 & 2.40 & 0.80 & 170 \\ 
 SFHo-135-135 & 1.00    & 0.26 & 0.54 & 3.3 & 256 & 1263 & 0.24 & 2.45 & 0.83 & 90 \\ 
 \hline
 SFHo-11-23  & 0.48    & 0.04 & 0.85 & 40.4 & 150 & 13175 & 0.01 & 3.09 & 0.82 & 260 \\ 
 \hline
 M3A8m1a5        & - & 0.23 & 0.58 & 24.7 & 296 & 4150 & 0.50 & 3.00 & 0.80 & 100 \\ 
 M3A8m3a5-v2 & -     & 0.24 & 0.56 & 70.1 & 177 & 2116 & 0.20 & 3.00 & 0.80 & 300 \\ 
 \hline  \hline 
 \end{tabular} 
\label{tab_astro_mods}
\end{table*}

Table~\ref{tab_astro_mods} lists the hydrodynamical models and their properties, including NS, BH or torus mass, spin, and ejected mass. 
For the models covering the dynamical ejecta, the values refer to those extracted at the end of the simulation time ($t\sim10$~ms after merging).
In the NS-NS simulations, the merger remnant did not yet collapse to form a BH, and the torus mass is estimated by determining which fraction of matter is rotationally supported, assuming a collapse took place (see \citet{Oechslin2007} for details). 
The assumption that the NS-NS merger remnant collapses within a reasonably short time is motivated by the fact that the total mass of the NS-NS models is high (2.7~\Msun). For such heavy systems, most EOSs predict a collapse of the HMNS once it loses enough angular momentum by turbulent viscosity and energy by neutrino emission and gravitational waves \citep{Bauswein2019,Baiotti2017,Radice2020b,Shibata2019}. However, for stiff EOSs, the NS remnant could be long-lived, but this case is not considered in our study.
As can be seen in Table~\ref{tab_astro_mods}, the total outflow mass of the dynamical ejecta in the NS-NS models are 7-20 times less massive than the BH-torus ejecta; hence the total r-process abundance distribution is dominated by the BH-torus component. In contrast, in the NS-BH system, the dynamical and secular ejecta masses are of the same order of magnitude.

\begin{figure}
\begin{center}
\includegraphics[width=\columnwidth]{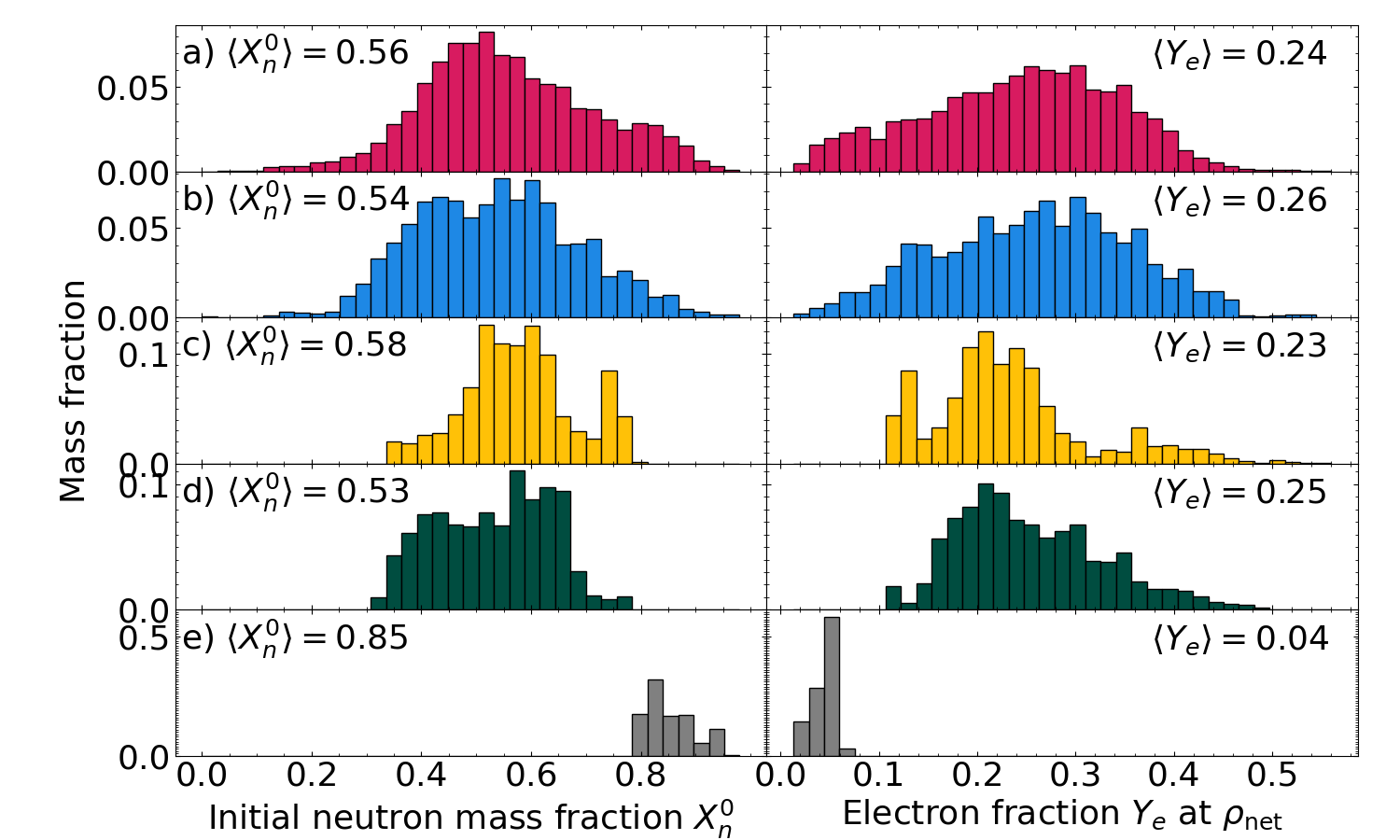}
\caption{(colour online)
Fractional mass distributions of the matter ejected as a function of the mass fraction of free neutrons $X_n^0$ (left column) and electron fraction $Y_e$ (right column) at the network initiation time. The mean electron fraction $\langle Y_e \rangle$ and the mean initial neutron mass fraction $\langle X_n^0 \rangle$ are given for each merger and BH-torus models: a) SFHo-125-145, b) SFHo-135-135, c) M3A8m1a5, d) M3A8m3a5-v2 and e) SFHo-11-23.
}
\label{fig_xn0_ye_distr}
\end{center}
\end{figure}

The ejecta distributions of the electron fraction $Y_e$ and mass fraction of free neutrons $X_n^0$ at the network initiation time are shown in Fig.~\ref{fig_xn0_ye_distr} for the three merger models (a,b,e) and the two remnant models (c,d), where the masses have been integrated over the simulated evolution time and normalized to the total ejected mass for each individual model. In the following, whenever discussing properties related to the r-process, ``initial'' indicates the network initiation time (defined in Sec.~\ref{sec_network_input}), not the conditions at the start of the hydrodynamical simulations.
$X_n^0$ provides a better rough indicator of the efficiency of the r-process than the $Y_e$ alone \citep[which is commonly used in the literature when discussing the conditions characterizing the r-process efficiency, e.g., see][]{Lemaitre2021}. 
By definition, a lower $Y_e$ gives a larger $X_n^0$, as can be seen in Fig.~\ref{fig_xn0_ye_distr}; however, a higher initial entropy will also give a larger $X_n^0$ due to the release of free neutrons through photodissociation \citep[see, for example,][for a discussion of the interplay between the $Y_e$, entropy, expansion time-scale and a successful r-process]{meyer1989,hoffman1997,Otsuki2000}.
For merger ejecta conditions with similar expansion time scales, it is only possible to produce the heaviest elements with a high $X_n^0$, while it is not always safe to conclude that heavy elements are only produced for $Y_e\lesssim 0.25$ because it is possible to have a successful r-process for $Y_e\gtrsim 0.25$ if the entropy is sufficiently high. 
As can be seen in Fig.~\ref{fig_xn0_ye_distr}(e),  the entire $X_n^0$ distribution lies above 0.8 for the NS-BH ejecta, which contrasts to the wide $X_n^0$ distribution obtained in the NS-NS simulations (Fig.~\ref{fig_xn0_ye_distr}(a,b)) that include neutrino interactions (see \citet{kullmann2021} for a comparison to a ``no-neutrino" case).  Therefore, the NS-BH scenario is well suited to study the impact of fission uncertainties on the final r-process abundance distribution since the r-process flow will reach the fissile region for most of the ejected material in this astrophysical scenario \citep[see][for a discussion of fission and the relation to the $X_n^0$ value]{Lemaitre2021}. 

\section{r-process calculations and nuclear input}
\label{sec_network_input}

The astrophysical simulations provide the detailed evolution of, among other variables, the temperature, density and entropy of the ejecta up to a few tens of ms or seconds, depending on the model. After that, the ejecta is assumed to expand according to a homologous expansion, i.e., with a constant velocity, and the densities of the ejecta clumps decrease proportionally to $1/t^3$.  The r-process calculations are performed in a post-processing step starting from the initial composition given by the astrophysical environment and use the detailed evolution of the ejected mass elements (i.e., trajectories) to follow the expansion and abundance evolution up to $\sim 1$ year. After this time, all unstable nuclei, except the long-lived $^{232}$Th and $^{238}$U cosmochronometers, are assumed to instantaneously decay to their stable descendants (either a stable nucleus, $^{232}$Th, or $^{238}$U). Thus, what we refer to as final abundances herein are those obtained after the decay of all radioactive species (except for the long-lived isotopes of Th and U).

We define the network initiation time to be the first time when both the temperature and density of a given trajectory drop below the threshold values of $T=10\times 10^{9}$ K and $\rho_\mathrm{drip}\simeq 4.2\times 10^{11}$~g~cm$^{-3}$, respectively. Depending on the specific history of the trajectory followed and the astrophysical scenario modelled, the temperature may already be lower than 10~GK at the first time step, and in these cases, the network calculation starts as soon as the density drops below the drip density $\rho_\mathrm{drip}$.
The alternative situation may also be found (in particular for the BH-torus models) where the densities start initially below $\rho_\mathrm{drip}$; in this case, the network is initiated when the temperature falls below 10~GK.
After the initiation time, the network includes the possible re-heating of the ejecta through $\beta$-decays, fission, and $\alpha$-decays so that the temperature evolution is calculated following the approach of \citet{meyer1989} using the laws of thermodynamics.
The initial abundance of the heavy nuclei for each trajectory is determined by nuclear statistical equilibrium (NSE) at the given density (referred to as $\rho_\mathrm{net}$),  temperature and electron fraction $Y_e$ at the initiation time. 

The nucleosynthesis is followed by a reaction network consisting of $\sim 5000$ species, ranging from protons up to $Z\simeq110$ and including all isotopes from the valley of beta stability to the neutron drip line\footnote{For the mass models adopted in this work the number of nuclei in the network ranges between 4263 and 4808.}. 
Elements with higher proton numbers than 110 are assumed to fission spontaneously with very short lifetimes. The production of elements with higher proton numbers with the nuclear models adopted here (see Sec.~\ref{subsec_fission} in particular) is unlikely and $Z=110$ is assumed to be a good termination point\footnote{Due to the excessive computational demand of varying the nuclear physics input for all hydrodynamical models, several of the calculations for the subset of NS-BH models were run with $Z=100$ as the maximum limit for the network. The impact of this limitation has been tested and found to be negligible while significantly reducing the CPU time.} of the r-process network (see also the discussion about the so-called ``fission roof'' for $Z=110$ in \citet{Lemaitre2021}).
For all these nuclei, many nuclear ingredients and processes are required to calculate the abundance evolution. 
Our r-process reaction network includes all charged particle fusion reactions on light and medium mass isotopes, photodisintegrations, beta-delayed neutron emission probabilities and the rates of radiative neutron capture, $\beta$-decay and $\alpha$-decay. 
If the network reaches trans-Pb species, we also include neutron induced, spontaneous and $\beta$-delayed fission and their corresponding fission fragment distributions for all fissile nuclei. Each fissioning parent is linked to the daughter nucleus, and the emitted neutrons may be recaptured by the nuclei present in the environment. If a fission fragment is located outside the neutron drip line, it is assumed to emit neutrons to reach the neutron drip line instantaneously.  
Whenever experimental data is available, it is included in the network. Reaction rates on light species are taken from the NETGEN library \citep{xu2013}, which includes the latest compilations of experimentally determined rates. 

Nuclear predictions are affected by systematic as well as parameter uncertainties. The former ones, also referred to as model uncertainties, are known to dominate for unstable nuclei \citep[e.g.,][]{Goriely2014a} since there is usually no or very little experimental information available to constrain the model on those nuclei. Whenever experimental data are available, either for rates or reaction model ingredients, they are considered in the theoretical modelling. In this case,  they also constrain the possible range of variation of the model parameters and reduce the impact on the model as well as parameter uncertainties. However, for exotic nuclei, models of different natures, ranging between macroscopic to microscopic approaches \citep[e.g.,][]{arnould2007,Goriely15b,Hilaire16, Goriely17a}, may provide rather different predictions.
For this reason, systematic global uncertainties affecting reaction or decay rates will be considered in the present study. Those systematic global uncertainties are propagated to the calculation of reaction rates and consistently applied to the nucleosynthesis simulations to estimate their impact on the r-process yields and decay heat.

\begin{figure}
\begin{center}
\includegraphics[width=\columnwidth]{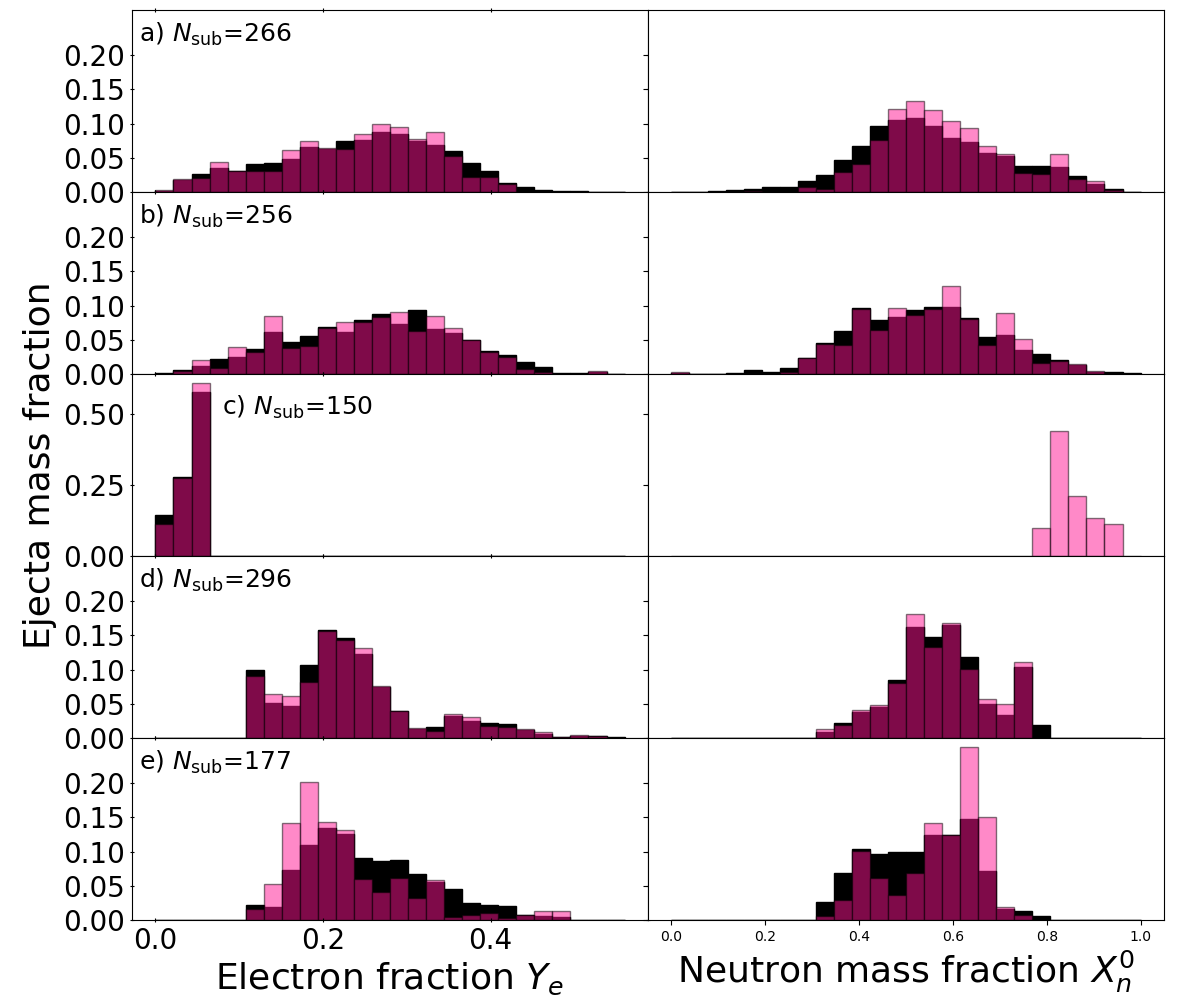}
\caption{(colour online)
Comparison of the electron fraction $Y_e$ (left) and $X_n^0$ (right) distributions for the selected trajectory subset (pink) with the complete set of trajectories (black) for models a) SFHo-125-145, b) SFHo-135-135, c) SFHo-11-23, d) M3A8m1a5, and e) M3A8m3a5-v2. Here, the black distributions have been normalized to the total ejected mass of each model (listed in Table~\ref{tab_astro_mods}), while each subset distribution has been normalized to the summed mass of the $N_\mathrm{sub}$ trajectories within the distribution. The $X_n^0$ values for model SFHo-11-23 have only been calculated for the subset.
}
\label{fig_NsubYeXn0}
\end{center}
\end{figure}

\begin{figure}
\begin{center}
\includegraphics[width=\columnwidth]{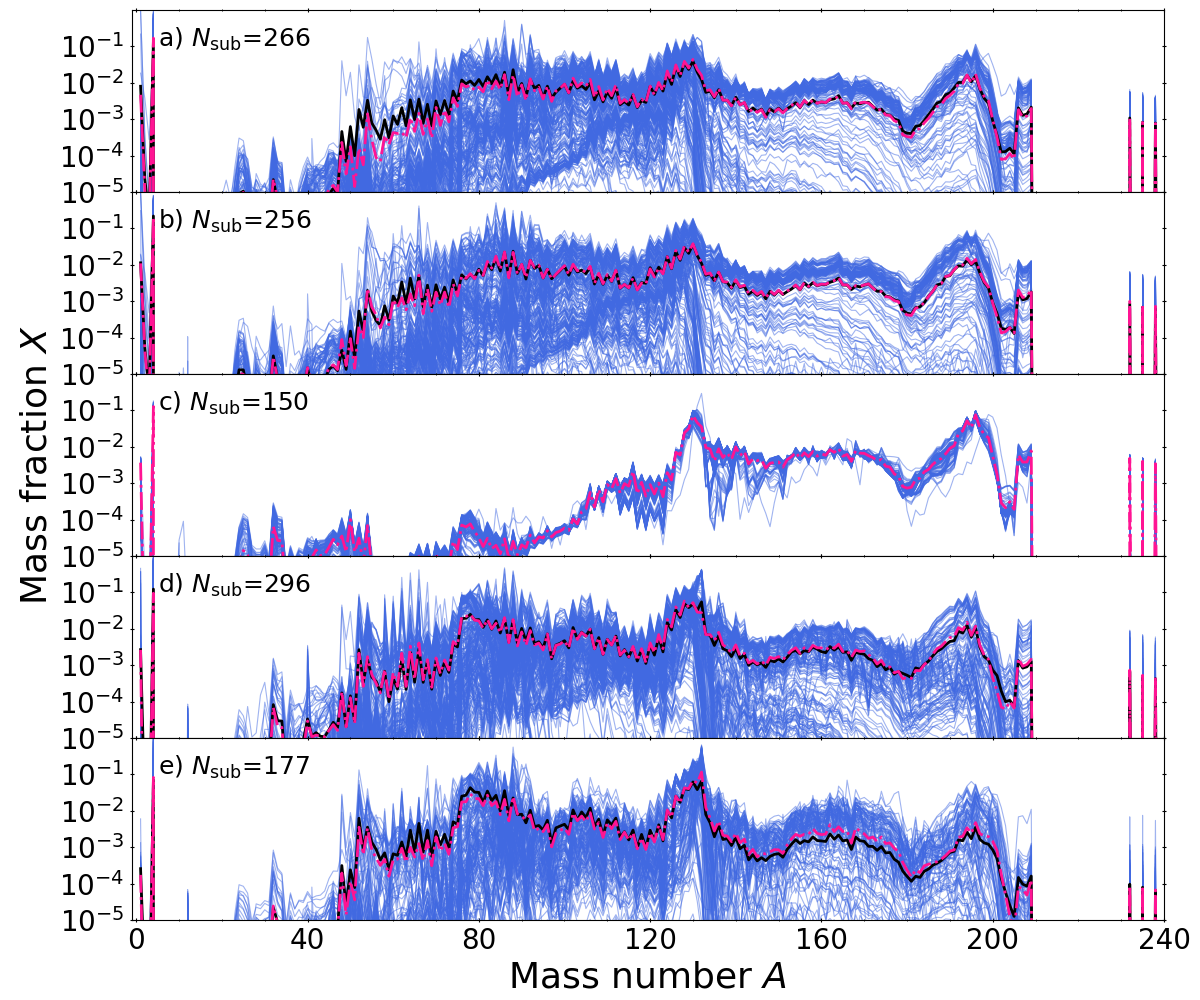}
\caption{(colour online)
Comparison of the final mass fractions $X$ for the selected trajectory subset with the complete set of trajectories for models a) SFHo-125-145, b) SFHo-135-135, c) SFHo-11-23, d) M3A8m1a5, and e) M3A8m3a5-v2. The blue lines are the mass fractions of the $N_\mathrm{sub}$ trajectories in the subset, the pink dashed-dotted line is the mass-averaged sum of the blue lines, and the black dashed line is the mass-averaged sum using the total number of trajectories.
}
\label{fig_NsubX}
\end{center}
\end{figure}

This paper aims to quantify the impact of nuclear physics uncertainties on the r-process nucleosynthesis. We have performed r-process nucleosynthesis calculations on all available trajectories for each hydrodynamical simulation using a default set of the nuclear input. When adopting a variation of the default input, we have calculated the nucleosynthesis abundances using a smaller subset of trajectories to reduce the computational demand. The subset of trajectories has been chosen in such a way that they contain 15 per cent or more\footnote{The NS-BH model uses only 1 per cent of the total ejecta mass.} of the total mass (see Table~\ref{tab_astro_mods}) while still reproducing the complete trajectory set in a satisfactory way. Here, `satisfactory' refers to the ability of a model subset to reproduce the shape of the $Y_e$, $X_n^0$, entropy and velocity mass distributions of the total ejecta, in addition to reproducing the final results, i.e., mass fraction distributions and heating rates, obtained with the full set. This is illustrated in Figs.~\ref{fig_NsubYeXn0} and \ref{fig_NsubX}, which show the $Y_e$ and $X_n^0$ mass distributions and the mass fraction distributions, respectively, for the subsets and when using the complete set of trajectories for the five hydrodynamical models. An even better correspondence is found between the heating rates of the trajectory subsets and the complete set. Such subsets are consequently considered as representative of the full set.


\begin{table}
\centering
\caption{
An overview of the nuclear physics input varied in the r-process nuclear network calculations in this work: model input set, mass model, direct capture (DC) component (yes if included), the model for the $\beta$-decay rates, fission barriers (Barr.) and fission fragment distributions (Frag.).  We use input set 1 as ``default'' in our calculations. See the text for the model references.}
\begin{tabular}{llccccc}
\hline \hline 
Set & Mass mod. & DC & $\beta$ mod. & Barr.  &  Frag. \\ 
\hline 
1 & BSkG2 & no & RHB+RQRPA & HFB-14 & SPY\\ 
2 & FRDM12 & no & RHB+RQRPA & HFB-14 & SPY\\ 
3 & WS4 & no & RHB+RQRPA & HFB-14 & SPY\\ 
4 & D1M & no & RHB+RQRPA & HFB-14 & SPY\\ 
5 & HFB-21 & no & RHB+RQRPA & HFB-14 & SPY\\ 
6 & HFB-31 & no & RHB+RQRPA & HFB-14 & SPY\\ 
\hline 
7 & HFB-31 & no & HFB21+GT2 & HFB-14 & SPY\\ 
8 & HFB-31 & no & TDA   & HFB-14 & SPY\\ 
9 & HFB-31 & no & FRDM+QRPA & HFB-14 & SPY\\ 
10 & FRDM12 & no & HFB21+GT2 & HFB-14 & SPY\\ 
11 & FRDM12 & no & TDA   & HFB-14 & SPY\\ 
12 & FRDM12 & no & FRDM+QRPA & HFB-14 & SPY\\
\hline
13 & HFB-21 & no & RHB+RQRPA & MS99  & GEF \\ 
14 & BSkG2 & no & RHB+RQRPA & MS99  & GEF \\ 
\hline 
15 & HFB-21    & yes & RHB+RQRPA & HFB-14 & SPY\\
\hline \hline 
\end{tabular} 
\label{tab_nuc_mods}
\end{table}

Table~\ref{tab_nuc_mods} lists the nuclear models considered for our uncertainty study, as described in the following sections. Unless otherwise specified, our calculations use input set 1 as default, and for $\alpha$-decay, the model of \citet{Koura2002} has been used for all calculations.
Note that all of the global models listed in Table~\ref{tab_nuc_mods} have proven their ability to describe experimental data with a high degree of accuracy, i.e., with a root mean square (rms)
deviation below 0.8~MeV with respect to the 2457 $Z, N \ge 8$ experimentally known masses \citep{Wang2021} \citep[this threshold of 0.8~MeV is difficult to reach within a given physical model and only a few global models have shown their capacity to reach such an accuracy, especially within the mean-field approach; see, e.g.,][]{Lunney03}. Models that do not fulfil such a necessary requirement are not considered for nuclear applications \citep{capote2009} and therefore excluded from our comparison study (Sec.~\ref{sec_compare}). 
However, accurately describing experimental data is a necessary but not a sufficient condition for a model to be applied to the r-process nucleosynthesis 
\citep[for more details see][]{arnould2007}. In summary, in order to best meet the nuclear-physics needs of the r-process, we require the applied models to be both accurate with respect to experimental observables but also as reliable as possible, i.e., to be based on a physically sound model that is as close as possible to a microscopic description of the nuclear systems.
The first criterion is an objective measure, while the latter criteria may be seen as more subjective, namely that nuclear models which are as close as possible to solving the nuclear many-body problem on the scale of the nuclear chart have a stronger predictive power. 
This second criterion is, however, fundamental for applications involving extrapolation away from experimentally known regions, in particular towards exotic neutron-rich nuclei of relevance for the r-process.
For example, the masses obtained with the Hartree-Fock-Bogolyubov (HFB) model based on the SLy4 Skyrme interaction \citep{Stoitsov2003} reproduce the complete set of experimental masses with an rms deviation larger than 5 MeV. This is to be compared with the few effective interactions which describe, within the HFB approach, all known masses with an rms deviation smaller than 0.8 MeV \citep[see, e.g.,][]{Goriely2016c}.
Another example regarding the latter criteria concerns some recent mass models based on machine learning algorithms \citep[see e.g.,][]{Shelley2021}, which essentially consider mathematical rather than physical approaches to extrapolate from the known to the unknown masses.


\subsection{Radiative neutron capture rates}
\label{subsec_ncap}

Many nuclear structure properties are required to calculate the radiative neutron capture rates, including mass model,  nuclear level density, nuclear potential, $\gamma$-strength function, and for the non-resonant region, even the energy, spin and parity of discrete states are needed. 
In the following discussion, we will focus on two different methods, namely the statistical Hauser-Feshbach method and the direct capture (DC) model for the estimation of radiative neutron capture rates.   

\subsubsection*{Compound nucleus and Direct capture}

The statistical Hauser-Feshbach reaction model is used as default in astrophysical applications to calculate the radiative neutron capture cross-sections for experimentally unknown nuclei. 
Within the Hauser-Feshbach method, the capture process is assumed to be a two-step process. First, a so-called compound nucleus (CN) is formed as an intermediate step. Second,  the nucleus de-excites to the ground state of the residual nucleus by emitting a particle or a $\gamma$-ray.
In this model, the CN is assumed to be in thermodynamic equilibrium so that the energy of the incident particle is shared uniformly by all the nucleons before de-excitation. This assumption is expected to hold if we assume that the level density of the compound nucleus at the energy of the incident energy is large, and then the compound nucleus has an average statistical continuum superposition of available resonances at this energy. 
Although the Hauser-Feshbach method has proven to accurately reproduce cross sections for medium- and heavy-mass nuclei, the model suffers from uncertainties originating from the underlying theoretical nuclear models, and the validity of its CN assumption should be questioned for some light and neutron-rich nuclei for which only a few or no resonant states are available. 

When the number of available states in the CN is relatively small, the neutron capture process may be dominated by the direct electromagnetic transition to a bound final state without forming a compound nucleus. 
The DC process has been shown to be non-negligible compared to the Hauser-Feshbach component and can even contribute up to 100 times more to the total cross-section for the most neutron-rich nuclei close to the neutron drip line \citep{sieja2021}. 
The DC model also suffers from large model uncertainties mainly due to the remarkable sensitivity of the cross-section to the few available final states, which for the very neutron-rich nuclei are unknown. Moreover, the energy, spin and parity of the discrete levels can modify the DC contribution by many orders of magnitude since the selection rules, which rely on the spin and parity differences, may switch on or off the DC component \citep{Xu2012}.

It is normally assumed that the statistical and direct processes may contribute to the radiative neutron capture rate in a non-exclusive way. It remains, however, fundamental to use the same set of nuclear-structure ingredients to estimate both contributions. In particular, the same optical potential needs to be adopted to ensure the same total reaction cross-section is found in both channels.
According to the study of \citet{Xu2014}, the experimental radiative neutron capture cross-sections are in good agreement with the DC model for the lightest nuclei, where the CN model overestimates the contribution.  For the experimentally unknown neutron-rich nuclei, the DC component contributes significantly to the total cross-section or even dominates over the CN contribution in some regions. 

\begin{figure}
\begin{center}
\includegraphics[width=\columnwidth,height=5cm]{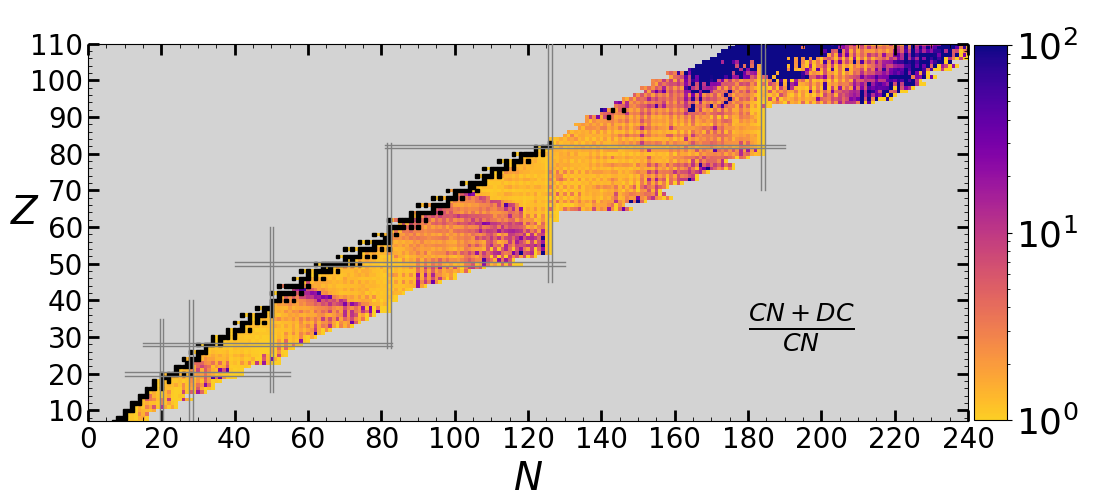}
\caption{(colour online)
Representation in the ($N$,$Z$) plane of the ratio between the Maxwellian-averaged $(n,\gamma)$ rates estimated within the CN statistical method only and when also including the DC component. The neutron capture rates are calculated with the TALYS code \citep{goriely2008,koning2012,Xu2014} using the HFB-21 mass model at $T = 10^9$ K for all nuclei from $Z=8$ up to $Z=110$ between the valley of $\beta$-stability and neutron drip line. The black squares correspond to the stable nuclei and the double solid lines depict the neutron and proton magic numbers.
Note that all values above $10^2$ are displayed in blue, which is relevant for the $Z>90$ region. 
}
\label{fig_ratesdiff}
\end{center}
\end{figure}

Both the DC and CN methods for calculating the theoretical radiative neutron capture rates are consistently included in the TALYS code \citep{goriely2008,koning2012,Xu2014}, which we use to estimate the rates. The reverse photoneutron emission rates, i.e., the $(\gamma,n)$ rates, are calculated using detailed balance, which includes an exponential dependence on the $Q$-value of the nuclei involved. Therefore, a large part of the sensitivity to the radiative neutron capture rates originates directly from uncertainties in the prediction of the neutron separation energy, hence of nuclear masses.  Deviations in the mass predictions on the order of several MeV (see Sec.  \ref{subsec_massmods}) translate into deviations in the neutron capture rates up to 3-5 orders of magnitude in certain regions of the nuclear chart.
In Fig.~\ref{fig_ratesdiff}, a comparison between the neutron capture rates obtained with the CN and CN plus DC component is displayed.  We can see that the largest contribution from the DC component is for the neutron-rich nuclei and in certain regions in between the nuclear magic numbers ($Z\sim 35-40$ and $Z\sim 50-70$).  There are also significant discrepancies in the fissile region ($Z>90$), where the CN+DC component is over $100$ times larger than the CN component. 

As default, all our r-process calculations are based on CN model rates, except for one variation of the nuclear inputs, which also includes the DC component (see Table~\ref{tab_nuc_mods}, input set 15).
In this paper, we have used the microscopic HFB plus combinatorial nuclear level densities  \citep{goriely2008} and the E1 and M1 D1M+QRPA  strength functions (with the inclusion of an empirical upbend) \citep{Goriely18a}.
The cross-section calculations are believed to be quite insensitive to varying the nuclear potential as long as they have the same volume integral per nucleon \citep{goriely1997b,Xu2012} and if the isovector imaginary potential remains large enough to ensure the neutron absorption \citep{Goriely07b}. We will therefore use the same nuclear potential, the Woods-Saxon-type optical potential \citep[KD;][]{koning2003} for all calculations in this paper.

\subsection{Nuclear mass models}
\label{subsec_massmods}


The mass of the nucleus is one of the most crucial ingredients  on which many nuclear properties depend.  Although great experimental progress has been made in the last decades to reach the exotic neutron-rich region and future large facilities such as the Facility for Rare Isotope Beams (FRIB) \citep{Surman18}, the Facility for Antiproton and Ion Research (FAIR) \citep{Walker2013} and the Radioactive Isotope Beam Factory (RIBF) at RIKEN \citep{Li2022} will certainly help to provide experimental masses deep inside the neutron-rich region,  it will not be feasible to measure the masses of the nuclei of most importance to the r-process in the years to come. Thus, our nucleosynthesis calculations will have to continue to rely on theoretical masses in the foreseeable future.

In order to compare mass models and measure their performance, it is common to calculate the rms deviation between a given mass model and all of the experimentally measured masses. 
Ideally, a mass model should be able to provide not only the mass but also the other nuclear properties like charge radii, quadrupole moments, fission barriers, shape isomers, as well as infinite nuclear matter properties. 
This way, it would be possible to evaluate the models performance not only based on its rms deviation from the known masses but also according to other nuclear properties and constraints. The mass models considered here have rms deviations from about 0.3 to 0.8 MeV on all the 2457 $Z, N \ge 8$  known masses  \citep{Wang2021}.

Many different approaches exist in order to calculate nuclear masses ranging from the first macroscopic classical models (e.g., liquid drop model) to microscopic models only relying on first principles (e.g., shell model). In between these two extremes, there are many semi-empirical approaches where a theoretical description of the nucleus is combined with free parameters which are fitted to the known masses (and sometimes other nuclear properties). 
In the following, we will focus on global mass models that try to reproduce the masses of all nuclei lying between the proton and neutron drip lines of relevance for astrophysical applications. For a review, see, e.g., \citet{Pearson00,Lunney03}.
Some of these models are described below.

\subsubsection*{Macroscopic-microscopic approach}

The classical liquid drop model describes the nucleus as consisting of nucleons that behave like the particles in a drop of liquid.  Based on this model, the semi-empirical mass formula (also known as the Bethe-Weizs\"acker formula) \citep{VonWeizsacker1935} was developed and shown to be quite successful in reproducing the general trends observed in nuclear data. However, it fails to describe quantum effects, and therefore several macroscopic-microscopic models have been proposed where microscopic corrections are added to the liquid drop part to account for the quantum shell and pairing correlation effects.
In this framework, the macroscopic and microscopic features are treated independently, both parts being connected exclusively by a parameter fit to experimental masses. The most sophisticated version of this macroscopic-microscopic mass formulas is the ``finite-range droplet model" (FRDM)  \citep{Moller2016}.  
The calculations are based on the finite-range droplet macroscopic model and the folded-Yukawa single-particle microscopic correction. The 31 independent mass-related parameters of the FRDM model are determined directly from a least-squares adjustment to  the ground-state masses on all the masses available at that time. 
The latest FRDM12 fit leads to a final rms deviation of 0.61~MeV for the 2457 $Z,N \geq 8$ nuclei with experimental masses.
Inspired by the Skyrme energy-density functional (see below), the so-called Weizs\"acker-Skyrme (WS) macroscopic-microscopic mass formula was proposed by \citet{Wang10,Wang2014,Liu11} with an rms deviation of about 0.3~MeV, from now on referred to as WS4. In such an approach, the mass formula is mathematically corrected by including a Fourier spectral analysis examining the deviations of nuclear mass predictions to the experimental data at the expense of a huge increase in the number of free parameters and potentially a decrease in the predictive power of the model.

Despite the great empirical success of the macroscopic-microscopic approach, it suffers from major shortcomings, such as the incoherent link between the macroscopic part and the microscopic correction, the instability of the mass prediction to different parameter sets, or the instability of the shell correction \citep{Pearson00,Lunney03}. 
The quality of the mass models available is traditionally estimated by the rms error obtained in the fit to experimental data and the associated number of free parameters. 
However, this overall accuracy does not imply a reliable extrapolation far away from the experimentally known region since models may achieve a small rms value by mathematically driven or unphysical corrections or can have possible shortcomings linked to the physics theory underlying the model.  
As discussed above, the reliability of the mass extrapolation should be considered as the second criterion of first importance when dealing with specific applications such as astrophysics, but also more generally for the predictions of experimentally unknown ground and excited state properties. For this reason, microscopic mass models have been developed, as discussed below.

\subsubsection*{Mean-field approach}

\begin{figure*}
\begin{center}
\includegraphics[width=\textwidth]{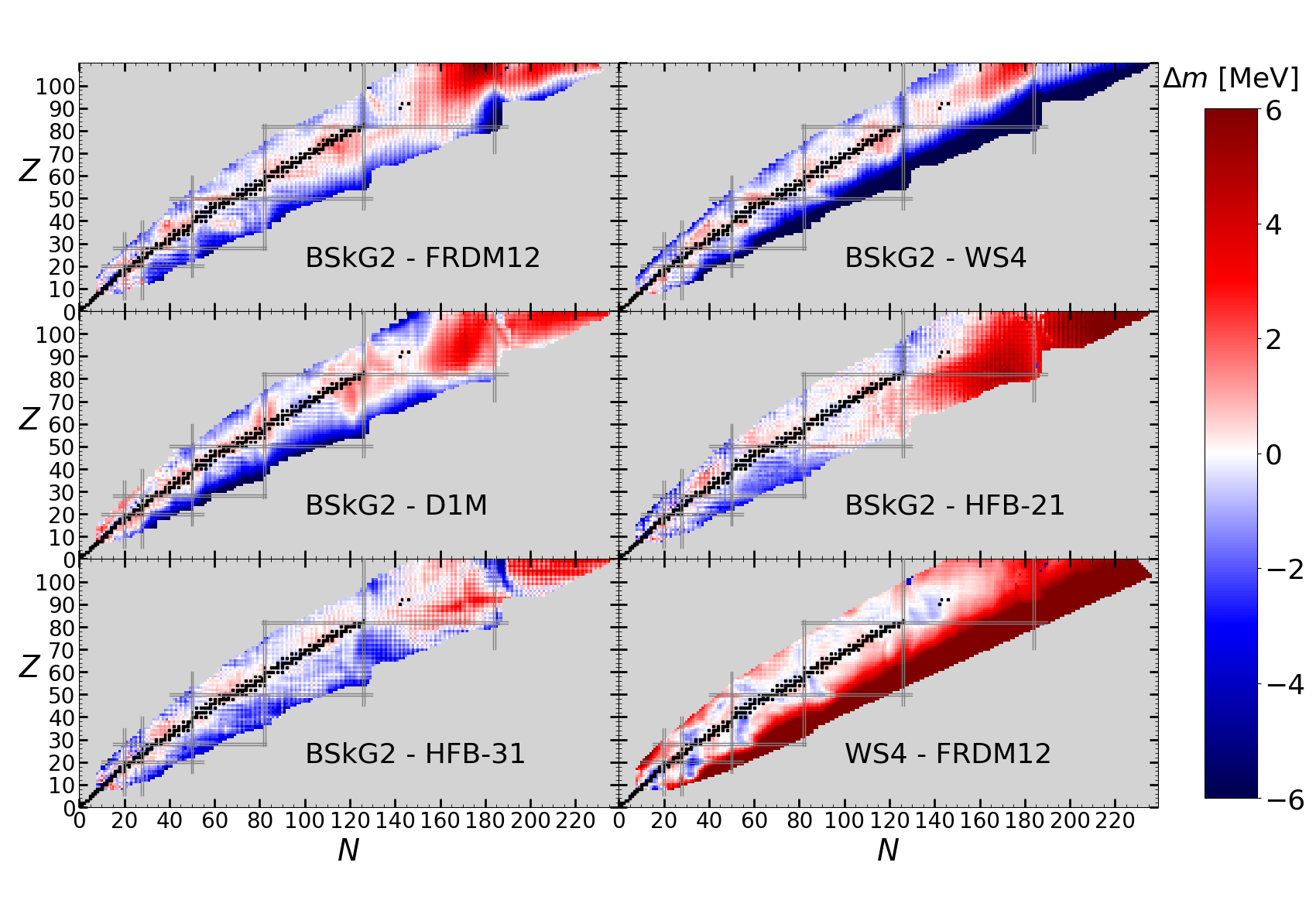}
\caption{(colour online) 
Similar to Fig.~\ref{fig_ratesdiff} for mass difference $\Delta m$ between different mass models listed in Table~\ref{tab_nuc_mods}.  Mass differences larger or smaller than the maximum or minimum values of the colour bar ($\pm 6$ MeV) are represented by red and blue, respectively. 
}
\label{fig_massdiff}
\end{center}
\end{figure*}

One of the most microscopic and successful models of the nucleus is the shell model, where the interaction between each individual nucleon (or each nucleon outside of a rigid core) is taken into account. However, due to the impossible task of solving the many-body problem for a large number of nucleons, e.g., due to the extreme computational demand, such a model can only be applied to light systems. Instead of calculating the interaction between all nucleons exactly, effective potentials describe the interaction of a nucleon with the mean-field generated by all the other nucleons. 
By using microscopic nuclear many-body models as a basis, relativistic or non-relativistic mean-field approaches based on the density functional theory can calculate all the masses of the entire nuclear chart.  However, fitting interaction parameters to essentially all mass data remains an extremely demanding task, and for this reason, most of these mean-field models have been adjusted on the properties of only a few nuclei. 
Despite their microscopic nature, some mean-field models may not reproduce the bulk of experimentally known masses and consequently not be suited for globally predicting masses. Such models were not developed for large-scale mass calculations and should therefore not be considered for r-process applications.

The underlying interactions of the microscopic models can also be used to calculate the EoS of infinite nuclear matter and neutron star material. This is an advantage for the mean-field  models since it gives another application of the theoretical framework and provides additional model constraints as, for example, the maximum NS mass or mass-radius relations \citep[e.g., see][]{fantina2013,pearson2018}.
In the long run, this approach may establish a consistency between the model of the network calculations and the EoS used in the astrophysical simulations, which is currently not considered.

In this work, we consider mean-field mass models that have been fitted to essentially all known masses and that are based on two different types of effective interactions, namely the Skyrme \citep{Vautherin72} and finite-range Gogny \citep{Gogny73} interactions. More specifically, we adopt the Gogny-D1M mass model \citep{Goriely2009b}, which takes into account all the quadrupole correlations self-consistently and microscopically and reproduces the 2457 experimental masses \citep{Wang2021} with an rms deviation of 0.81 MeV.
Many versions of the Brussels-Montr\'eal Skyrme-HFB mass models have been developed in a series of continuous improvements. 
We consider here the HFB-21 \citep{goriely2010} and HFB-31 \citep{Goriely2016c} mass models with an rms deviation of 0.59~MeV, since both have been used previously in nucleosynthesis calculations. In addition, we also consider the BSkG2 mass model \citep{Ryssens2022}, which is obtained through a three-dimensional coordinate-space representation of the single-particle wave functions allowing for both axial and triaxial deformations and treats nuclei with odd number of nucleons in the same way as the even-even nuclei by breaking time-reversal symmetry. Its rms deviation amount to 0.68 MeV with respect to the 2457 known masses. Unless otherwise specified, we use BSkG2 as our default mass model.

In Fig.~\ref{fig_massdiff}, the mass differences between BSkG2 and the other five mass models, as well as between both droplet-type parametrizations,  are displayed. The mass difference can be over 6 MeV in some regions, where the largest deviations generally lie close to the drip lines or in regions further from the neutron magic numbers for some models.

\subsection{$\beta$-decay rates }
\label{subsec_betarates}

$\beta$-decay rates remain crucial for the r-process nucleosynthesis since they define the time scale for the flow of abundance from one $Z$ to the next.
At later times, after freeze-out, the $\beta$-decay and $\beta$-delayed neutron emission probabilities play an essential role in determining the flow back to the valley of $\beta$-stability and also the energy release relevant for the kilonova light curve. 
Although $\beta$-delayed neutron emission occurs throughout the duration of the r-process, it is in particular important during the late phases when the competition between neutron captures and $\beta$-decays shape the final r-process abundance distribution, in particular around the r-process peaks. 

As with the other nuclear input properties required by the r-process, the experimental $\beta$-decay rates are included when available \citep{kondev2021}, but for almost all nuclei produced during the neutron irradiation, we have to apply theoretical rates from one of the few available global models. 
Similar to the mass models, microscopic shell model calculations for $\beta$-decay exist. However, due to the computational cost, they are restricted to nuclei near closed shells and cannot yet tackle the heavy, neutron-rich nuclei required for r-process applications.
One widely used model is Gross Theory, which allows for fast and reasonably accurate computation of the half-lives given the input of appropriate $Q_\beta$-values. The Gross Theory is based on $\beta$-decay one-particle strength functions to describe the general behaviour of $\beta$-decay. The one-particle strength functions and the pairing scheme has been improved in its GT2 version \citep{tachibana1990} based on the HFB-21 $Q_\beta$-values.

Several global semi-microscopic models are based on an effective interaction to describe the nuclear structure and adopt the random phase approximation (RPA) or its quasiparticle (QRPA) extension to include pairing interactions to describe allowed Gamow-Teller decay.
The FRDM+QRPA \citep{moller2003} model uses the FRDM $Q$-values and a separable residual interaction for the QRPA description of the Gamow-Teller transitions and also includes the Gross Theory for the first-forbidden transitions.
One of the most reliable models available is the spherical relativistic Hartree-Bogoliubov (RHB) model plus relativistic QRPA (RQRPA) \citep{Marketin2016}  based on the DC3* residual interaction, which also includes first-forbidden transitions.  

Finally, the Tamm-Dancoff approximation (TDA) \citep{klapdor1984} is another global model available that provides a simple analytical solution to calculate the $\beta$-decay strength distribution by using a simplified Gamow-Teller residual interaction, neglecting the influence of first-forbidden transitions on the half-lives. 

\begin{figure*}
\begin{center}
\includegraphics[width=\textwidth]{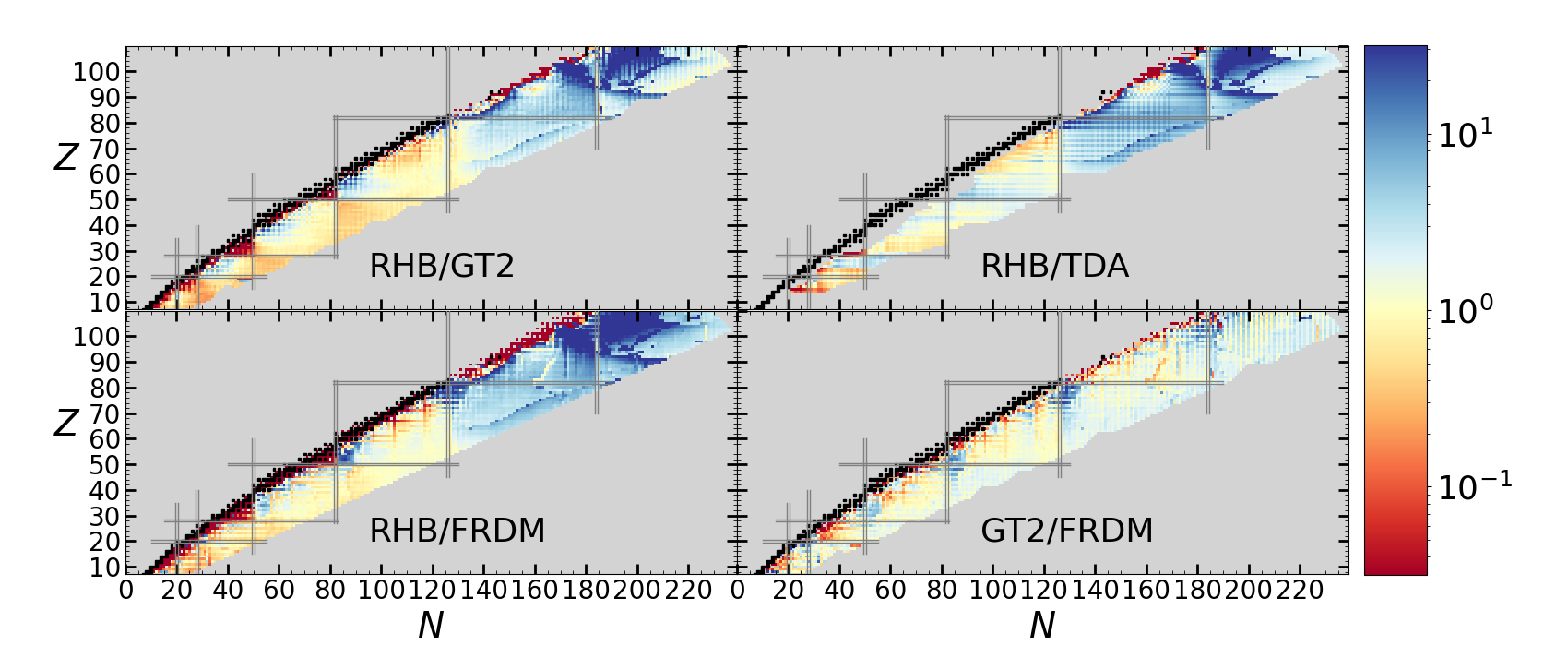}
\caption{(colour online)
Similar to Fig.~\ref{fig_ratesdiff} where the ratio between the $\beta$-decay rates predicted by the four models listed in Table~\ref{tab_nuc_mods} are plotted in the ($N$,$Z$) plane.  Differences larger or smaller than the maximum or minimum values of the colour bar ($10^{\pm 1.5}$ MeV) are represented by blue and red, respectively.  
}
\label{fig_betadiff}
\end{center}
\end{figure*}

In Fig.~\ref{fig_betadiff}, the ratio between the $\beta$-decay rates estimated by the relativistic mean-field (RMF), i.e., RHB+RQRPA, and the HFB21+GT2, TDA and FRDM+QRPA as well as the ratio between the HFB21+GT2 and FRDM+QRPA predictions are shown.  
Major differences can be seen for exotic neutron-rich nuclei, in particular for $N>126$, where, in particular, the RHB+RQRPA model is seen to predict larger rates compared to the other models.
We apply all four $\beta$-decay models in Fig.~\ref{fig_betadiff} in our r-process calculations, where the RHB+RQRPA model is used as default (see Table~\ref{tab_nuc_mods}). 
Note that when we change the applied mass model for our r-process calculations, it directly impacts the estimated neutron capture rates but not the $\beta$-decay rates since they are estimated separately by the models discussed above.  Thus, due to the limited set of available global $\beta$-decay models, we often have an inconsistency between the mass model applied in the r-process calculations and the masses used to calculate the $\beta$-decay rates. Therefore, a future prospective is to develop fully consistent models for all nuclear physics input applied in r-process calculations.

\subsection{Fission models for r-process applications}
\label{subsec_fission}

In sufficiently neutron-rich astrophysical scenarios ($X_n^0$ typically larger than 0.7), the r-process can reach nuclei in the $Z \ga 90$ region, where fission may become the dominant mode of decay. Therefore, if efficient, fission terminates the r-process because it stops the abundances from reaching even further up in the super-heavy region of the nuclear chart and recycles material back to the $100\lesssim A\lesssim 180$ region. Given sufficient time before freeze-out,  these daughter nuclei may again be involved in a series of neutron captures and $\beta$-decays before reaching the fissile region for the second time. This is often referred to as ``fission recycling'' \citep{goriely2011,goriely2013,goriely2015,goriely2015b,Mendoza-Temis2015,Vassh2019}.
In addition to producing daughter nuclei, fission also contributes to the r-process by releasing neutrons that may be recaptured by the various nuclei in the ejecta. This late neutron production can boost the neutron capture phase and impact the final abundance distribution after freeze-out. 
Another essential aspect of fission during the r-process is that it releases heat which can impact the kilonova light curve. 
In particular, the spontaneous fission of ${}^{254}$Cf with a half-life of 60 days has been shown to be important for energy generation and to impact the light curve at $t>100$ days \citep{wanajo2018,wu2019,zhu2018}.

\begin{figure*}
\begin{center}
\includegraphics[width=\textwidth]{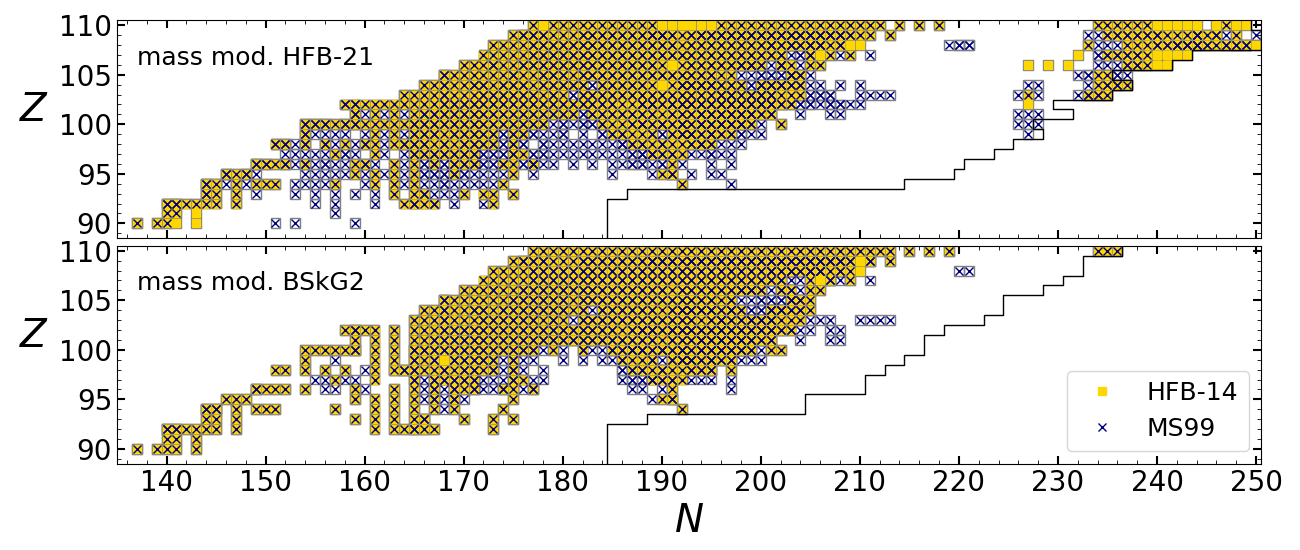}
\caption{(colour online)
The fission-dominated regions using the HFB-14 (gold squares) or MS99 (dark blue crosses) fission barriers, as used for input sets 1, 6, 15 and 17 in Table~\ref{tab_nuc_mods}. For each nucleus, a marker is drawn whenever the probability (or rate) for fission is larger than any other decay or reaction mode (i.e., whenever the probability (or rate) for spontaneous fission or $\beta$-induced fission is larger than $\beta$-decay, or neutron-induced fission is larger than neutron capture). The top panel uses mass model HFB-21, and the bottom panel BSkG2 for calculating the neutron capture rates, where the corresponding neutron drip line is shown as a black line.
}
\label{fig_fiss}
\end{center}
\end{figure*}

Fission is a very complex nuclear process, and a variety of nuclear structure inputs are required to achieve a successful model. For the r-process, we need models describing the probability for a nucleus to undergo neutron-induced, spontaneous and $\beta$-delayed fission, but also fission fragment distributions to determine the daughter nuclei and the number of free neutrons released. 
Photo-induced fission is assumed to be unimportant since the temperatures have typically fallen below 1 GK by the time the r-process network reaches the fissile region.
In this work, we have used either the HFB-14 fission paths based on the BSk14 Skyrme interaction\footnote{Ideally, the fission paths should be calculated based on the same interaction as the mass models applied in the r-process calculations (i.e.,  HFB-21 and so on,  see Table~\ref{tab_nuc_mods}).  However, fission barrier calculations are very expensive; therefore, only a few global models suitable for astrophysical applications are available.}
 \citep{goriely2010} or the Thomas-Fermi model \citep[][hereafter MS99]{Myers1999} to estimate the fission rates. 
Fig.~\ref{fig_fiss} shows the regions where fission processes dominate over other decay modes or neutron capture for the HFB-14 and MS99 fission barriers using either mass model HFB-21 or BSkG2 for the calculation of the neutron capture rates.
The MS99 barriers are usually lower than the predictions of the HFB model, so more nuclei are found to be affected by fission processes when the MS99 barriers are adopted. Note that the MS99 barriers are not always available for very exotic n-rich nuclei close to the neutron-drip line, especially for $Z \geq 106$. In these cases, the HFB-14 barriers are adopted. 
During the freeze-out phase, if we adopt the HFB-14 fission barriers, fission takes place around $Z \simeq 101 -102$ along the abundant $A \simeq 278$ isobar, while the lower MS99 barriers lead to fission already around the $Z \simeq 95 - 97$ isotopes.
The applied mass model impacts the balance between neutron-induced fission and neutron capture through the calculation of the neutron capture rates but also through the position of the neutron drip line since this determines which nuclei are available to fission. This is illustrated in Fig.~\ref{fig_fiss} where we can see that mass model HFB-21 allows for fission in a large region for $N>230$, while mass model BSkG2 predicts this region to be unstable for neutron emission. 

For the fission fragment distributions we adopt either the renewed microscopic Scission Point Yield (SPY) model \citep{Lemaitre2019,Lemaitre2021} or the 2018 version of the semi-empirical GEF model (''GEneral description of Fission observables'')  \citep{Schmidt2016}. Unless otherwise specified, we use the HFB-14 barriers and the SPY fission fragment distributions as default for our calculations.

\section{Impact on r-process nucleosynthesis}
\label{sec_nucuncert}

The following section will present the nucleosynthesis results obtained when varying the nuclear input discussed in Sec.~\ref{sec_network_input}. First, we show the r-process nucleosynthesis distributions and heating rates obtained using the astrophysical models for the dynamical and BH-torus ejecta separately before we discuss the results for the combined ejecta and the cosmochronometric age estimates for six metal-poor r-process-enhanced stars.
For each nuclear input set, all quantities (except those presented in Fig.~\ref{fig_rpro_onetraj}) are mass averaged over representative trajectories for each hydrodynamical model (see Sec.~\ref{sec_network_input} and Table~\ref{tab_astro_mods}). 

\subsection{Dynamical ejecta}

\subsubsection{NS+NS merger}

\begin{figure}
\begin{center}
\includegraphics[width=\columnwidth]{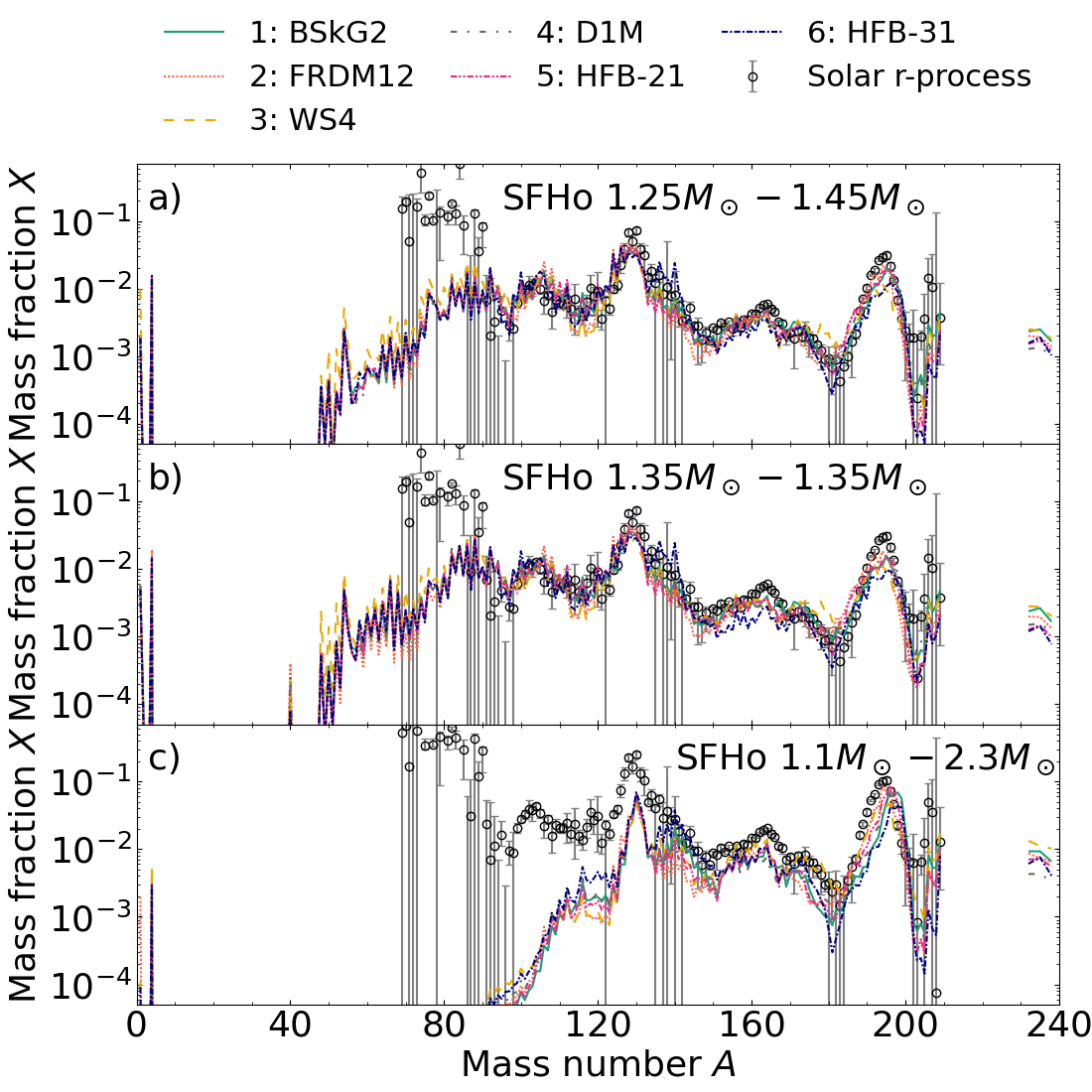}
\caption{(colour online) 
Final mass fractions of stable nuclei (and long-lived Th and U) of the material ejected as a function of the atomic mass $A$ for our three NS--NS and NS--BH dynamical ejecta merger models: a) SFHo-125-145, b) SFHo-135-135, and c) SFHo-11-23 when varying the applied mass model (i.e., input sets 1-6 in Table~\ref{tab_nuc_mods}).  See the text for references and details about the different mass models. 
The solar system r-abundance distribution (open circles) \citep{goriely1999} is shown for comparison and arbitrarily normalized at the third r-process peak using input set 1 in Table~\ref{tab_nuc_mods} of model SFHo-135-135 in a) and b), and SFHo-11-23 in c).
}
\label{fig_rpro_u1-7}
\end{center}
\end{figure}

\begin{figure}
\begin{center}
\includegraphics[width=\columnwidth]{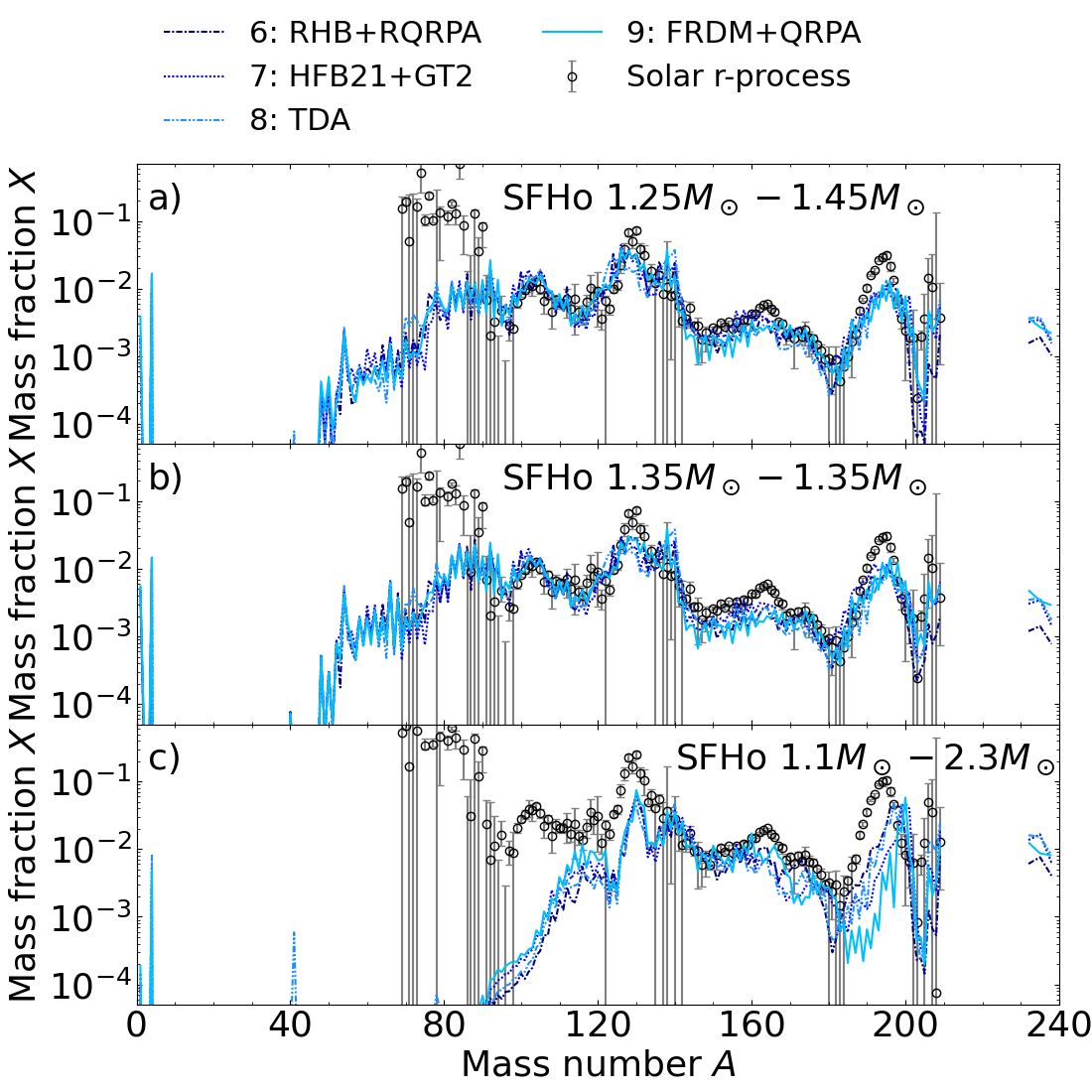}
\caption{(colour online)
Same as Fig.~\ref{fig_rpro_u1-7} when varying the four models for the $\beta$-decay rates, but using the same HFB-31 mass model (i.e., input sets 6-9 in Table~\ref{tab_nuc_mods}). See the text for references and details about the different models. 
The solar system abundance distribution is normalized as in Fig.~\ref{fig_rpro_u1-7}.
}
\label{fig_rpro_u7-10}
\end{center}
\end{figure}

\begin{figure}
\begin{center}
\includegraphics[width=\columnwidth]{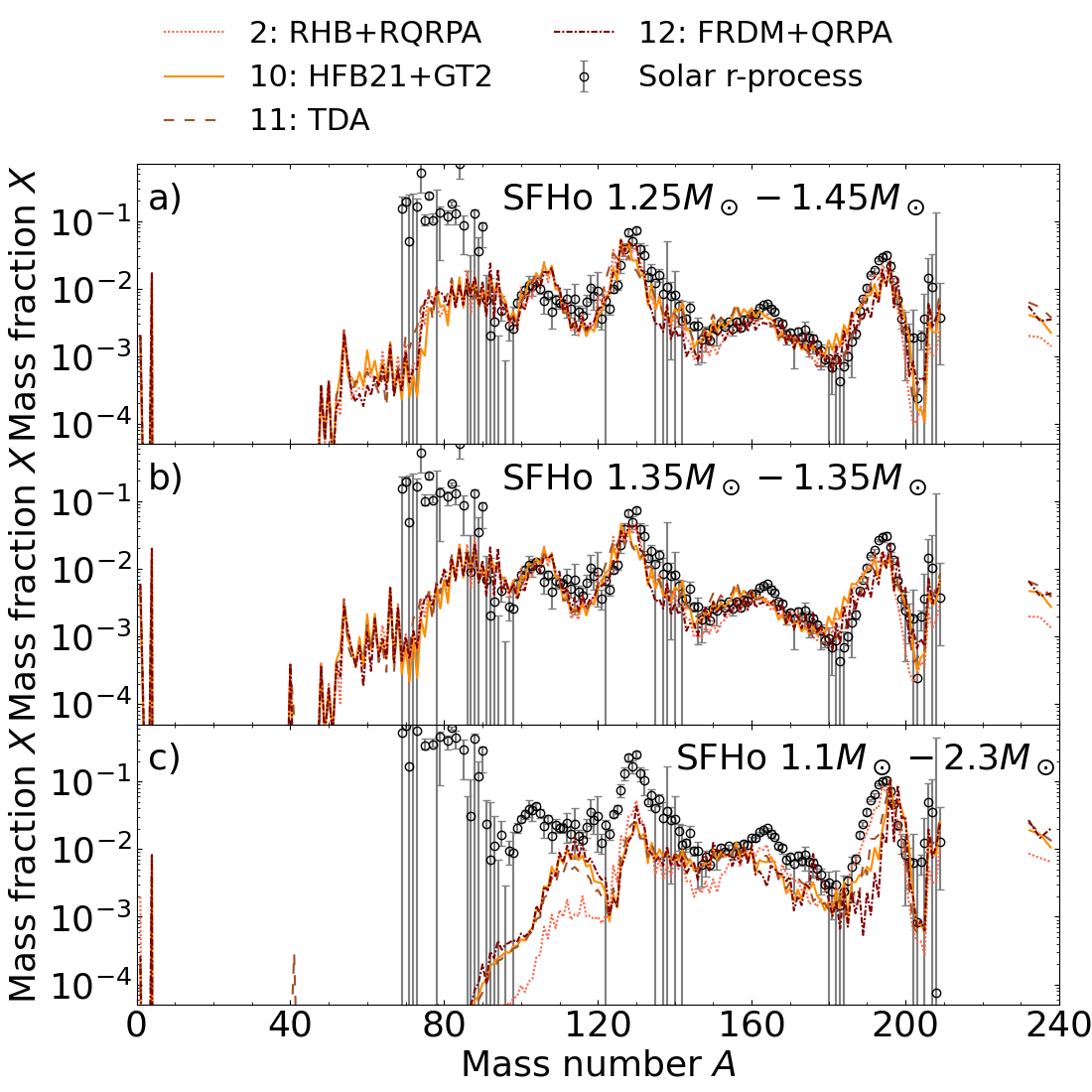}
\caption{(colour online)
Same as Fig.~\ref{fig_rpro_u1-7} when varying the four models for the $\beta$-decay rates, but using the same FRDM12 mass model (i.e., input sets 2 and 10-12 in Table~\ref{tab_nuc_mods}). See the text for references and details about the different models. 
The solar system abundance distribution is normalized as in Fig.~\ref{fig_rpro_u1-7}.
}
\label{fig_rpro_u3-13}
\end{center}
\end{figure}

\begin{figure}
\begin{center}
\includegraphics[width=\columnwidth]{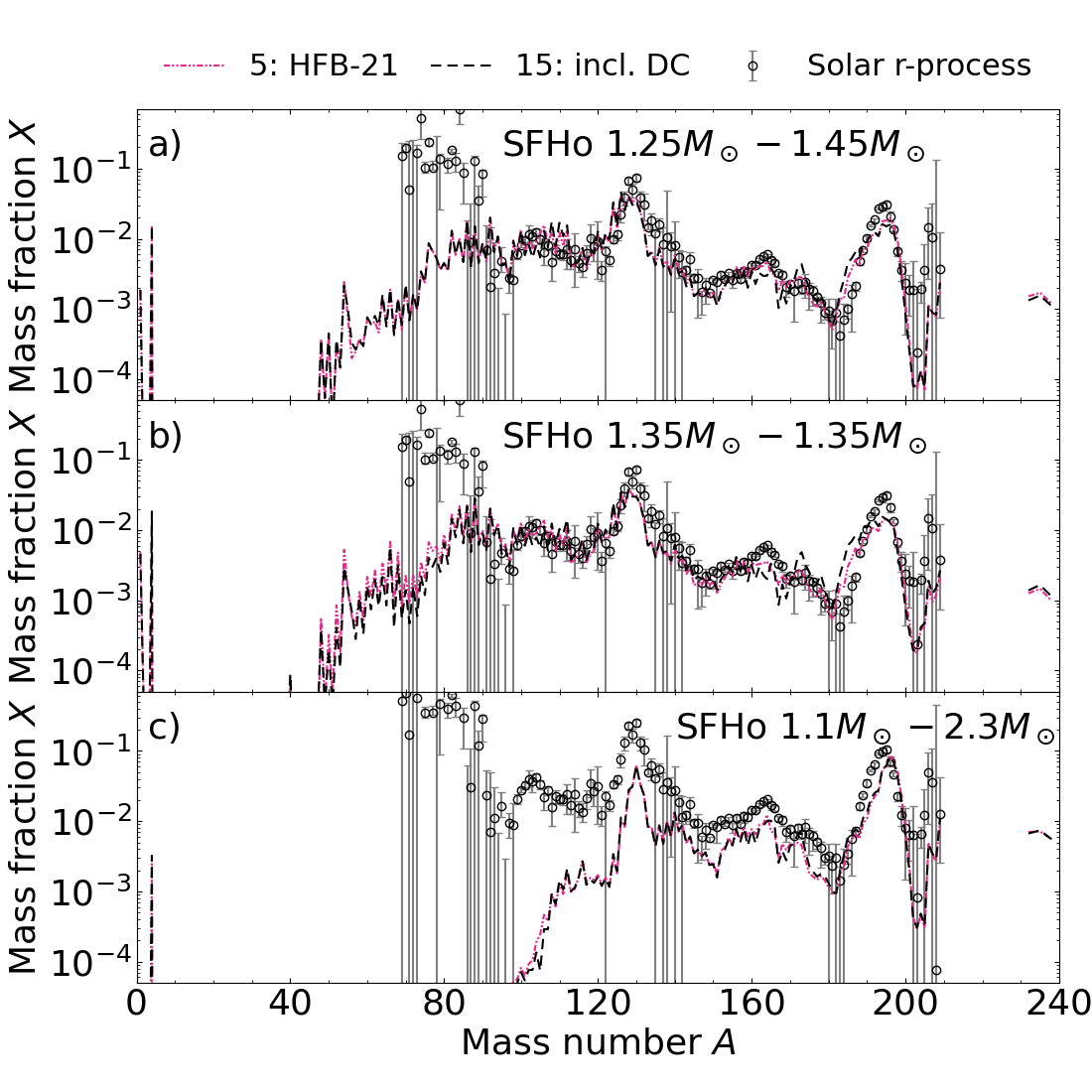}
\caption{(colour online)
Same as Fig.~\ref{fig_rpro_u1-7} when including the DC component for the radiative neutron capture rates, but using the same mass model, the HFB-21 model (i.e., input sets 5 and 15 in Table~\ref{tab_nuc_mods}). See the text for references and details about the different models. 
The solar system abundance distribution is normalized as in Fig.~\ref{fig_rpro_u1-7}.
}
\label{fig_rpro_uDC}
\end{center}
\end{figure}

\begin{figure}
\begin{center}
\includegraphics[width=\columnwidth]{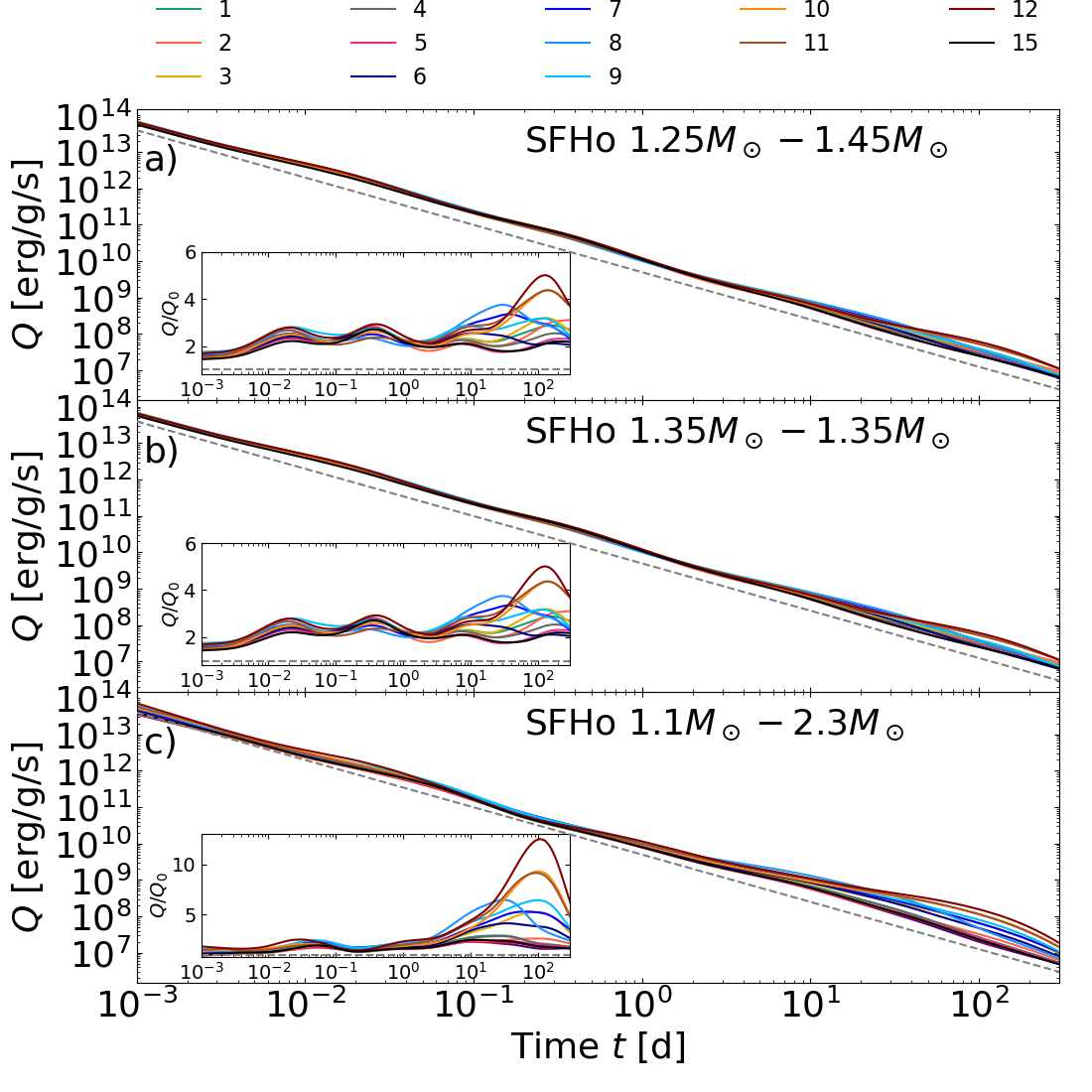}
\caption{(colour online)
Time evolution of the radioactive heating rate $Q$ (before thermalization) for our NS–NS and NS-BH dynamical ejecta merger models: a) SFHo-125-145, b) SFHo-135-135, and c) SFHo-11-23 when varying the nuclear input between all models listed in Table~\ref{tab_nuc_mods}.  
The grey dashed line corresponds to the global trend followed by $Q_0=10^{10}[t/1~\mathrm{day}]^{-1.3}$~erg/g/s. The ratio $Q/Q_0$ is also shown as an insert, where the grey dashed line indicates $y=1$.
}
\label{fig_Qt_all_dyn}
\end{center}
\end{figure}

In Fig.~\ref{fig_rpro_u1-7}(a,b) we show the r-process abundance distributions for the SFHo symmetric and asymmetric merger models, respectively, computed with six different mass models (see Table~\ref{tab_nuc_mods}). 
Except for two mass models, HFB-31 around $A\sim 130$ and WS4 around $A\sim 180$, the other mass models agree globally well and follow the solar system distribution for $A>90$.  
All models produce a significant amount of actinides, ranging from 0.2-0.7\% of the total abundance. Local differences, in particular around the third r-process peak, can be observed in relation to the prediction of the strength of the $N=126$ closed shell.

The calculations based on the HFB-31 mass model (Fig.~\ref{fig_rpro_u7-10}), shows an `extra'' peak around $A\sim 130-140$. 
This over-abundance stems from a significantly stronger odd-even effect predicted around $N=90$ for HFB-31 compared to the other mass models considered.
Similar to what can be observed at closed neutron shells,  the larger $S_{2n}$ values at $N=90$ lead to    abundance accumulating during the r-process simulation at this isotone, which is later transformed into the $A\sim 130-140$ peak after neutron freeze-out.

Similarly, Figs.~\ref{fig_rpro_u7-10}(a,b) and \ref{fig_rpro_u3-13}(a,b) give the abundance distributions calculated using four different models for the $\beta$-decay rates applying the two mass models HFB-31 and FRDM12, respectively (see Table~\ref{tab_nuc_mods}).  Compared to the mass models, the differences are larger between the $\beta$-decay models, particularly for $A>130$.  
The actinide production is larger when applying the FRDM12 masses (Fig.~\ref{fig_rpro_u3-13}), ranging from $1.3$ to $1.7$\%, compared to using HFB-31 masses, which result in $0.4$ to $1.1$\% of actinides.

Fig.~\ref{fig_rpro_uDC} compares the difference between the abundance distributions when the radiative neutron capture rates are estimated with and without the DC component for hydrodynamical models SFHo-125-145 and SFHo-135-135. The most considerable differences between both sets of rates are found around $A\sim150$, where the model including the DC component has a peak structure (double lanthanide peak), and in particular, for model SFHo-125-145, the third r-process peak is also broader when the DC is included.

The heating rate is defined here as the radioactive energy release rate per unit of mass due to $\beta$-decay, $\alpha$-decay, and fission:
\begin{equation}
Q = Q_\beta + Q_\alpha + Q_\mathrm{fis},
\end{equation}
not including the contribution lost into neutrino emission $Q_\nu$, and 
globally follow the approximated trend given by
\begin{equation}
Q_0=10^{10}[t/1~\mathrm{day}]^{-1.3} \mathrm{erg/g/s}.
\end{equation}
The effective heating rate $Q_\mathrm{heat}$ measures the fraction of radioactive energy that is converted into thermal energy in the expanding ejecta, i.e., the energy that actually powers the kilonova, defined as:
\begin{equation}
Q_\mathrm{heat} = f_\mathrm{therm}\cdot (Q + Q_\nu),
\label{eq_Q_heat}
\end{equation}
where the total thermalization efficiency $f_\mathrm{therm}$ is calculated as in  \citet{Just2021} (which is based on \citet{Rosswog2017} and \citet{barnes2016}).

Fig.~\ref{fig_Qt_all_dyn}(a-b) shows the time evolution of the heating rate ($Q$) for all combinations of nuclear models (see Table~\ref{tab_nuc_mods}) for the dynamical ejecta of the NS-NS systems. 
All nuclear models predict roughly the same r-process heating. 
After $t>30$ days, the contribution to the heating rate from fission differs between the nuclear input cases, where the largest heating is found when using mass model FRDM12 and either FRDM+QRPA, HFB21+GT2 or TDA $\beta$-decay rates. 

\subsubsection{NS-BH merger}

\begin{figure}
\begin{center}
\includegraphics[width=\columnwidth]{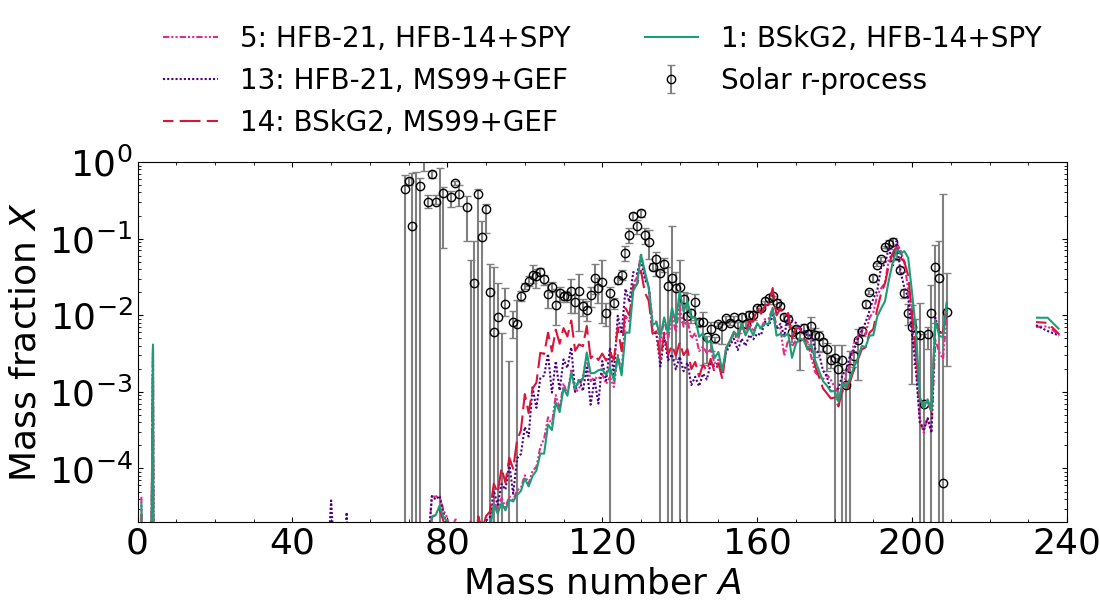}
\caption{(colour online)
Same as Fig.~\ref{fig_rpro_u1-7}c when varying the fission barriers and fragment distributions, using mass model HFB-31 or BSkG2 (i.e., input sets 1, 5, 13 and 14 in Table~\ref{tab_nuc_mods}) for the material ejected by our NS-BH merger model SFHo-11-23.  See the text for references and details about the different nuclear models. 
The solar system abundance distribution is normalized as in Fig.~\ref{fig_rpro_u1-7}c.
}
\label{fig_rpro_ufiss}
\end{center}
\end{figure}

Fig.~\ref{fig_rpro_u1-7}c displays the composition estimated when varying the six mass models for the NS-BH merger model SFHo-11-23. We can see that compared to the dynamical ejecta of the NS-NS merger systems (panels a and b), the NS-BH merger ejecta undergoes a strong r-process and follows the solar system r-process distribution for $A\ga 140$ (compared to $A \ga 90$ for NS-NS systems). The astrophysical conditions lead to a weak second r-process peak and an additional structure around $A\sim140$ due to fission processes (see below); however, the strength of this structure varies between the mass models and is found to be particularly strong for mass model HFB-31.

When varying the models for the $\beta$-decay rates in Fig.~\ref{fig_rpro_u7-10}c (using mass model HFB-31), the structure around $A\sim 140$ is present with the same strength, which was not always the case for the NS-NS merger models. 
The material ejected from the NS-BH merger model has initial conditions that favour a larger production of fissile material compared with the NS-NS merger models considered here and are therefore well suited to study the impact of varying the fission properties.
In Fig.~\ref{fig_Qt_all_dyn}c, we can see that the impact of fission on the heating rate at $t>30$d is more prominent than in the NS-NS merger models.  

Fig.~\ref{fig_rpro_ufiss} shows that fission barriers and fragment distributions significantly affect the predicted abundance distributions (and proportionally more than varying the mass model). For example, all models predict about the same amount of actinides (2--2.5\%); however, BSkG2+MS99+GEF (brown line) predicts a sharper peak at $A\sim 165$ than the other models.  The asymmetric nature of the fission fragment distribution for HFB-14+SPY, particularly around $A\simeq 278$ isobars, is also found to impact the final abundance distribution in the $A\simeq 140$ region, as already pointed out by \citet{goriely2013}.


\subsection{BH-torus ejecta}

\begin{figure}
\begin{center}
\includegraphics[width=\columnwidth]{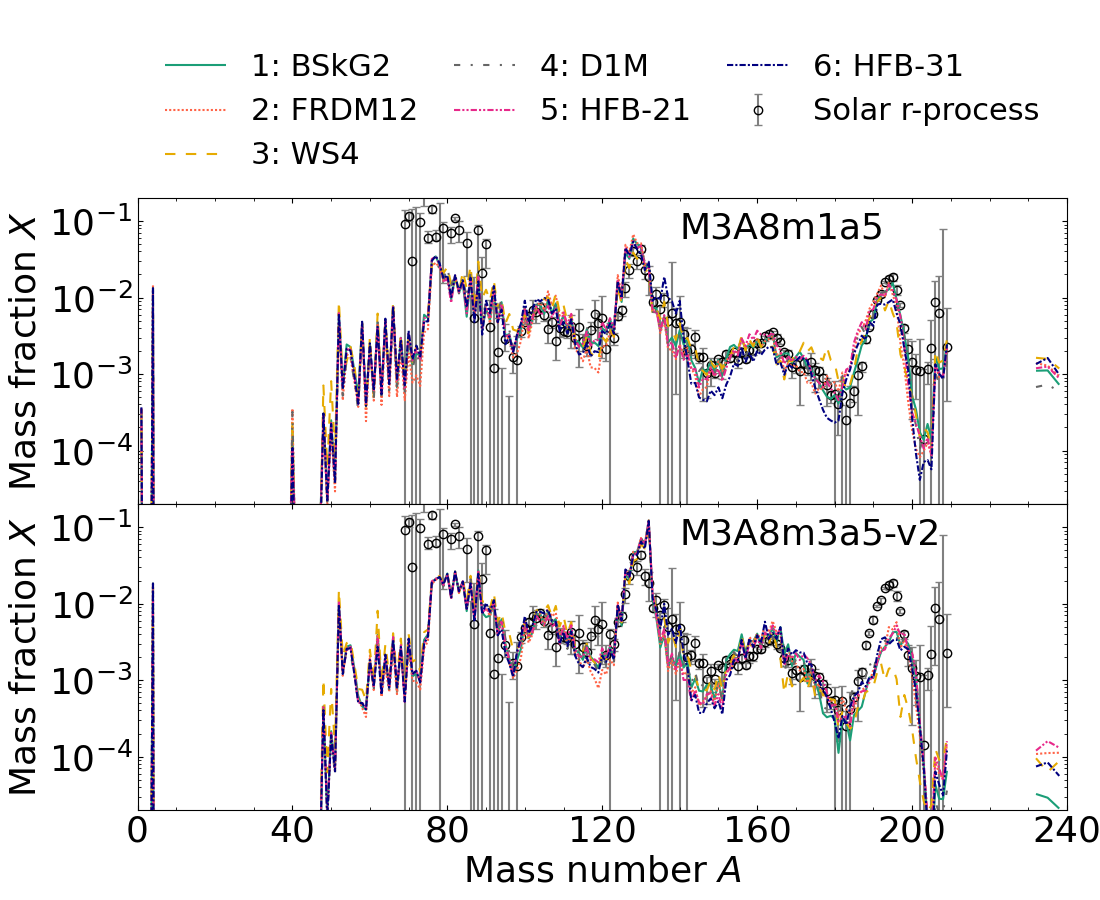}
\caption{(colour online)
Final mass fractions of the material ejected as a function of the atomic mass $A$ for our two BH-torus models (M3A8m1a5 and M3A8m3a5-v2) when varying between six mass models (i.e., input sets 1-6 in Table~\ref{tab_nuc_mods}). See the text for references and details about the different models. 
The solar system r-abundance distribution (open circles) from \citet{goriely1999} is shown for comparison and arbitrarily normalized at the third r-process peak of model M3A8m1a5.
}
\label{fig_rprowind_u1-7}
\end{center}
\end{figure}

\begin{figure}
\begin{center}
\includegraphics[width=\columnwidth]{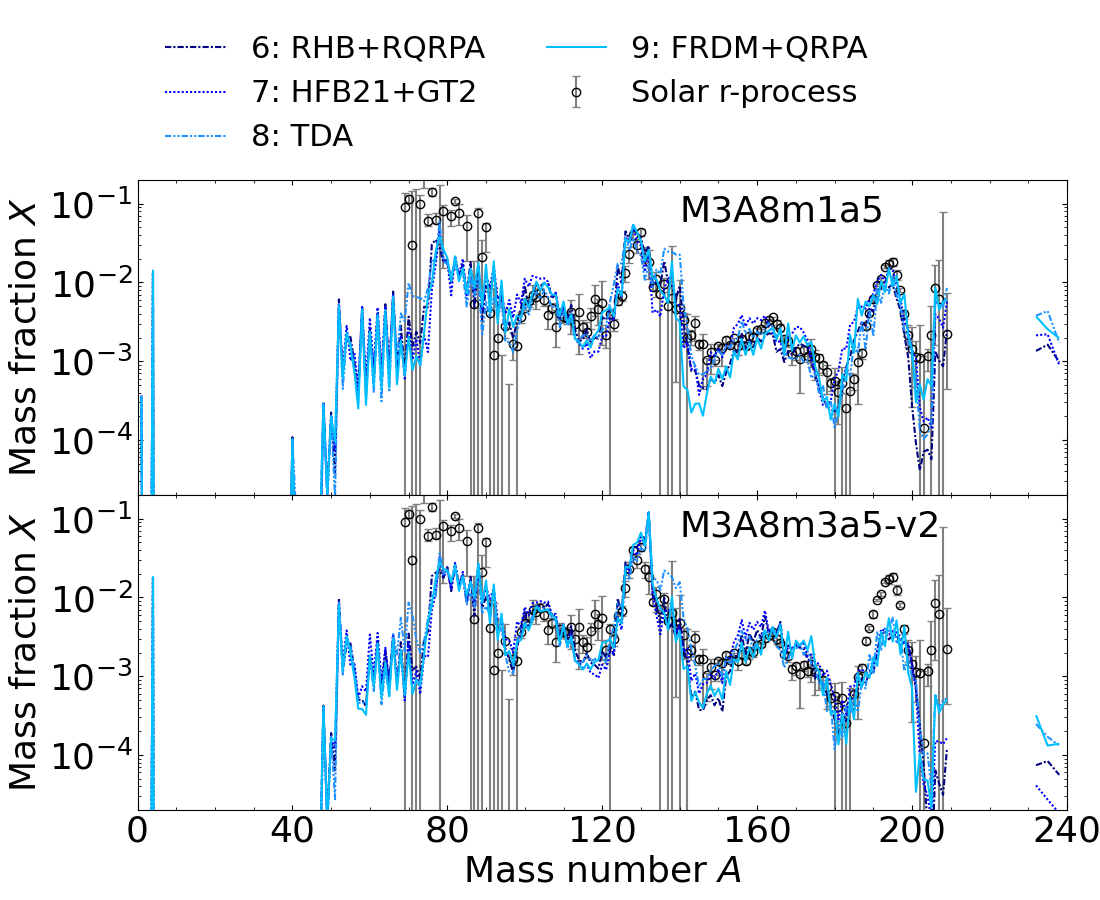}
\caption{(colour online)
Same as Fig.~\ref{fig_rprowind_u1-7} when varying the models for the $\beta$-decay rates, but using the same HFB-31 mass model (i.e., input sets 6-9 in Table~\ref{tab_nuc_mods}). See the text for references and details about the different models. 
The solar system abundance distribution is normalized as in Fig.~\ref{fig_rprowind_u1-7}.
}
\label{fig_rprowind_u7-10}
\end{center}
\end{figure}

\begin{figure}
\begin{center}
\includegraphics[width=\columnwidth]{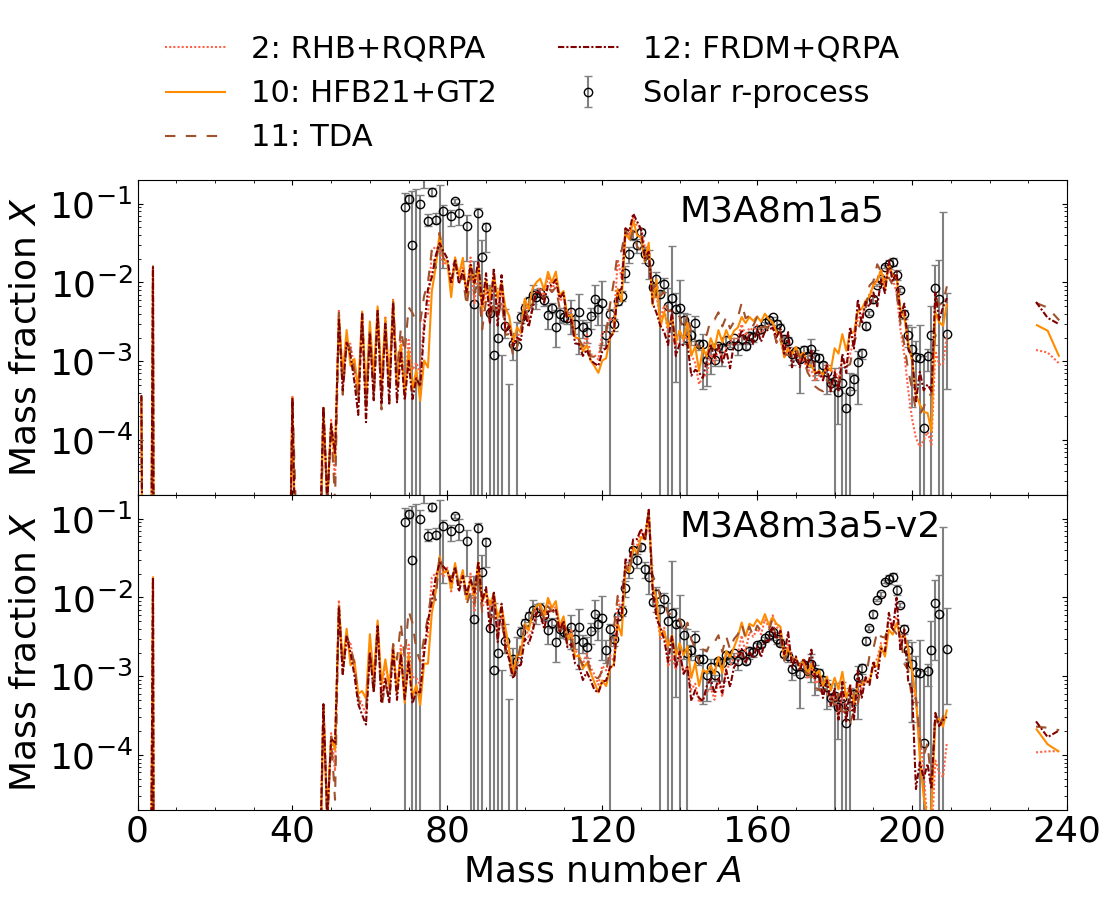}
\caption{(colour online)
Same as Fig.~\ref{fig_rprowind_u1-7} when varying the models for the $\beta$-decay rates, but using the same FRDM12 mass model (i.e., input sets 2 and 10-12 in Table~\ref{tab_nuc_mods}). See the text for references and details about the different models. 
The solar system abundance distribution is normalized as in Fig.~\ref{fig_rprowind_u1-7}.
}
\label{fig_rprowind_u3-13}
\end{center}
\end{figure}

\begin{figure}
\begin{center}
\includegraphics[width=\columnwidth]{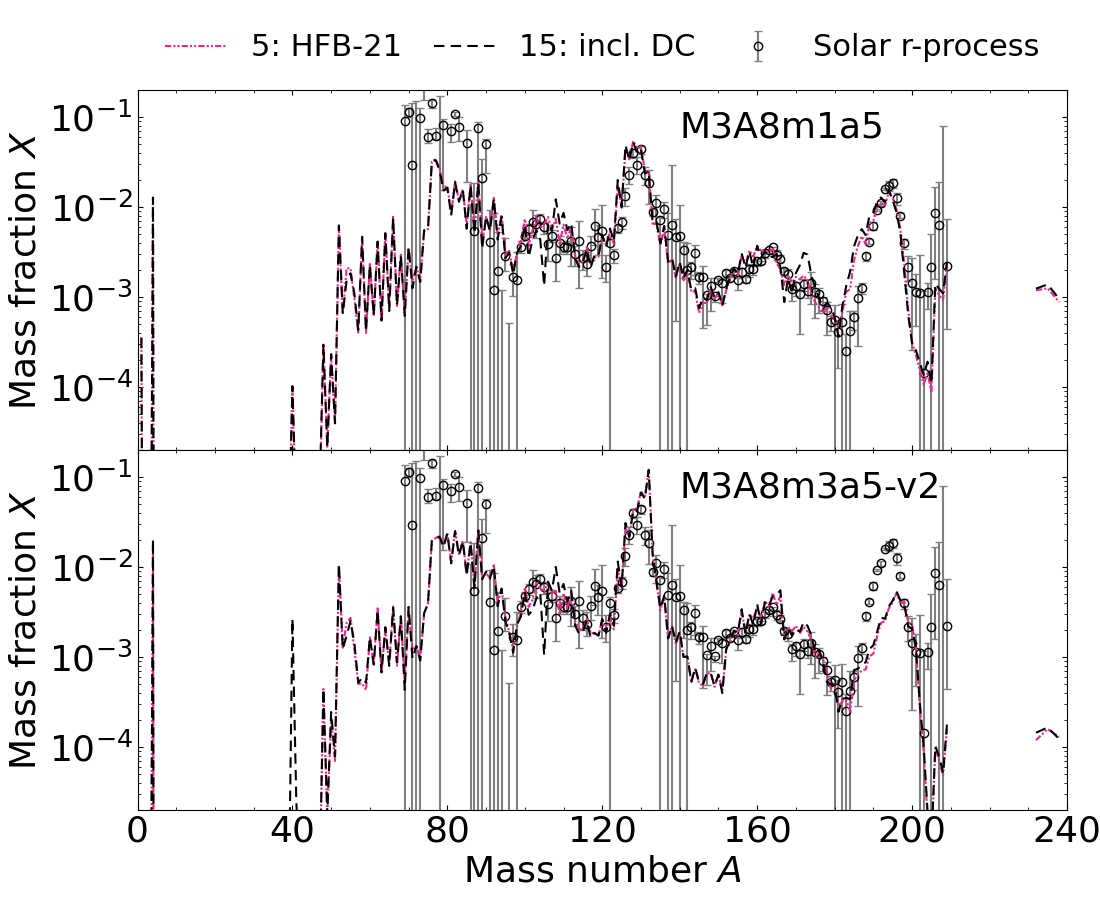}
\caption{(colour online)
Same as Fig.~\ref{fig_rprowind_u1-7} when including the DC component for the radiative neutron capture rates, but using the same mass model, the HFB-31 model (i.e., input sets 5 and 15 in Table~\ref{tab_nuc_mods}). See the text for references and details about the different models. 
The solar system abundance distribution is normalized as in Fig.~\ref{fig_rprowind_u1-7}.
}
\label{fig_rprowind_DC}
\end{center}
\end{figure}

\begin{figure}
\begin{center}
\includegraphics[width=\columnwidth]{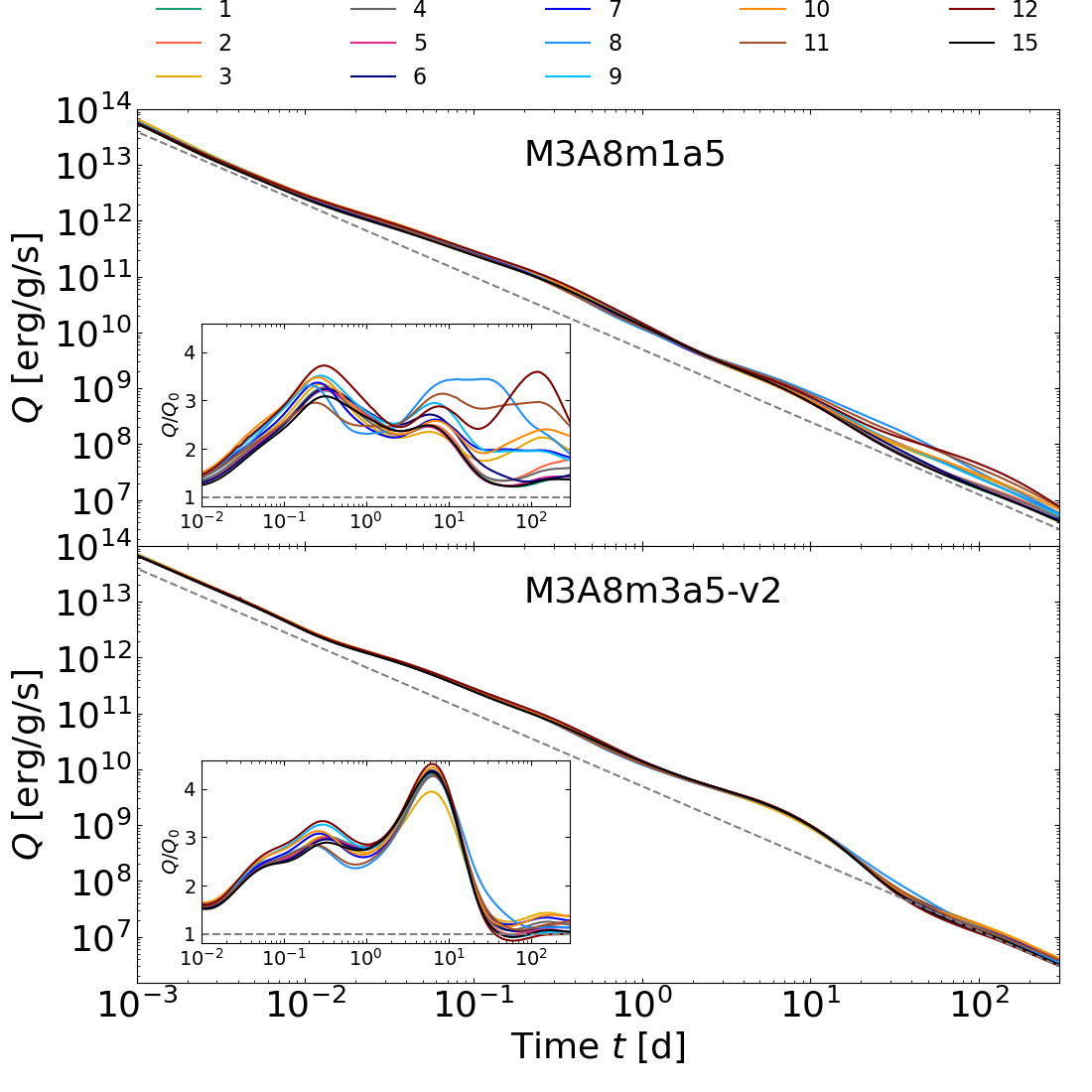}
\caption{(colour online)
Time evolution of the radioactive heating rate $Q$ (before thermalization) for the two hydrodynamical BH-torus models M3A8m1a5 and M3A8m3a5-v2 when varying the nuclear input between all models listed in Table~\ref{tab_nuc_mods}.
The grey dashed line corresponds to the approximation $Q_0=10^{10}[t/1~\mathrm{day}]^{-1.3}$~erg/g/s The ratio $Q/Q_0$ is also shown as an insert, where the grey dashed line indicates $y=1$.
}
\label{fig_Qt_all_wind}
\end{center}
\end{figure}

The material ejected in the post-merger phase from the torus surrounding the BH has quite different conditions than the dynamical ejecta (see Fig.~\ref{fig_xn0_ye_distr}). Therefore, the r-process distribution differs in its shape, as seen in Figs.~\ref{fig_rprowind_u1-7}--\ref{fig_rprowind_u3-13} for the various mass and $\beta$-decay models.  In general, the disk wind models produce more nuclei in the $A\sim80$ region, and in particular, model M3A8m3a5-v2 produces less of third peak elements and actinides, which is also where the most considerable differences between the mass models can be seen.  
The shape of the peak around $A\sim132$ also differs,  where a very narrow structure can be seen for the M3A8m3a5-v2 model\footnote{This feature is also observed in \citet{Wu2016}. It is related to trajectories, in which material falls back towards the torus and gets re-heated before it ultimately becomes ejected.}, compared to the wider peak for model M3A8m1a5 and the dynamical ejecta. 

Figs.~\ref{fig_rprowind_u7-10}-\ref{fig_rprowind_u3-13} show the r-process results when varying the four models for the $\beta$-decay rates and the two mass models HFB-31 and FRDM12, respectively.  Just as for the dynamical ejecta,  the HFB-31 mass model combined with the TDA or GT2 $\beta$-decay models overproduce abundances for nuclei around $A\sim 130-140$. 

The time evolution of the heating rate is shown in Fig.~\ref{fig_Qt_all_wind} for all the 15 nuclear models. We can see that the heating rate varies less for different nuclear models in the BH-torus ejecta than in the dynamical ejecta (Fig.~\ref{fig_Qt_all_dyn}).
The nuclear mass models only start to differ significantly at $t>10$ d for model M3A8m1a5 due to the more significant production of heavy and fissile r-process elements. Model M3A8m3a5-v2 peaks at around 0.2 and 10~d due to the $\beta$-decays of $A\sim80-90$ and $\sim140$ nuclei, respectively, and has a negligible contribution from fission and $\alpha$-decay at late times ($t>20$~d).  

\subsection{Combining dynamical and BH-torus ejecta}
\label{subsec_combej}

\begin{figure*}
\begin{center}
\includegraphics[width=\textwidth]{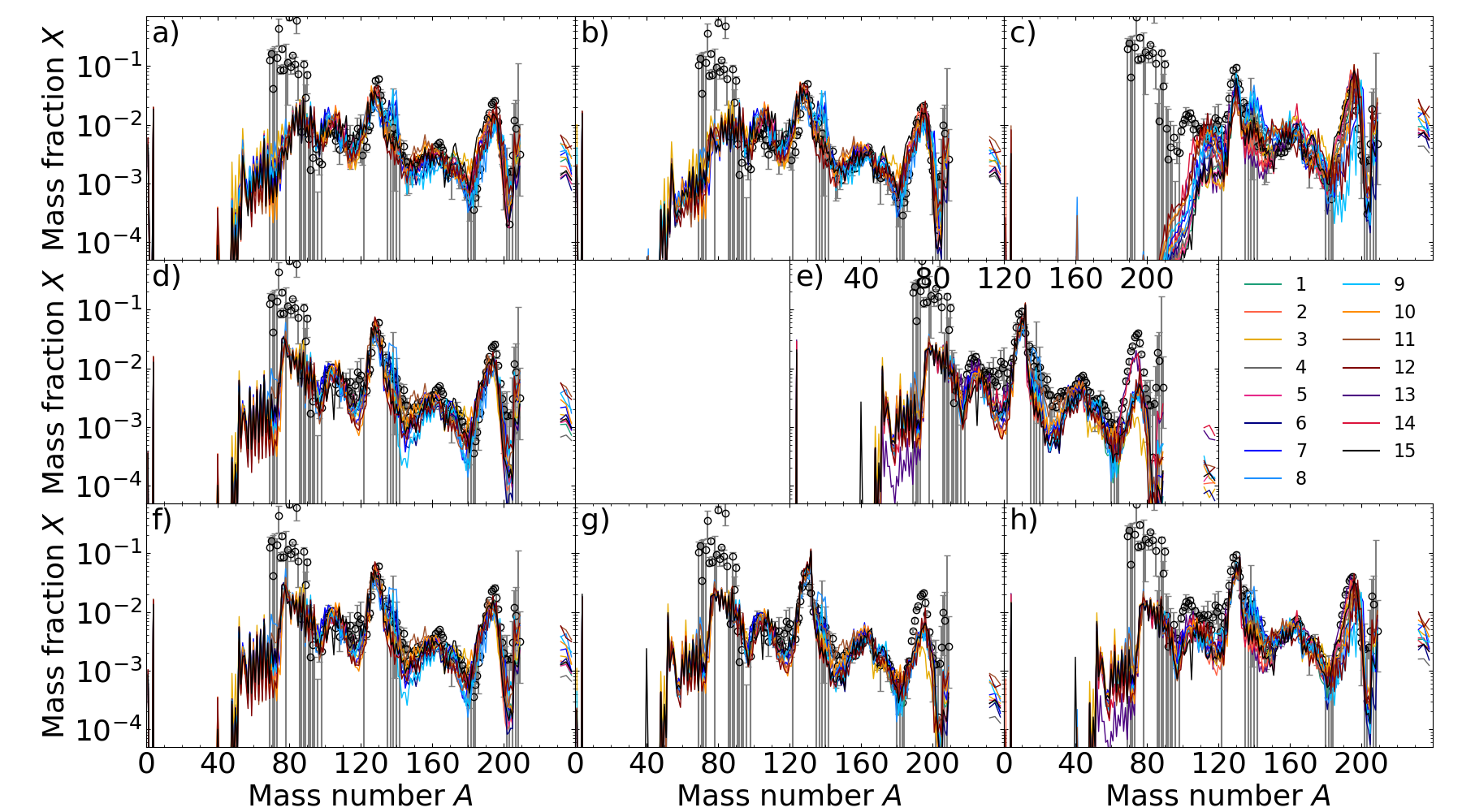}
\caption{(Colour online)
Final mass fractions of the material ejected as a function of the atomic mass $A$ when varying all nuclear models listed in Table~\ref{tab_nuc_mods} for each hydrodynamical model: a) SFHo-135-135, b) SFHo-125-145, c) SFHo-11-23, d) M3A8m1a5 and e) M3A8m3a5-v2. The total ejecta (bottom panels) are calculated by summing the dynamical ejecta (top panels) with the BH-torus ejecta (middle panels), weighted by their respective ejected masses. For example panel f) is calculated as $X_\mathrm{f}=(X_\mathrm{a} M_\mathrm{a}+X_\mathrm{d} M_\mathrm{d})/(M_\mathrm{a}+M_\mathrm{d})$, and similarly panel g) by combining b) and e), and h) by combining c) and e). 
The legend refers to the input sets in Table~\ref{tab_nuc_mods}.
The solar system r-abundance distribution (open circles) from \citet{goriely1999} is shown for comparison and arbitrarily normalized to the combined ejecta models (i.e., panel a, d use the normalisation from f, panel b from g, and panel c and e from h).
}
\label{fig_rpro_comb_all}
\end{center}
\end{figure*}

\begin{figure}
\begin{center}
\includegraphics[width=\columnwidth]{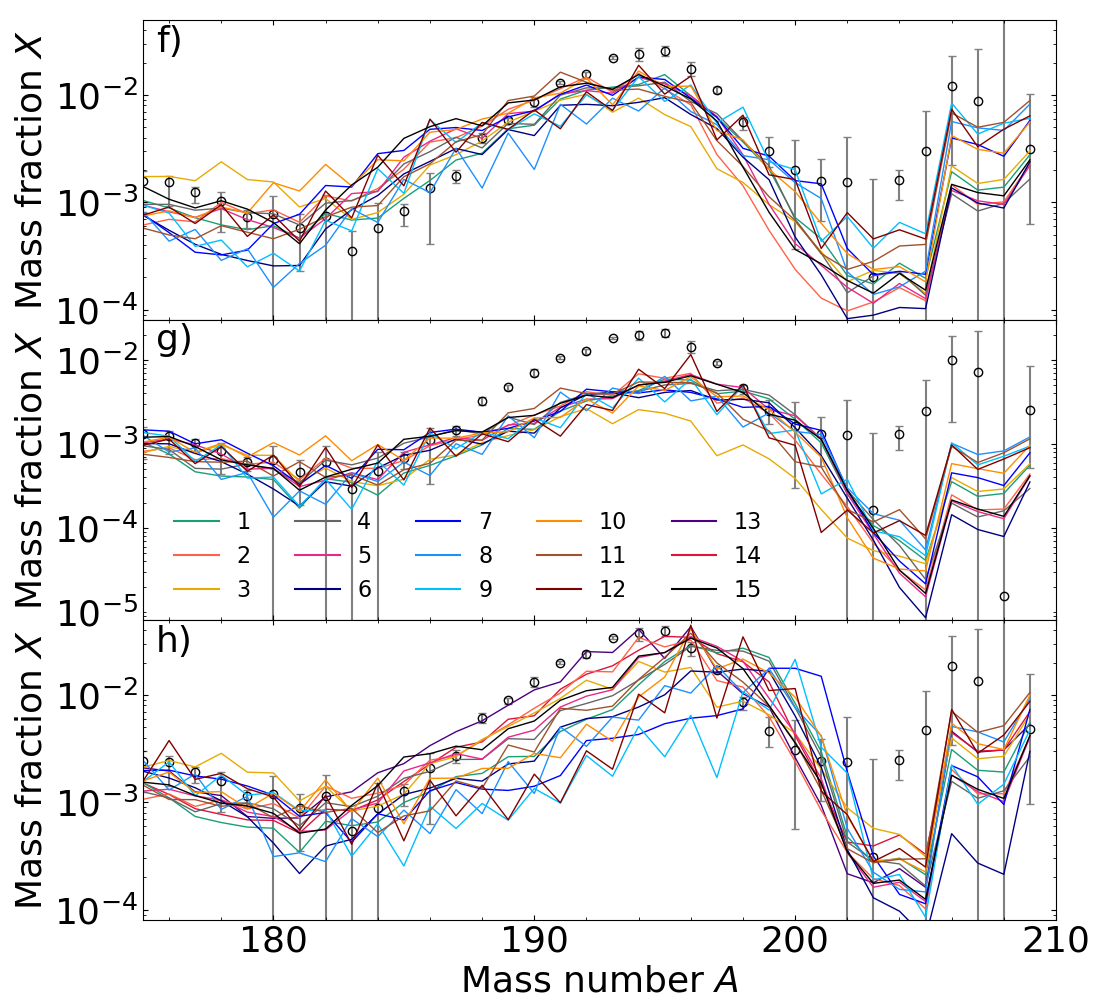}
\caption{(colour online)
Same as Fig.~\ref{fig_rpro_comb_all} panel f), g) and h) for the combined ejecta models zoomed in on the third r-process peak.}
\label{fig_rpro_comb_zoom}
\end{center}
\end{figure}

\begin{figure*}
\begin{center}
\includegraphics[width=0.92\textwidth]{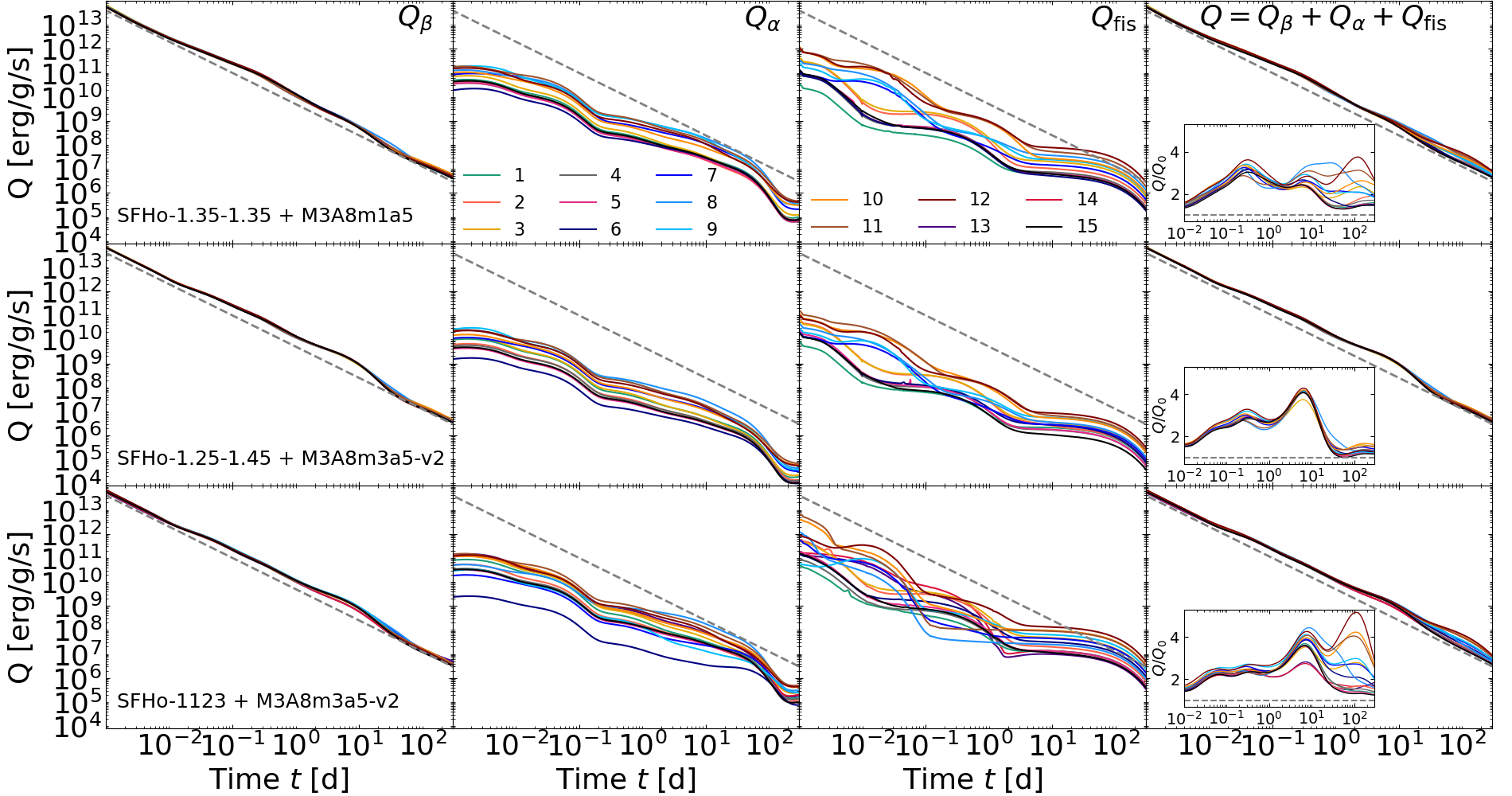}
\caption{(colour online)
Time evolution of the radioactive heating rate $Q$ (before thermalization) for the combined dynamical and secular ejecta for model SFHo-135-135 plus M3A8m1a5 (top), SFHo-125-145 plus M3A8m3a5-v2 (middle) and SFHo-11-23 plus M3A8m3a5-v2 (bottom) when varying the nuclear input between all models listed in Table~\ref{tab_nuc_mods}.
The columns displays the heating generated by $\beta$-decay ($Q_\beta$), $\alpha$-decay ($Q_\alpha$), fission ($Q_\mathrm{fis}$) and the total heat from all or the decay modes ($Q=Q_\beta + Q_\alpha + Q_\mathrm{fis}$).
The grey dashed line corresponds to the approximation $Q_0=10^{10}[t/1~\mathrm{day}]^{-1.3}$~erg/g/s. The ratio $Q/Q_0$ is also shown as an insert in the right column, where the grey dashed line indicates $y=1$.
The legend refers to the input sets in Table~\ref{tab_nuc_mods}.
}
\label{fig_Qmodes}
\end{center}
\end{figure*}

\begin{figure*}
\begin{center}
\includegraphics[width=0.92\textwidth]{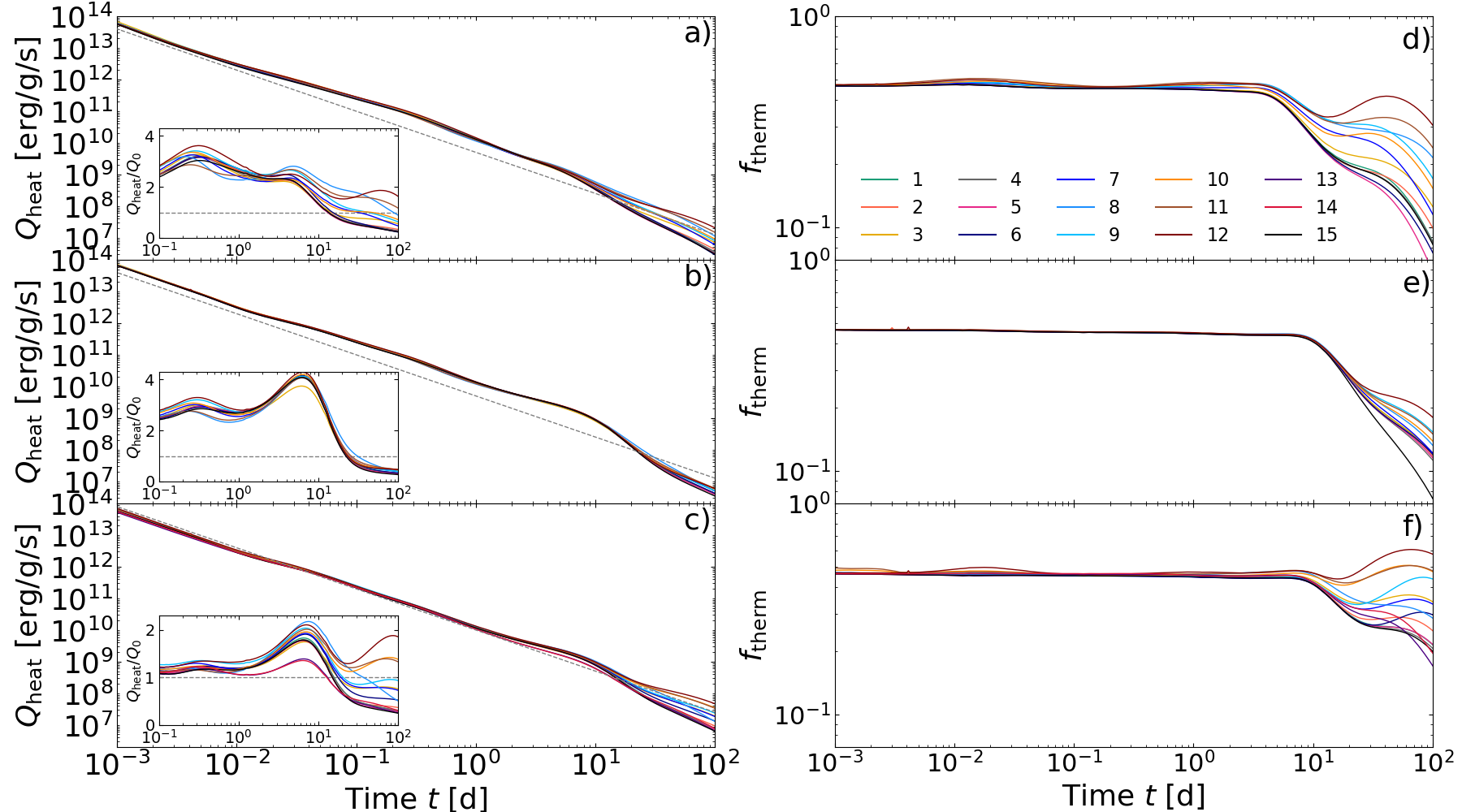}
\caption{(colour online)
Time evolution of the effective heating rate $Q_\mathrm{heat}$ (left column) and the total thermalization efficiency (right column) as defined in eq.~\ref{eq_Q_heat} for the combined dynamical and secular ejecta for model SFHo-135-135 plus M3A8m1a5 (top), SFHo-125-145 plus M3A8m3a5-v2 (middle) and SFHo-11-23 plus M3A8m3a5-v2 (bottom) when varying the nuclear input between all models listed in Table~\ref{tab_nuc_mods} (the legend refers to the input sets of this table). 
The grey dashed line in the left panels corresponds to the approximation $Q_0=10^{10}[t/1~\mathrm{day}]^{-1.3}$~erg/g/s. The ratio $Q_\mathrm{heat}/Q_0$ is also shown as an insert for $t>0.1$~d in the left panels, where the grey dashed line indicates $y=1$.
}
\label{fig_Qheat}
\end{center}
\end{figure*}

The sensitivity to the nuclear physics input when we combine the material from the dynamical ejecta with the post-merger BH-torus ejecta are displayed in Figs.~\ref{fig_rpro_comb_all}, \ref{fig_rpro_comb_zoom}, \ref{fig_Qmodes} and \ref{fig_Qheat}. Since the ejected mass of the BH-torus systems dominates over the mass ejected from the dynamical component for the NS-NS systems (see Table~\ref{tab_astro_mods}), the total r-process abundance distribution (and its uncertainty due to the nuclear physics input) in Figs.~\ref{fig_rpro_comb_all}(f-g) mostly resembles that of the disk ejecta. However, the dynamically ejected mass from the NS-BH merger system (model SFHo-11-23) is of the same order of magnitude as the one ejected from the BH-torus model M3A8m3a5-v2. In this case, the total ejecta given in Fig.~\ref{fig_rpro_comb_all}h shows a composition where both components play a significant role and which agrees with the solar system distribution fairly well for $A>90$.
See Fig.~\ref{fig_rpro_comb_zoom} for a zoom in on the third r-process peak for the combined models shown in Fig.~\ref{fig_rpro_comb_all}.

The radioactive heat generated from the separate decay modes $Q_\beta$, $Q_\alpha$ and $Q_\mathrm{fis}$ and the total heating rate $Q$ are displayed in Fig.~\ref{fig_Qmodes}. 
The heating rate is generally dominated by heat stemming from $\beta$-decays. However, at $t>10$~d, the contributions from fission and $\alpha$-decay become significant.  Overall, we observe small relative variations ($\lesssim2$) for the $\beta$-decay heat, whereas the heating rates produced by $\alpha$-decay and fission can vary by up to 1-2 orders of magnitude.

The impact of varying the nuclear physics input on the shape and magnitude of $Q_\alpha$ is directly linked to the number of heavy r-process elements produced and the detailed decay path taken back to the stable or long-lived nuclei, while it is not affected by the involved $\alpha$-decay rates since these are experimentally known.
Model SFHo-125-145 plus M3A8m3a5-v2 generates fewer trans-Pb species than models SFHo-135-135 plus M3A8m1a5 and SFHo-11-23 plus M3A8m3a5-v2,  and therefore $Q_\alpha$ has an insignificant impact on the total heating rate for this model. 
All nuclear input sets generate heat from the $\alpha$-decay chain starting from $^{224}$Ra; however, only a few input sets have non-negligible contributions from other decay chains\footnote{See the supplemental material of \citet{wu2019} for a complete list of the most important $\alpha$-decay chains contributing to the heating rate and also the discussion in  \citet{kullmann2021} for the dynamical ejecta models applied in this work.}.
In particular, for input sets 9, 10, 12 and 13 that apply the TDA or FRDM+QRPA $\beta$-decay models (see Table~\ref{tab_nuc_mods}),  the heat generated by the $\alpha$-decay chains starting from $^{223}$Ra, $^{225}$Ac or $^{221}$Fr leads to a large increase in $Q_\alpha$. 
This is related to the faster $\beta$-decay rates for the most neutron deficient $Z>83$ nuclei (see Fig.~\ref{fig_betadiff})
for models TDA and FRDM+QRPA, but also to the larger production of heavy r-process elements for these models. 

The spontaneous fission of $^{254}$Cf completely dominates the $Q_\mathrm{fis}$ curve at late times; no other fissile nuclei generate any considerable heat.
Therefore, the impact on $Q_\mathrm{fis}$ arising when varying the nuclear physics input is related to the ability of a particular input set to produce the heaviest elements, which is not only sensitive to the various fission models but all of the nuclear physics properties considered in this work. 

At $t\sim 10$~d, another enhancement can be seen in the heating rate, which is particularly strong for the combined ejecta models involving the BH-torus model M3A8m3a5-v2 (i.e., the two bottom panels of Fig.~\ref{fig_Qmodes}). 
This structure in the heating rate is related to the relatively high abundance peak at $A=132$ seen in Fig.~\ref{fig_rprowind_u1-7} and discussed earlier. 
The heat produced by the $\beta$-decay of the nuclei along the $A=132$ isobar dominates at late times (note that the contribution from fission and $\alpha$-decay are subdominant for this model). In particular, it is $^{132}$Te which sets the time scale with a half-life of 3.2~d, while the subsequent $\beta$-decay of $^{132}$I contributes the most to the heating rate among all nuclei due to its significant $Q_\beta$ value of 3.5~MeV.

Fig.~\ref{fig_Qheat} shows the effective heating rate $Q_\mathrm{heat}$ as well as the total thermalization efficiency $f_\mathrm{therm}$ relevant for kilonova light cure modelling.  Note that $f_\mathrm{therm}$ starts well below 1 due to the energy lost to neutrinos (contained in $Q_\nu$).
For the astrophysical models studied here, we can see that the biggest impact on the thermalization efficiency comes when varying the $\beta$-decay rates and fission properties. More specifically, a significant enhancement is found when adopting the FRDM+QRPA, TDA or HFB21+GT2 models (input sets 7-12) compared to the RHB+RQRPA rates, in particular for model SFHo-135-135 plus M3A8m1a5.  
As discussed above, varying the $\beta$-decay model leads to variations in $Q_\alpha$ and $Q_\mathrm{fis}$ (see Fig.~\ref{fig_Qmodes}) which in turn contributes to the $f_\mathrm{therm}$ variations seen for $t>10$~d in Fig.~\ref{fig_Qheat}. 
Varying the fission properties also impacts the thermalization efficiency, however, to a lesser extent than varying the $\beta$-decay model.  
The masses have a secondary impact on the thermalization efficiency, though results with FRDM (input set 2) and WS4 (input set 3) mass models deviate significantly from the other models (input sets 1, 4-6). 

Table~\ref{tab_rpro} summarizes several r-process properties, 
including the mass fraction of strontium, lanthanides plus actinides ($X_\mathrm{LA}$), $A>69$ and $A>183$ nuclei, and the heating rate ($Q$) at 1~d for all models studied in this work.
We can see that only minor changes in the amount of heavy r-process elements and lanthanides plus actinides are found between the different hydrodynamical models.
The lanthanide and actinide content of the ejecta and the heating rates are listed here for completeness, because they are particularly relevant for kilonova modelling, which is outside the scope of the present work.

\begin{table*}
\centering
\caption{ 
The strontium mass fraction $X_\mathrm{Sr}$, lanthanide plus actinide mass fraction $X_\mathrm{LA}$, the mass fractions of r-process nuclei ($A>69$) and third-r-process peak nuclei ($A>183$), and the radioactive heating rate before thermalization $Q$ at $t=1$~d for the three dynamical, two BH-torus and the two combined ejecta models considered in the present study.
For each quantity, the minimum (min), maximum (max), mean and standard deviation from the mean ($\sigma$) arising from the variation of the nuclear physics input sets in Table~\ref{tab_nuc_mods} are shown. An extended version of this table is available in the supplementary material. 
}
 \begin{tabular}{lcccccc}
\hline
\hline
              & & $X_\mathrm{Sr}$ & $X_\mathrm{LA}$ & $X_{A>69}$ & $X_{A>183}$ & $Q(t=1\mathrm{d})$  \\
              & &          &          &            &             & [$10^{10}$ erg/g/s] \\
\hline
SFHo-125-145 & min  & 0.01 & 0.09 & 0.95 & 0.11 & 0.99 \\
     & max & 0.04 & 0.14 & 0.97 & 0.17 & 1.22 \\
     & mean      & 0.02 & 0.12 & 0.97 & 0.14 & 1.11 \\
     & $\sigma $ & 0.005 & 0.015 & 0.005 & 0.021 & 0.054 \\
\hline
SFHo-135-135 & min  & 0.03 & 0.09 & 0.93 & 0.10 & 1.00 \\
     & max & 0.05 & 0.14 & 0.95 & 0.17 & 1.21 \\
     & mean      & 0.04 & 0.11 & 0.95 & 0.14 & 1.10 \\
     & $\sigma $ & 0.004 & 0.015 & 0.006 & 0.021 & 0.047 \\
\hline
   SFHo-11-23 & min  & $4\cdot10^{-5}$ & 0.21 & 0.99 & 0.19 & 0.78 \\
     & max & $3\cdot10^{-4}$ & 0.38 & 1.00 & 0.52 & 1.15 \\
     & mean      & $8\cdot10^{-5}$ & 0.30 & 0.99 & 0.43 & 0.91 \\
     & $\sigma $ & $6\cdot10^{-5}$ & 0.056 & 0.002 & 0.100 & 0.112 \\
\hline
 M3A8m3a5-v2 & min  & 0.01 & 0.06 & 0.93 & 0.01 & 1.15 \\
     & max & 0.04 & 0.10 & 0.96 & 0.13 & 1.42 \\
     & mean      & 0.03 & 0.08 & 0.95 & 0.05 & 1.32 \\
     & $\sigma $ & 0.008 & 0.012 & 0.007 & 0.032 & 0.084 \\
\hline
    M3A8m1a5 & min  & 0.01 & 0.06 & 0.94 & 0.09 & 1.15 \\
     & max & 0.03 & 0.11 & 0.95 & 0.15 & 1.48 \\
     & mean      & 0.02 & 0.08 & 0.95 & 0.12 & 1.32 \\
     & $\sigma $ & 0.005 & 0.015 & 0.004 & 0.021 & 0.083 \\
\hline
SFHo-125-145 & min  & 0.02 & 0.07 & 0.93 & 0.02 & 1.18 \\
+ M3A8m3a5-v2 & max & 0.04 & 0.11 & 0.95 & 0.05 & 1.40 \\
     & mean      & 0.03 & 0.08 & 0.95 & 0.05 & 1.32 \\
     & $\sigma $ & 0.005 & 0.013 & 0.005 & 0.007 & 0.060 \\
\hline
SFHo-135-135 & min  & 0.02 & 0.07 & 0.94 & 0.09 & 1.14 \\
  + M3A8m1a5 & max & 0.03 & 0.12 & 0.95 & 0.15 & 1.45 \\
     & mean      & 0.02 & 0.08 & 0.95 & 0.12 & 1.29 \\
     & $\sigma $ & 0.005 & 0.015 & 0.004 & 0.020 & 0.077 \\
\hline
   SFHo-11-23 & min  & 0.01 & 0.12 & 0.95 & 0.09 & 1.06 \\
+ M3A8m3a5-v2 & max & 0.02 & 0.20 & 0.98 & 0.27 & 1.32 \\
     & mean      & 0.02 & 0.16 & 0.96 & 0.19 & 1.17 \\
     & $\sigma $ & 0.005 & 0.025 & 0.005 & 0.046 & 0.066 \\
\hline
\hline
 \end{tabular} 
\label{tab_rpro}
\end{table*}

\subsection{Impact on cosmochronometers}

\begin{table*}
\centering
\caption{ 
Age estimates of six metal-poor r-process-enhanced stars based on the Th/U cosmochronometry (see eq.~\ref{eq_cosmo}). The observational ratios, $(\mathrm{Th/U})_\mathrm{obs}$, and their corresponding uncertainty are from the references listed, and the $(\mathrm{Th/U})_\mathrm{r}$ ratios are taken directly from our r-process calculations using our three astrophysical models for the combined ejecta. 
The minimum, maximum and mean of the Th/U ratio, as well as the standard deviation from the mean ($\sigma_{\log(\mathrm{Th/U})_\mathrm{r}}$) and the stellar age ($\sigma_{t^*_\mathrm{r}}$) arising from varying the nuclear input sets listed in Table~\ref{tab_nuc_mods} are given for the three astrophysical models. An extended version of this table is available in the supplementary material.
}
 \begin{tabular}{lccccccccc}
\hline \hline
 & $\log(\mathrm{Th/U})_\mathrm{obs}$ & $\sigma_{t^*,\mathrm{obs}}$ & ref. & SFHo-135-135 & SFHo-125-145 & SFHo-11-23 \\
 & & & & + M3A8m1a5 & + M3A8m3a5-v2 & + M3A8m3a5-v2 \\
\hline
$ \log (\mathrm{Th/U})_\mathrm{r,min}$ & - & - & - & 0.04 & 0.07 & 0.08 \\
$ \log (\mathrm{Th/U})_\mathrm{r,max}$ & - & - & - & 0.31 & 0.37 & 0.34 \\
$ \log (\mathrm{Th/U})_\mathrm{r,mean}$ & - & - & - & 0.17 & 0.21 & 0.18 \\
$\sigma_{\log(\mathrm{Th/U})_\mathrm{r}}$ & - & - & - &0.09 & 0.10 & 0.08 \\
$\sigma_{t^*_\mathrm{r}}$ & - & - & - &2.0 Gyr & 2.1 Gyr & 1.7 Gyr \\
\hline
CS22892-052 & 0.73 $\pm 0.22$ & 4.9 Gyr & \citet{sneden2003c} & 12.1 Gyr & 11.2 Gyr & 12.1 Gyr \\
CS29497-004 & 1.04 $\pm 0.33$ & 7.2 Gyr & \citet{Hill2017} & 18.9 Gyr & 18.0 Gyr & 18.8 Gyr \\
CS31082-001 & 0.94 $\pm 0.21$ & 4.7 Gyr & \citet{Mello2013} & 16.7 Gyr & 15.8 Gyr & 16.6 Gyr \\
HE1523-0901 & 0.86 $\pm 0.13$ & 2.8 Gyr & \citet{frebel2007} & 14.9 Gyr & 14.1 Gyr & 14.9 Gyr \\
 J0954+5246 & 0.82 $\pm 0.22$ & 4.9 Gyr & \citet{holmbeck2018} & 14.1 Gyr & 13.2 Gyr & 14.0 Gyr \\
 J2038-0023 & 0.90 $\pm 0.20$ & 4.4 Gyr & \citet{Placco2017} & 15.8 Gyr & 14.9 Gyr & 15.8 Gyr \\
\hline \hline
 \end{tabular} 
\label{tab_cosmo}
\end{table*}

Based on the sensitivity analysis performed above, we can estimate the impact of the nuclear uncertainties on the production of the Th and U cosmochronometers and the age of specific metal-poor stars for which the surface abundances of Th and U have been determined.
Table~\ref{tab_cosmo} lists the mean estimated ages ($t^*$) for six metal-poor stars, assuming the natal gas clouds from which the stars were made have been initially polluted by the combined ejecta corresponding to our models for the NS-NS or NS-BH merger.  
The age is calculated by using eq.~\ref{eq_cosmo} with the r-process abundance (molar fraction) ratio of (Th/U)$_\mathrm{r}$ consistently obtained from our three combined ejecta models and the 15 different nuclear inputs and the observed abundance ratios (Th/U)$_\mathrm{obs}$ from the literature (see Table~\ref{tab_cosmo} for the references).  
In addition, the standard deviation ($\sigma$) from the mean arising when varying the nuclear physics input is listed for the (Th/U)$_\mathrm{r}$ ratio and stellar ages $t^*_r$.
Note that $\sigma_{t^*_\mathrm{r}}$ is identical for all stars within each hydrodynamical model since a change in the observed Th/U ratio only leads to a shift in the estimated age.
If the references do not provide the uncertainty of $\log(\mathrm{Th/U})_\mathrm{obs}$, we calculate it as the square root of the quadratic sum of the individual, observational Th and U uncertainties given and propagate it to the age estimates ($\sigma_{t^*_\mathrm{obs}}$).
The average ages obtained for all different nuclear physics inputs vary from 11.2 to 18.9 Gyr depending on the star and astrophysical model applied. 
The theoretical Th/U ratio for combined models SFHo-135-135+M3A8m1a5 and SFHo-11-23+M3A8m3a5-v2 are similar, leading to almost identical age estimates, while model SFHo-125-145+M3A8m3a5-v2 has a larger Th/U ratio giving in general smaller age estimates. It is also model SFHo-125-145+M3A8m3a5-v2 which has the largest spread in the calculated values when the nuclear physics inputs are varied. For example, the age of star CS22892-052 ranges between 7.9 and 14.4~Gyr for the minimum and maximum ages, respectively. It is always input set 7 (which applies mass model HFB-31 and $\beta$-decay model HFB21+GT2, see Table~\ref{tab_nuc_mods}) that gives rise to the minimum age estimate. The maximum age is found with input sets 5, 4 and 15 (using mass models HFB-21 or D1M) for models SFHo-135-135+M3A8m1a5, SFHo-125-145+M3A8m3a5-v2 and SFHo-11-23+M3A8m3a5-v2, respectively.
The observational uncertainties for the Th/U ratios and the age estimates ($\sigma_{t^*,\mathrm{obs}}$) are also listed in Table~\ref{tab_cosmo}. We calculate the observational uncertainty of the Th/U ratios as the square root of the quadratic sum of the Th and U abundance uncertainties provided by the references listed in Table~\ref{tab_cosmo} and propagate the observational uncertainty of the ratios to the age estimate ($\sigma_{t^*,\mathrm{obs}}$).
We can see that the observational uncertainties are, in general, larger than the uncertainties stemming from the nuclear physics input ($\sigma_{t^*_\mathrm{r}}$).



\subsection{Comparison with other works}
\label{sec_compare}

\begin{table*}
\centering
\caption{
Comparison with previous works: summary of the adopted ejecta conditions, number of trajectories used ($N_\mathrm{traj}$), the origin of the trajectories (either hydrodynamical simulations or parametrized trajectories), the method used to vary the nuclear physics input (systematic or statistical), and an indication if the rms value of the applied mass models on known masses are all below 0.8~MeV.
The ejecta components refer to either the dynamical ejecta of NS-NS or NS-BH mergers (dyn.) or secular wind ejecta in BH-torus remnant systems (wind). 
Note that the work of \citet{Mumpower2016} summarizes the works of several MC studies. 
}
\begin{tabular}{cccccc}
\hline 
Ref. & Ejecta comp. & $N_{\mathrm{traj}}$ & Traj. origin & Variation method & rms~$\lesssim 0.8$~MeV \\ 
\hline 
This work & wind, dyn. & 150-296 & hydro & systematic & yes \\ 
\citet{Barnes2021,Zhu2021} & wind & 1 & param. & systematic & some \\ 
\citet{Mendoza-Temis2015} & dyn. & 528 & hydro & systematic & yes \\ 
\citet{Eichler2015} & dyn. & 30 & hydro & systematic & some \\ 
\citet{Vassh2019} & dyn. & 1 (\& 30) & hydro & systematic & some \\ 
\citet{Giuliani2020} & wind, dyn. & 1 & hydro \& param. & systematic & yes \\ 
\citet{Caballero2014} & dyn. & 1 & hydro & systematic & yes \\
\citet{Marketin2016} & dyn. & 1 & hydro & systematic & yes \\
\citet{Nikas2020} & wind, dyn. & 1 & hydro & statistical & some \\ 
\citet{Mumpower2016} & wind, dyn. & 1 & hydro \& param. & statistical & yes \\ 
\citet{Lund2022} & wind & 1 & param. & systematic & some \\
\hline 
\end{tabular} 
\label{tab_comp}
\end{table*}

Many r-process studies in the last decade have focused on the impact of the nuclear physics uncertainties on the r-process yields \citep{Goriely92,Goriely2001,Schatz2002,surman2014,Caballero2014,Mendoza-Temis2015,Eichler2015,goriely2015,Martin2016,Liddick2016,Mumpower2016,Nishimura2016,Bliss2017,Denissenkov2018,Eichler19,Holmbeck19,Barnes2021, Vassh2019,Nikas2020,sprouse2020a,McKay2020,Giuliani2020, Zhu2021,Barnes2021,Lund2022}.
Often, sensitivity studies aim to identify specific nuclei or regions in the nuclear chart where the r-process has the largest sensitivity to the experimentally unknown nuclear properties. If such nuclei are identified, they can be targeted by experimental campaigns as long as they are within reach of the given facility \citep[e.g.,][]{Surman18}.
Another aim of r-process sensitivity studies is to estimate the magnitude of the nuclear uncertainties so that they can be compared to other sources of uncertainty, like those arising from hydrodynamical modelling. This is particularly important for applications like cosmochronometry, galactic chemical evolution or kilonova models, which require r-process yields from a given site or ejecta component as input.
A large range of astrophysical conditions has been applied in various r-process sensitivity studies in the literature, making detailed comparisons between results difficult. In the following we will, in a quantitative way, compare our results to other sensitivity studies that apply the same or similar nuclear models and ejecta components as we do and as summarized in Table~\ref{tab_comp}.

\begin{figure*}
\begin{center}
\includegraphics[width=\textwidth]{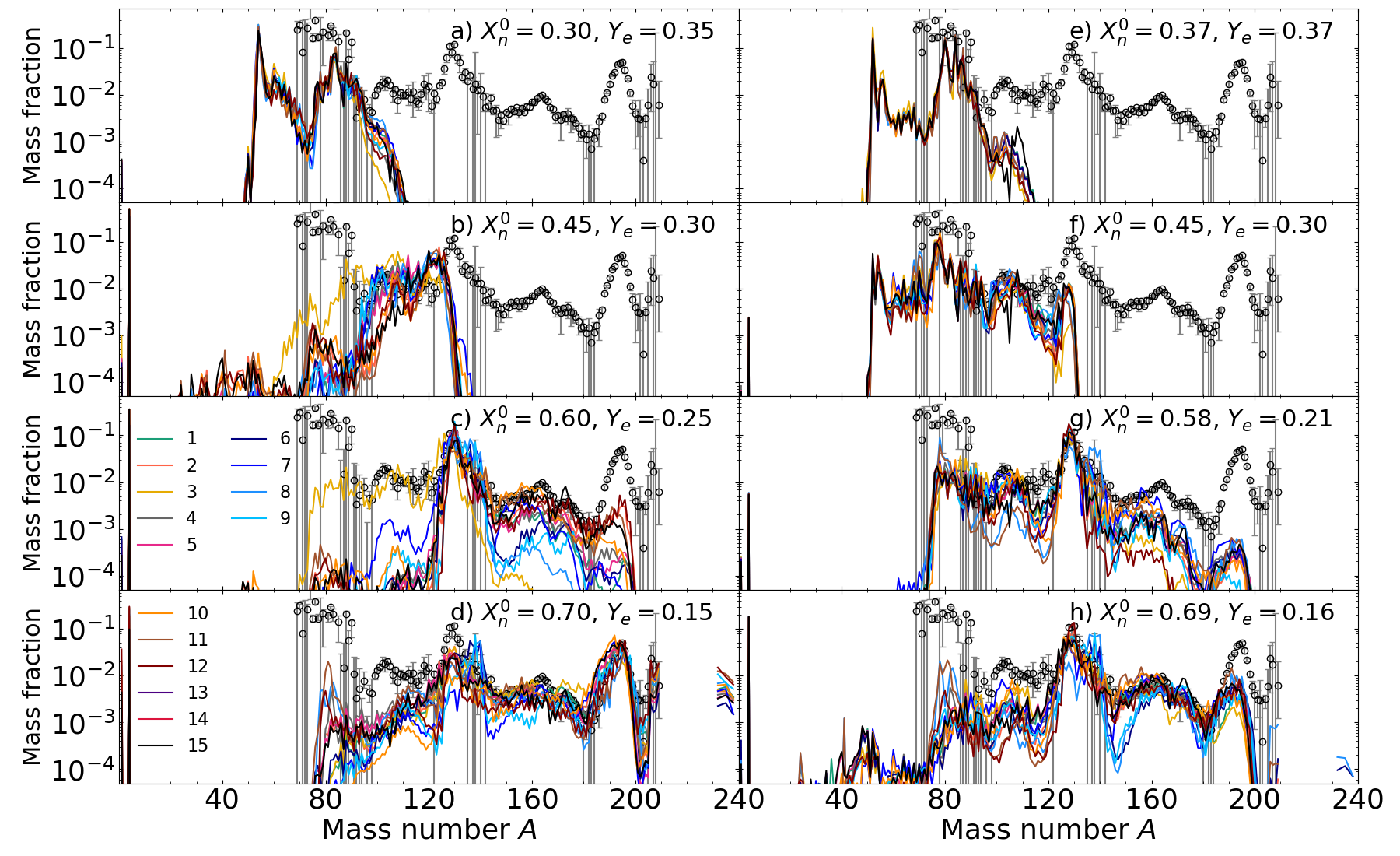}
\caption{(colour online)
Abundance distributions obtained in eight individual trajectories with different initial neutron mass fractions (or initial $Y_e$) from model SFHo-135-135 (left column) and M3A8m1a5 (right column) using all of the nuclear input sets listed in Table~\ref{tab_nuc_mods}. 
See for example Figs.~\ref{fig_rprowind_u1-7}, \ref{fig_rprowind_u7-10}, \ref{fig_rprowind_u3-13} and \ref{fig_rprowind_DC} for the colour labels for each nuclear input combination. 
}
\label{fig_rpro_onetraj}
\end{center}
\end{figure*}

Since the r-process is very sensitive to the conditions of the environment, it is essential to keep in mind which assumptions and simplifications have been applied for the nucleosynthesis calculations. 
Basically, three approaches to modelling the r-process conditions exist for a given astrophysical site or ejecta component: 1) parametrized trajectories which typically only depend on the initial electron fraction, entropy and expansion time scale, or 2) a single or a few ($<10$) trajectories extracted from a hydrodynamical simulation, or 3) the complete set, or a significant representative sample, of trajectories representing all the mass elements ejected, as given by a hydrodynamical simulation. 
The first two methods have the obvious benefit that the required computing time is significantly shorter than in the third method, which may require up to several thousand r-process calculations or more depending on the hydrodynamical model.
However, the central assumption adopted when using methods 1) or 2) is that just a single trajectory (or very few trajectories) are representative of the overall set and in particular of the various conditions found in the ejecta. This approach assumes that the different ejecta components show minor variations between trajectories in terms of the evolution of quantities such as electron fraction, temperature and density.
However, this is not what most hydrodynamical simulations predict \citep{wanajo2014,Mendoza-Temis2015,Just2015,foucart2016,goriely2016b,radice2018a,Siegel2018,ardevol-pulpillo2019}. 
See also the approach of \citet{Lund2022}, which uses a weighted $Y_e$-distribution, i.e. an analytical probability function, to combine single trajectories to mimic the non-uniform composition of the ejecta.

Fig.~\ref{fig_rpro_onetraj} displays the r-process abundance distribution for four individual trajectories from models SFHo-135-135 and M3A8m1a5 with $Y_e$ values ranging from 0.15 to 0.37 using all of the nuclear input sets. 
For the four trajectories shown here, the impact of the nuclear physics uncertainties strongly depends on the adopted trajectory and its neutron richness. For example, the predicted amount of heavy r-process nuclei  ($A\gtrsim 90$) can easily vary by a factor of 100 and more between the adopted nuclear physics input sets when a single or a few trajectories are considered. However, when the r-process abundance distribution is calculated from an ensemble of trajectories representing a range of conditions, the uncertainties due to the nuclear physics input shrink significantly, to at most a factor of about 20 for some specific individual $A>90$ nuclei (see Fig.~\ref{fig_rpro_comb_all} which is based on $\sim150-300$ hydrodynamical trajectories). 
The comparison of Figs.~\ref{fig_rpro_onetraj} and \ref{fig_rpro_comb_all}\footnote{Similar trends can also be seen in Figs.~2a and b in \cite{Zhu2021}.} shows that uncertainty studies based on a single or a few trajectories may artificially exacerbate the impact of nuclear physics uncertainties, particularly in cases of $Y_e\sim0.2-0.3$ close to the threshold of lanthanide production. Therefore, one should be careful when drawing conclusions regarding the total r-process yields of a given site or ejecta component from single-trajectory studies. 

Another critical point when comparing sensitivity analyses concerns how nuclear uncertainties are propagated to nucleosynthesis calculations. A popular technique used to propagate the nuclear uncertainties to the final r-process results is to increase or decrease, for example, the nuclear mass or the neutron capture rate of a single nucleus by a given factor \citep[e.g., see ][who applied factors of 5, 10, 50 or 100 to the $(n,\gamma)$ or $(\alpha,n)$ rates]{Surman18,Bliss2017}. Then, for each variation or change of, for example, the nuclear mass of a given nucleus, the r-process abundances are re-calculated and compared to a base calculation with fixed nuclear (and astrophysical) input. After multiple variations, i.e., r-process calculations, the nuclei for which the abundances are the most sensitive with respect to changes in a given nuclear property are revealed.
This technique can be seen as a variant of the MC method \citep[see][for details about various implementations]{Mumpower2016,Rauscher2020} since it applies random variations of quantities, such as mass, $\beta$-decay half-lives, neutron capture rates or $\beta$-delayed neutron emission, pulled from a distribution which represents the assumed nuclear uncertainty of the property investigated.

When considering uncertainties arising from (theoretical) nuclear physics inputs such as, for example, the nuclear mass or neutron capture rates, there are two sources of uncertainty, namely statistical and systematic errors. The latter are often referred to as model errors. 
MC studies adopt a given nuclear model for the baseline calculation (i.e., they have to choose the mean of the uncertainty distribution) and consider uncorrelated variations around that baseline. Therefore, by design, MC methods and similar techniques can only probe uncorrelated parameter errors which in essence are of a different origin compared to model-correlated systematic uncertainties. For this reason, the magnitude of the parameter uncertainties cannot be extracted from  deviations obtained with different models, but rather from parameter variations within a given model, provided the parameter change remains somehow constrained by experimental data for nuclei where measurements are available. If not constrained by experimental data, the magnitude of the parameter error, i.e. the width of the uncertainty distributions considered in MC studies, may be overestimated leading to an overestimate of their impact on abundance calculations.
As discussed in Sec.~\ref{sec_intro}, we only consider here systematic uncertainties for which theoretical physical models are responsible for the nuclear correlations between various nuclei involved as well as between the different properties of interest. For the neutron-rich nuclei relevant to the r-process, for which no experimental information is available, systematic uncertainties have been shown to dominate over the statistical errors due to variations of model parameters \citep[see in particular][who discuss the extrapolation uncertainties of mass models]{Goriely2014a}. However, since we neglect here the parameter uncertainties, our approach may underestimate the impact of nuclear physics uncertainties on the nucleosynthesis calculations.

As discussed in Sec.~\ref{sec_network_input}, not all available global nuclear models are suitable for astrophysical applications. For example, mass models that have a relatively high rms value with respect to available experimental data should not be included when probing the nuclear uncertainties of the r-process. In this paper we only consider mass models with rms values around $0.8$~MeV or below. Works that use mass models with larger rms values than this (indicated in Table~\ref{tab_comp}) might achieve larger uncertainty bands in their r-process sensitivity studies. For each study, Table~\ref{tab_comp} lists the type of ejecta conditions adopted, the number of and origin of the trajectories used, as well as the the type of uncertainties, systematic or statistical, considered.
For instance, the works of \citet{Barnes2021}, \citet{Zhu2021} and also Sec.~3.1 of \citet{Lund2022} used single parametrized trajectories
for a given $Y_e$ value, fixed expansion time scale and fixed entropy to mimic disk wind conditions. 
They applied an extensive set of the existing mass models in their input variations, which led to a large uncertainty band spanning up to a factor of 100, in particular for the low-$Y_e$ case shown in Fig.~2 in \citet{Zhu2021}. However, if we exclude the nuclear mass models with rms deviations higher than 0.8~MeV, such as UNEDF1, SLy4 and TF, their results get closer to ours in terms of the magnitude of the uncertainty stemming from the nuclear mass models. Similarly, for the heating rate shown in Fig.~3 of \citet{Zhu2021}, the factor between the maximum and minimum values can be as large as 1000 for the low-$Y_e$ case if all their applied mass models are included but less than a factor of 100 if only WS3, HFB-22, HFB-27 and FRDM12 mass models are considered. 
Similar results are found in Fig.~1 of \citet{Barnes2021} for the nuclear uncertainties on the r-process abundance distribution and heating rates. 

Considering now studies using a larger set of trajectories, \citet{Eichler2015} considered 30 trajectories based on a Newtonian hydrodynamical simulation of the low-$Y_e$ dynamical ejecta for their r-process calculations, and \citet{Mendoza-Temis2015} considered 528 low-$Y_e$ relativistic SPH trajectories for the dynamical ejecta. Both studied the impact of systematic uncertainties associated with a few mass models rather similar to those considered here. 
Their results, shown in Figs.~5 and 6 in \citet{Eichler2015} and \citet{Mendoza-Temis2015}, respectively, are similar to ours (see in particular Fig.~\ref{fig_xn0_ye_distr}e corresponding to the somewhat similar low-$Y_e$ conditions) in that they predict relatively small abundance variations. 
This is related to the similar nuclear physics input adopted and the fact that they adopted a relatively large ensemble of trajectories instead of single trajectories.

\citet{Eichler2015} also investigated the impact of four fission fragment distribution models on the final r-process abundance distribution. However, they did not implement the SPY or GEF distributions, as we did, but the empirical models of \citet{Kodama1975} and \citet{Panov2001,Panov2008} and the statistical approach by the ABLA07 code \citep{Kelic2008,Kelic2009}. Though a direct comparison is not straightforward, Fig.~4b of \citet{Eichler2015} shows an impact in the $A\sim140-150$ region similar to the one observed in our Fig.~\ref{fig_rpro_ufiss} between the GEF and SPY models. 
Similarly, Fig.~8 in \citet{Vassh2019} based on a single $Y_e<0.02$ dynamical ejecta trajectory, as well as the upper and middle panels of Fig.~3 in \citet{Giuliani2020} based on a similar low-$Y_e$ single trajectory, show a significant impact around $A\sim150$ when varying between different models for the fission fragment distributions. 
The largest divergence can be found between the symmetric ($A_\mathrm{frag} = \frac{1}{2}A_\mathrm{fission}$) approximation for the fragments and the GEF fission fragment distributions in \citet{Vassh2019}, or the HFB-14 and FRDM+TF fission barriers in \citet{Giuliani2020}, both of which are of the same order of magnitude as the deviations we observe between SPY and GEF. 
Figs.~8 and 9 of \citet{Zhu2021} show the impact of assuming symmetric fragments versus a double-peaked Gaussian fission fragment distribution on the heating rates. In order to compare trajectories with the same $Y_e$ and the same nuclear mass model, cases 3 and 4 (blue lines) should be compared, where the relative contribution to the heating rate from $\alpha$-decays and fission changes between the cases, leading to a discrepancy of about a factor of 2, which also impacts the final kilonova light curve shown in their Fig.~11.
Similarly, in Figs~11 and 17 in \citet{Vassh2019}, a three-order-of-magnitude discrepancy in the heating generated by neutron-induced fission is found when varying between symmetric distributions and the GEF fission fragment distributions.
In summary,  the r-process studies that varied the fission properties in their nuclear physics input showed a significant impact on the shape of the r-process abundance distribution. 
For the works mentioned here, the magnitude and shape of the r-process abundance distribution are generally found to be more robust with respect to changes in the nuclear physics input when an extensive set of trajectories is considered and the input is limited to models that have proven their capacity to reproduce experimental data accurately. 

Most r-process sensitivity studies have focused on the variation of mass models and neutron capture rates; however, a few have also estimated the impact of varying the theoretical $\beta$-decay rates. In particular, \citet{Eichler2015} implemented the RHB+RQRPA and FRDM+QRPA rates also applied in our work. 
Their Figs. 14a and b show that the $\beta$-decay rates can impact the shape of the third r-process peak in a way comparable to what we find in our NS-BH scenario (in Figs.~\ref{fig_rpro_u3-13} and \ref{fig_rpro_u7-10}). The RHB+RQRPA rates reproduce the solar distribution for the low-$A$ end of the peak, while the FRDM+QRPA rates cause deviations from the solar distribution and give rise to a narrower peak. The same trends can also be observed in the lower panel of Fig.~14 in \citet{Marketin2016}, Fig.~2 of \citet{Caballero2014} and Fig.~16 of \citet{Vassh2019}.
\citet{Lund2022} studied the impact associated with the RHB+RQRPA, FRDM+QRPA and the recent $\beta$-decay rates of \citet{Ney2020} on the r-process heating rates, shown for three single-$Y_e$ trajectories in their Fig.~4. Similar to our results in Fig.~\ref{fig_Qmodes}, their $\beta$-decay (fission) heating rates are the least (most) sensitive to variations in the applied nuclear mass or $\beta$-decay models.

While our approach to estimating the nuclear uncertainties on the r-process yields only considers theoretical models which accurately reproduce experimental data (see Sec.~\ref{sec_nucuncert} for more details),  some limitations are present. First of all, due to the limited number of available global nuclear physics models, most of the input sets applied in our r-process calculations are not consistent in terms of, for example, the nuclear mass model and the masses used to calculate the $Q_\beta$-values in the $\beta$-decay models, or the masses used to calculate the fission properties. 
These inconsistencies need to be further improved in future nuclear modelling. In addition, as already mentioned, for completeness, parameter uncertainties need to be included around each of the model uncertainties and correspondingly propagated.



\section{Summary and conclusions}
\label{sec_concl}

In this paper, we report an extensive study of the systematic uncertainties related to the theoretical nuclear models, which are used as input to r-process nuclear network calculations, and propagate them to the r-process nucleosynthesis to estimate their impact on the abundance distribution and the resulting radioactive heating rate.
Our r-process calculations are based on detailed hydrodynamical models, which describe the time evolution of the dynamical and post-merger ejecta, including neutrino interactions and viscosity. Our hydrodynamical models do not cover all possible evolutionary paths but include three representative cases of the NS-NS or NS-BH systems, and two subsequent models for the BH-torus system which is expected to form after the merger. These two ejecta components are then combined into the total ejecta. For the NS-NS merger system, we use one system with equal-mass NSs, SFHo-135-135, and one asymmetrical model, SFHo-125-145, which are combined with BH-torus models M3A8m1a5 and M3A8m3a5-v2, respectively. Additionally, we employ one model to describe the dynamical ejecta of an NS-BH merger, model SFHo-11-23, which we combine with model M3A8m3a5-v2 (see Table~\ref{tab_astro_mods}).
With this setup, we study the impact of varying the nuclear physics input on the r-process nucleosynthesis, the heating rate and thermalization efficiency, as well as the predicted age of six metal-poor stars by Th/U cosmochronometry. 
We vary six nuclear mass models, two frameworks for calculating the radiative neutron capture rates, four $\beta$-decay models and two sets of fission barriers and fragment distributions, as detailed in Table~\ref{tab_nuc_mods}. 
Only global mass models that have an rms value smaller than 0.8~MeV are included in this work. 

For each nuclear input set, between 150 and 300 representative trajectories of the complete set of $\sim1000-13000$ trajectories (depending on the hydrodynamical model applied) are used to run the r-process nuclear network calculations (see Sec.~\ref{sec_network_input} for more details). This way, our nucleosynthesis trajectories consistently sample a wide range of thermodynamical conditions encountered at different ejection angles and merger phases.
This is in contrast to sensitivity studies, which often only used one single trajectory, or very few trajectories, to represent entire ejecta components (see Sec.~\ref{sec_compare} and Table~\ref{tab_comp}).

The main conclusions from our study can be summarized as follows:
\begin{itemize}
\item The systematic uncertainties connected to the nuclear physics input have a minor impact on the global shape of the r-process abundance distribution for the astrophysical scenarios studied here. Our combined ejecta models (consisting of the dynamical NS-NS or NS-BH merger plus the secular BH-torus ejecta) reproduce the solar system distribution well for $A>90$ and yield a significant amount of Th and U, irrespective of the adopted nuclear physics model. In addition, only small changes in the amount of heavy r-process elements ($X_{A>69}$) and lanthanides plus actinides ($X_\mathrm{LA}$) are found (see Table~\ref{tab_rpro}) for all model variations.
However, when studying the detailed distribution shape, a variation between different nuclear physics models can have a significant (local) impact (see Fig.~\ref{fig_rpro_comb_zoom}). 
When fission plays an important role (in particular, for the NS-BH case), the most considerable impact on the abundance distribution and the heating rate is found in connection with the fission barrier height and fragment distribution. 
\item The position, shape and width of the r-process peaks vary with the nuclear physics input used but also with the ejecta model. In particular, when mass model HFB-31 is applied, an additional structure at $A\sim 130-140$ appears. For BH-torus model M3A8m3a5-v2, a very narrow peak around $A\sim 132$ is formed, which is not observed for the other BH-torus or dynamical models. For the high-$X_n^0$ (e.g., low-$Y_e$) conditions in the NS-BH merger, the width of the third r-process peak is very sensitive to the adopted $\beta$-decay rates, where the RHB+RQRPA rates of \citet{Marketin2016} lead to better agreement with the solar r-process distribution, as also found in other studies (see Sec.~\ref{sec_compare}).
\item While globally, all nuclear models give rise to a relatively similar ejecta composition, deviations up to a factor $\sim20$ can be found for some specific individual abundances of elements with $A>90$, in particular around the second and third r-process peaks (the latter being most prominent for the NS-BH merger scenario). The factor between the largest and smallest yields of actinides ranges between 5 and 7. This conclusion holds for both the dynamical and wind ejecta, the latter dominating the overall mass ejected in the present modelling. 
 \item The radioactive heating rate before thermalization is found to be relatively insensitive (within a factor of $\lesssim2$) to variations of the nuclear physics input at early times ($t \la 10$~d). However, more significant deviations are found in particular related to the contribution from fission in cases where the heaviest elements are produced (i.e., for the NS-BH case at $t \ga 5$~d).
The thermalization efficiency introduces an additional spread between different nuclear physics models, particularly at late times when $\alpha$-decay and spontaneous fission become important.
\item We find a similar order of magnitude for the uncertainties due to the nuclear physics input when we compare our results to the works of others which, also varied the nuclear input between global models similar to the ones applied here. 
A larger impact, up to a factor of 100 on individual $A>90$ r-process nuclei and 1000 on the heating rates, can be found when applying global models that have not been developed to reproduce experimental data accurately
\citep{Barnes2021,Zhu2021}, or when considering MC-type methods with large statistical variations \citep[e.g.,][and references therein]{Mumpower2016}.
\item For the astrophysical scenarios studied here, the nuclear physics uncertainties are typically small compared to variations related to the different ejecta components.
Moreover, they are also small (or at most similar) compared to the variations encountered when using different disk masses, BH masses, and BH spins in BH-torus simulations \citep[e.g.,][]{Wu2016,Just2021c} or different EoSs in NS-NS merger simulations \citep[e.g.,][]{kullmann2021,sekiguchi2015,radice2018a}. This circumstance strengthens the credibility of nucleosynthesis analyses based on hydrodynamical models of these systems, and it lends further support to the notion that NS mergers can be significant sources of r-process elements in the Universe.
\item For the stellar Th/U cosmochronometry age estimates, the systematic uncertainties associated with the nuclear physics input are still larger than the variations associated with changes in the hydrodynamical model, at least for the cases studied here. The mean age inferred by our method for the six metal-poor stars varies between $11\pm 2$ and $19\pm2$ Gyr for stars CS22892-052 and CS29497-004, respectively.
\end{itemize}

It is difficult to pin down a single nucleus (or a few nuclei) having a dominant impact on the nuclear r-process uncertainties. First, when changing the nuclear input from one model to another, many nuclei are affected in a systematic way (in contrast to sensitivity studies that neglect systematic correlations and change, for example, the neutron capture rate of individual reactions randomly within a given range). 
Second, because the r-process conditions can vary between the astrophysical scenarios, the significance of a nuclear process can also vary, which would change the sensitivity resulting from a specific nuclear mass region.
Therefore, continued efforts on the experimental and theoretical side have to systematically improve the amount of data available and the description of nuclear structure, reactions and the radioactive decay of neutron-rich nuclei in the foreseeable future.
In particular, a future aim should be to develop fully consistent models for all nuclear physics properties (e.g., masses, $\beta$-decay, fission) required as input for r-process calculations.
Similarly, the push towards improving hydrodynamical simulations of the NS-NS or NS-BH merging systems covering the dynamical as well as post-merging phases will significantly improve our understanding of the conditions for the r-process.
In particular, questions regarding the mass, velocity, entropy, and neutron-richness of the various ejecta components need to be further resolved.
With continuous advancements on the nuclear and astrophysical side, systematic studies of the nuclear physics impact will continue to be essential to quantify the uncertainties of r-process yields and their corresponding production of decay heat.

\section*{Acknowledgments}
SG acknowledges financial support from F.R.S.-FNRS (Belgium). This work has been supported by the Fonds de la Recherche Scientifique (FNRS, Belgium) and the Research Foundation Flanders (FWO, Belgium) under the EOS Project nr O022818F. 
The present research benefited from computational resources made available on the Tier-1 supercomputer of the Fédération Wallonie-Bruxelles, infrastructure funded by the Walloon Region under the grant agreement n$^\circ$1117545 and the Consortium des Équipements de Calcul Intensif (CÉCI), funded by the Fonds de la Recherche Scientifique de Belgique (F.R.S.-FNRS) under Grant No. 2.5020.11 and by the Walloon Region. 
This work was supported in part by the National Science Foundation under Grant No. PHY-1430152 (JINA Center for the Evolution of the Elements).
OJ and  AB acknowledge support by the European Research Council (ERC) under the European Union's Horizon 2020 research and innovation programme under grant agreement No. 759253. AB was supported by Deutsche Forschungsgemeinschaft (DFG, German Research Foundation) - Project-ID 279384907 - SFB 1245 and - Project-ID 138713538 - SFB 881 (``The Milky Way System'', subproject A10) and acknowledges the support by the State of Hesse within the Cluster Project ELEMENTS. OJ acknowledges computational support by the HOKUSAI supercomputer at RIKEN, by the Max Planck Computing and Data Facility (MPCDF), and by the VIRGO cluster at GSI.
At Garching, funding by the European Research Council through Grant ERC-AdG No.~341157-COCO2CASA and by the Deutsche Forschungsgemeinschaft (DFG, German Research Foundation)
through Sonderforschungsbereich (Collaborative Research Centre) SFB-1258 ``Neutrinos and Dark Matter in Astro- and Particle Physics (NDM)'' and under Germany's Excellence Strategy through
Cluster of Excellence ORIGINS (EXC-2094)---390783311 is acknowledged.

\section*{Data availability}
The data underlying this article will be shared on reasonable request to the corresponding author.

\bibliographystyle{mnras}
\bibliography{references.bib}

\begin{thebibliography}{}
\makeatletter
\relax
\def\mn@urlcharsother{\let\do\@makeother \do\$\do\&\do\#\do\^\do\_\do\%\do\~}
\def\mn@doi{\begingroup\mn@urlcharsother \@ifnextchar [ {\mn@doi@}
  {\mn@doi@[]}}
\def\mn@doi@[#1]#2{\def\@tempa{#1}\ifx\@tempa\@empty \href
  {http://dx.doi.org/#2} {doi:#2}\else \href {http://dx.doi.org/#2} {#1}\fi
  \endgroup}
\def\mn@eprint#1#2{\mn@eprint@#1:#2::\@nil}
\def\mn@eprint@arXiv#1{\href {http://arxiv.org/abs/#1} {{\tt arXiv:#1}}}
\def\mn@eprint@dblp#1{\href {http://dblp.uni-trier.de/rec/bibtex/#1.xml}
  {dblp:#1}}
\def\mn@eprint@#1:#2:#3:#4\@nil{\def\@tempa {#1}\def\@tempb {#2}\def\@tempc
  {#3}\ifx \@tempc \@empty \let \@tempc \@tempb \let \@tempb \@tempa \fi \ifx
  \@tempb \@empty \def\@tempb {arXiv}\fi \@ifundefined
  {mn@eprint@\@tempb}{\@tempb:\@tempc}{\expandafter \expandafter \csname
  mn@eprint@\@tempb\endcsname \expandafter{\@tempc}}}

\bibitem[\protect\citeauthoryear{Abbott et~al.}{Abbott
  et~al.}{2017a}]{abbott2017c}
Abbott B.~P.,  et~al., 2017a, \mn@doi [Physical Review Letters]
  {10.1103/PhysRevLett.119.161101}, 119, 30

\bibitem[\protect\citeauthoryear{Abbott et~al.}{Abbott
  et~al.}{2017b}]{abbott2017d}
Abbott B.~P.,  et~al., 2017b, \mn@doi [The Astrophysical Journal]
  {10.3847/2041-8213/aa91c9}, 848, L12

\bibitem[\protect\citeauthoryear{{Aguilera-Miret}, {Vigan{\`o}}  \&
  {Palenzuela}}{{Aguilera-Miret} et~al.}{2022}]{Aguilera-Miret2022}
{Aguilera-Miret} R.,  {Vigan{\`o}} D.,   {Palenzuela} C.,  2022, \mn@doi
  [\apjl] {10.3847/2041-8213/ac50a7}, \href
  {https://ui.adsabs.harvard.edu/abs/2022ApJ...926L..31A} {926, L31}

\bibitem[\protect\citeauthoryear{{Arcones}, {Janka}  \& {Scheck}}{{Arcones}
  et~al.}{2007}]{Arcones2007}
{Arcones} A.,  {Janka} H.,   {Scheck} L.,  2007, \aap, 467, 1227

\bibitem[\protect\citeauthoryear{{Ardevol-Pulpillo}, Janka, Just  \&
  Bauswein}{{Ardevol-Pulpillo} et~al.}{2019}]{ardevol-pulpillo2019}
{Ardevol-Pulpillo} R.,  Janka H.~T.,  Just O.,   Bauswein A.,  2019, \mn@doi
  [Monthly Notices of the Royal Astronomical Society] {10.1093/mnras/stz613},
  485, 4754

\bibitem[\protect\citeauthoryear{{Argast}, {Samland}, {Thielemann}  \&
  {Qian}}{{Argast} et~al.}{2004}]{Argast2004}
{Argast} D.,  {Samland} M.,  {Thielemann} F.~K.,   {Qian} Y.~Z.,  2004, \mn@doi
  [\aap] {10.1051/0004-6361:20034265}, \href
  {https://ui.adsabs.harvard.edu/abs/2004A&A...416..997A} {416, 997}

\bibitem[\protect\citeauthoryear{Arnould, Goriely  \& Takahashi}{Arnould
  et~al.}{2007}]{arnould2007}
Arnould M.,  Goriely S.,   Takahashi K.,  2007, \mn@doi [Physics Reports]
  {10.1016/j.physrep.2007.06.002}, 450, 97

\bibitem[\protect\citeauthoryear{{Artemova}, {Bjoernsson}  \&
  {Novikov}}{{Artemova} et~al.}{1996}]{Artemova1996}
{Artemova} I.~V.,  {Bjoernsson} G.,   {Novikov} I.~D.,  1996, \mn@doi [\apj]
  {10.1086/177084}, \href
  {https://ui.adsabs.harvard.edu/abs/1996ApJ...461..565A} {461, 565}

\bibitem[\protect\citeauthoryear{{Baiotti} \& {Rezzolla}}{{Baiotti} \&
  {Rezzolla}}{2017}]{Baiotti2017}
{Baiotti} L.,  {Rezzolla} L.,  2017, \mn@doi [Reports on Progress in Physics]
  {10.1088/1361-6633/aa67bb}, \href
  {https://ui.adsabs.harvard.edu/abs/2017RPPh...80i6901B} {80, 096901}

\bibitem[\protect\citeauthoryear{Barnes \& Kasen}{Barnes \&
  Kasen}{2013}]{barnes2013}
Barnes J.,  Kasen D.,  2013, The Astrophysical Journal, p.~9

\bibitem[\protect\citeauthoryear{Barnes, Kasen, Wu  \&
  {Mart'inez-Pinedo}}{Barnes et~al.}{2016}]{barnes2016}
Barnes J.,  Kasen D.,  Wu M.-R.,   {Mart'inez-Pinedo} G.,  2016, \mn@doi [The
  Astrophysical Journal] {10.3847/0004-637X/829/2/110}, 829, 1

\bibitem[\protect\citeauthoryear{Barnes, Zhu, Lund, Sprouse, Vassh, McLaughlin,
  Mumpower  \& Surman}{Barnes et~al.}{2021}]{Barnes2021}
Barnes J.,  Zhu Y.~L.,  Lund K.~A.,  Sprouse T.~M.,  Vassh N.,  McLaughlin
  G.~C.,  Mumpower M.~R.,   Surman R.,  2021, \mn@doi [The Astrophysical
  Journal] {10.3847/1538-4357/ac0aec}, 918, 44

\bibitem[\protect\citeauthoryear{Bauswein \& Stergioulas}{Bauswein \&
  Stergioulas}{2019}]{Bauswein2019}
Bauswein A.,  Stergioulas N.,  2019, \mn@doi [J. Phys. G]
  {https://doi.org/10.1016/j.aop.2019.167958}, 46, 113002

\bibitem[\protect\citeauthoryear{Bauswein, Goriely  \& Janka}{Bauswein
  et~al.}{2013}]{bauswein2013}
Bauswein A.,  Goriely S.,   Janka H.~T.,  2013, \mn@doi [Astrophysical Journal]
  {10.1088/0004-637X/773/1/78}, 773

\bibitem[\protect\citeauthoryear{{Bauswein}, {Ardevol Pulpillo}, {Janka}  \&
  {Goriely}}{{Bauswein} et~al.}{2014}]{Bauswein2014}
{Bauswein} A.,  {Ardevol Pulpillo} R.,  {Janka} H.~T.,   {Goriely} S.,  2014,
  \mn@doi [\apjl] {10.1088/2041-8205/795/1/L9}, \href
  {https://ui.adsabs.harvard.edu/abs/2014ApJ...795L...9B} {795, L9}

\bibitem[\protect\citeauthoryear{Bliss, Arcones, Montes  \& Pereira}{Bliss
  et~al.}{2017}]{Bliss2017}
Bliss J.,  Arcones A.,  Montes F.,   Pereira J.,  2017, \mn@doi [Journal of
  Physics G: Nuclear and Particle Physics] {10.1088/1361-6471/aa63bd}, 44,
  054003

\bibitem[\protect\citeauthoryear{Burbidge, Burbidge, Fowler  \& Hoyle}{Burbidge
  et~al.}{1957}]{burbidge1957}
Burbidge E.~M.,  Burbidge G.~R.,  Fowler W.~A.,   Hoyle F.,  1957, \mn@doi
  [Review of Modern Physics] {10.1103/RevModPhys.29.547}, 29

\bibitem[\protect\citeauthoryear{{Butcher}}{{Butcher}}{1987}]{Butcher1987}
{Butcher} H.~R.,  1987, \mn@doi [\nat] {10.1038/328127a0}, \href
  {https://ui.adsabs.harvard.edu/abs/1987Natur.328..127B} {328, 127}

\bibitem[\protect\citeauthoryear{{Caballero}, {Arcones}, {Borzov}, {Langanke}
  \& {Martinez-Pinedo}}{{Caballero} et~al.}{2014}]{Caballero2014}
{Caballero} O.~L.,  {Arcones} A.,  {Borzov} I.~N.,  {Langanke} K.,
  {Martinez-Pinedo} G.,  2014, arXiv e-prints, \href
  {https://ui.adsabs.harvard.edu/abs/2014arXiv1405.0210C} {p. arXiv:1405.0210}

\bibitem[\protect\citeauthoryear{Cameron}{Cameron}{1957}]{cameron1957}
Cameron A. G.~W.,  1957, Publications of the Astronomical Society of the
  Pacific, 69, 201

\bibitem[\protect\citeauthoryear{Cameron}{Cameron}{2001}]{Cameron2001}
Cameron A.,  2001, Astrophys. J, 562, 456

\bibitem[\protect\citeauthoryear{Cameron}{Cameron}{2003}]{cameron2003}
Cameron A.,  2003, Astrophys. J, 587, 327

\bibitem[\protect\citeauthoryear{Capote et~al.,}{Capote
  et~al.}{2009}]{capote2009}
Capote R.,  et~al., 2009, \mn@doi [Nuclear Data Sheets]
  {10.1016/j.nds.2009.10.004}, 110, 3107

\bibitem[\protect\citeauthoryear{{Cayrel} et~al.,}{{Cayrel}
  et~al.}{2001}]{Cayrel01}
{Cayrel} R.,  et~al., 2001, Nature, 409, 691

\bibitem[\protect\citeauthoryear{{Christie}, {Lalakos}, {Tchekhovskoy},
  {Fern{\'a}ndez}, {Foucart}, {Quataert}  \& {Kasen}}{{Christie}
  et~al.}{2019}]{Christie2019}
{Christie} I.~M.,  {Lalakos} A.,  {Tchekhovskoy} A.,  {Fern{\'a}ndez} R.,
  {Foucart} F.,  {Quataert} E.,   {Kasen} D.,  2019, \mn@doi [\mnras]
  {10.1093/mnras/stz2552}, \href
  {https://ui.adsabs.harvard.edu/abs/2019MNRAS.490.4811C} {490, 4811}

\bibitem[\protect\citeauthoryear{{Ciolfi} \& {Kalinani}}{{Ciolfi} \&
  {Kalinani}}{2020}]{Ciolfi2020}
{Ciolfi} R.,  {Kalinani} J.~V.,  2020, \mn@doi [\apjl]
  {10.3847/2041-8213/abb240}, \href
  {https://ui.adsabs.harvard.edu/abs/2020ApJ...900L..35C} {900, L35}

\bibitem[\protect\citeauthoryear{C{\^o}t{\'e} et~al.,}{C{\^o}t{\'e}
  et~al.}{2019}]{cote2019}
C{\^o}t{\'e} B.,  et~al., 2019, \mn@doi [The Astrophysical Journal]
  {10.3847/1538-4357/ab10db}, 875, 106

\bibitem[\protect\citeauthoryear{Cowan, Sneden, Lawler, Aprahamian, Wiescher,
  Langanke, Mart{\'\i}nez-Pinedo  \& Thielemann}{Cowan
  et~al.}{2021}]{cowan2021}
Cowan J.,  Sneden C.,  Lawler J.,  Aprahamian A.,  Wiescher M.,  Langanke K.,
  Mart{\'\i}nez-Pinedo G.,   Thielemann F.-K.,  2021, Rev. Mod. Phys., 93,
  015002

\bibitem[\protect\citeauthoryear{{Cowperthwaite} et~al.,}{{Cowperthwaite}
  et~al.}{2017}]{Cowperthwaite2017}
{Cowperthwaite} P.~S.,  et~al., 2017, \mn@doi [\apjl]
  {10.3847/2041-8213/aa8fc7}, \href
  {https://ui.adsabs.harvard.edu/abs/2017ApJ...848L..17C} {848, L17}

\bibitem[\protect\citeauthoryear{Denissenkov et~al.,}{Denissenkov
  et~al.}{2018}]{Denissenkov2018}
Denissenkov P.,  et~al., 2018, \mn@doi [Journal of Physics G: Nuclear and
  Particle Physics] {10.1088/1361-6471/aabb6e}, 45, 055203

\bibitem[\protect\citeauthoryear{{Drout} et~al.,}{{Drout}
  et~al.}{2017}]{Drout2017}
{Drout} M.~R.,  et~al., 2017, \mn@doi [Science] {10.1126/science.aaq0049},
  \href {https://ui.adsabs.harvard.edu/abs/2017Sci...358.1570D} {358, 1570}

\bibitem[\protect\citeauthoryear{Dvorkin, Daigne, Goriely, Vangioni  \&
  Silk}{Dvorkin et~al.}{2020}]{dvorkin2020}
Dvorkin I.,  Daigne F.,  Goriely S.,  Vangioni E.,   Silk J.,  2020,
  arXiv:2010.00625 [astro-ph]

\bibitem[\protect\citeauthoryear{Eichler et~al.,}{Eichler
  et~al.}{2015}]{Eichler2015}
Eichler M.,  et~al., 2015, \mn@doi [The Astrophysical Journal]
  {10.1088/0004-637x/808/1/30}, 808, 30

\bibitem[\protect\citeauthoryear{Eichler, Sayar, Arcones  \& Raucher}{Eichler
  et~al.}{2019}]{Eichler19}
Eichler M.,  Sayar W.,  Arcones A.,   Raucher T.,  2019, Astrophys. J, 879, 47

\bibitem[\protect\citeauthoryear{{Etienne}, {Faber}, {Liu}, {Shapiro},
  {Taniguchi}  \& {Baumgarte}}{{Etienne} et~al.}{2008}]{Etienne2008}
{Etienne} Z.~B.,  {Faber} J.~A.,  {Liu} Y.~T.,  {Shapiro} S.~L.,  {Taniguchi}
  K.,   {Baumgarte} T.~W.,  2008, \mn@doi [\prd] {10.1103/PhysRevD.77.084002},
  \href {https://ui.adsabs.harvard.edu/abs/2008PhRvD..77h4002E} {77, 084002}

\bibitem[\protect\citeauthoryear{{Etienne}, {Liu}, {Shapiro}  \&
  {Baumgarte}}{{Etienne} et~al.}{2009}]{Etienne2009}
{Etienne} Z.~B.,  {Liu} Y.~T.,  {Shapiro} S.~L.,   {Baumgarte} T.~W.,  2009,
  \mn@doi [\prd] {10.1103/PhysRevD.79.044024}, \href
  {https://ui.adsabs.harvard.edu/abs/2009PhRvD..79d4024E} {79, 044024}

\bibitem[\protect\citeauthoryear{{Even} et~al.,}{{Even}
  et~al.}{2020}]{Even2020}
{Even} W.,  et~al., 2020, \mn@doi [\apj] {10.3847/1538-4357/ab70b9}, \href
  {https://ui.adsabs.harvard.edu/abs/2020ApJ...899...24E} {899, 24}

\bibitem[\protect\citeauthoryear{{Fahlman} \& {Fern{\'a}ndez}}{{Fahlman} \&
  {Fern{\'a}ndez}}{2022}]{Fahlman2022}
{Fahlman} S.,  {Fern{\'a}ndez} R.,  2022, \mn@doi [\mnras]
  {10.1093/mnras/stac948}, \href
  {https://ui.adsabs.harvard.edu/abs/2022MNRAS.513.2689F} {513, 2689}

\bibitem[\protect\citeauthoryear{Fantina, Chamel, Pearson  \& Goriely}{Fantina
  et~al.}{2013}]{fantina2013}
Fantina A.,  Chamel N.,  Pearson J.,   Goriely S.,  2013, \mn@doi [A\&A]
  {10.1051/0004-6361/201321884}, 559, A128

\bibitem[\protect\citeauthoryear{{Fern{\'a}ndez}, {Foucart}, {Kasen},
  {Lippuner}, {Desai}  \& {Roberts}}{{Fern{\'a}ndez}
  et~al.}{2017}]{Fernandez2017}
{Fern{\'a}ndez} R.,  {Foucart} F.,  {Kasen} D.,  {Lippuner} J.,  {Desai} D.,
  {Roberts} L.~F.,  2017, \mn@doi [Classical and Quantum Gravity]
  {10.1088/1361-6382/aa7a77}, \href
  {https://ui.adsabs.harvard.edu/abs/2017CQGra..34o4001F} {34, 154001}

\bibitem[\protect\citeauthoryear{{Fern{\'a}ndez}, {Tchekhovskoy}, {Quataert},
  {Foucart}  \& {Kasen}}{{Fern{\'a}ndez} et~al.}{2019}]{Fernandez2019}
{Fern{\'a}ndez} R.,  {Tchekhovskoy} A.,  {Quataert} E.,  {Foucart} F.,
  {Kasen} D.,  2019, \mn@doi [\mnras] {10.1093/mnras/sty2932}, \href
  {https://ui.adsabs.harvard.edu/abs/2019MNRAS.482.3373F} {482, 3373}

\bibitem[\protect\citeauthoryear{{Fern{\'a}ndez}, {Foucart}  \&
  {Lippuner}}{{Fern{\'a}ndez} et~al.}{2020}]{Fernandez2020}
{Fern{\'a}ndez} R.,  {Foucart} F.,   {Lippuner} J.,  2020, \mn@doi [\mnras]
  {10.1093/mnras/staa2209}, \href
  {https://ui.adsabs.harvard.edu/abs/2020MNRAS.497.3221F} {497, 3221}

\bibitem[\protect\citeauthoryear{{Fischer}, {Whitehouse}, {Mezzacappa},
  {Thielemann}  \& {Liebend{\"o}rfer}}{{Fischer} et~al.}{2010}]{Fischer2010}
{Fischer} T.,  {Whitehouse} S.~C.,  {Mezzacappa} A.,  {Thielemann} F.~K.,
  {Liebend{\"o}rfer} M.,  2010, \mn@doi [Astronomy and Astrophysics]
  {10.1051/0004-6361/200913106}, \href
  {https://ui.adsabs.harvard.edu/abs/2010A&A...517A..80F} {517, A80}

\bibitem[\protect\citeauthoryear{Foucart et~al.,}{Foucart
  et~al.}{2016}]{foucart2016}
Foucart F.,  et~al., 2016, \mn@doi [Physical Review D]
  {10.1103/PhysRevD.93.044019}, 93, 044019

\bibitem[\protect\citeauthoryear{Fowler \& Hoyle}{Fowler \&
  Hoyle}{1960}]{Fowler60}
Fowler W.,  Hoyle F.,  1960, Ann. Phys., 10, 280

\bibitem[\protect\citeauthoryear{Frebel, Christlieb, Norris, Thom, Beers  \&
  Rhee}{Frebel et~al.}{2007}]{frebel2007}
Frebel A.,  Christlieb N.,  Norris J.~E.,  Thom C.,  Beers T.~C.,   Rhee J.,
  2007, \mn@doi [The Astrophysical Journal] {10.1086/518122}, 660, L117

\bibitem[\protect\citeauthoryear{Freiburghaus \& Rosswog}{Freiburghaus \&
  Rosswog}{1999}]{Freiburghaus1999}
Freiburghaus C.,  Rosswog S.,  1999, Astrophys. J. Lett., 525, 4

\bibitem[\protect\citeauthoryear{Fujibayashi, Kiuchi, Nishimura, Sekiguchi  \&
  Shibata}{Fujibayashi et~al.}{2018}]{Fujibayashi2018}
Fujibayashi S.,  Kiuchi K.,  Nishimura N.,  Sekiguchi Y.,   Shibata M.,  2018,
  \mn@doi [The Astrophysical Journal] {10.3847/1538-4357/aabafd}, 860, 64

\bibitem[\protect\citeauthoryear{{Fujibayashi}, {Shibata}, {Wanajo}, {Kiuchi},
  {Kyutoku}  \& {Sekiguchi}}{{Fujibayashi} et~al.}{2020a}]{Fujibayashi2020b}
{Fujibayashi} S.,  {Shibata} M.,  {Wanajo} S.,  {Kiuchi} K.,  {Kyutoku} K.,
  {Sekiguchi} Y.,  2020a, \mn@doi [\prd] {10.1103/PhysRevD.101.083029}, \href
  {https://ui.adsabs.harvard.edu/abs/2020PhRvD.101h3029F} {101, 083029}

\bibitem[\protect\citeauthoryear{{Fujibayashi}, {Shibata}, {Wanajo}, {Kiuchi},
  {Kyutoku}  \& {Sekiguchi}}{{Fujibayashi} et~al.}{2020b}]{Fujibayashi2020}
{Fujibayashi} S.,  {Shibata} M.,  {Wanajo} S.,  {Kiuchi} K.,  {Kyutoku} K.,
  {Sekiguchi} Y.,  2020b, \mn@doi [\prd] {10.1103/PhysRevD.102.123014}, \href
  {https://ui.adsabs.harvard.edu/abs/2020PhRvD.102l3014F} {102, 123014}

\bibitem[\protect\citeauthoryear{{Giuliani}, {Mart{\'\i}nez-Pinedo}, {Wu}  \&
  {Robledo}}{{Giuliani} et~al.}{2020}]{Giuliani2020}
{Giuliani} S.~A.,  {Mart{\'\i}nez-Pinedo} G.,  {Wu} M.-R.,   {Robledo} L.~M.,
  2020, \mn@doi [\prc] {10.1103/PhysRevC.102.045804}, \href
  {https://ui.adsabs.harvard.edu/abs/2020PhRvC.102d5804G} {102, 045804}

\bibitem[\protect\citeauthoryear{Gogny}{Gogny}{1973}]{Gogny73}
Gogny D.,  1973, in de Boer J.,  Mang H.~J.,  eds, Proceedings of the
  International Conference on Nuclear Physics. p.~48

\bibitem[\protect\citeauthoryear{{Goriely}}{{Goriely}}{1997}]{goriely1997b}
{Goriely} S.,  1997, \aap, \href
  {https://ui.adsabs.harvard.edu/abs/1997A&A...325..414G} {325, 414}

\bibitem[\protect\citeauthoryear{Goriely}{Goriely}{1999}]{goriely1999}
Goriely S.,  1999, Astronomy and Astrophysics, 342, 881

\bibitem[\protect\citeauthoryear{Goriely}{Goriely}{2015a}]{Goriely15b}
Goriely S.,  2015a, Eur. Phys. J. A, 51, 172

\bibitem[\protect\citeauthoryear{Goriely}{Goriely}{2015b}]{goriely2015}
Goriely S.,  2015b, \mn@doi [European Physical Journal A]
  {10.1140/epja/i2015-15022-3}, 51, 1

\bibitem[\protect\citeauthoryear{Goriely \& Arnould}{Goriely \&
  Arnould}{1992}]{Goriely92}
Goriely S.,  Arnould M.,  1992, Astron. Astrophys., 262, 73

\bibitem[\protect\citeauthoryear{Goriely \& Arnould}{Goriely \&
  Arnould}{2001}]{Goriely2001}
Goriely S.,  Arnould M.,  2001, Astron. Astrophys., 379, 1113

\bibitem[\protect\citeauthoryear{Goriely \& Capote}{Goriely \&
  Capote}{2014}]{Goriely2014a}
Goriely S.,  Capote R.,  2014, Phys. Rev. C, 89, 054318

\bibitem[\protect\citeauthoryear{Goriely \& Delaroche}{Goriely \&
  Delaroche}{2007}]{Goriely07b}
Goriely S.,  Delaroche J.-P.,  2007, Phys. Lett. B, 653, 178

\bibitem[\protect\citeauthoryear{Goriely \& Janka}{Goriely \&
  Janka}{2016}]{Goriely2016d}
Goriely S.,  Janka H.-T.,  2016, MNRAS, 459, 4174

\bibitem[\protect\citeauthoryear{Goriely \& Mart{\'\i}nez~Pinedo}{Goriely \&
  Mart{\'\i}nez~Pinedo}{2015}]{goriely2015b}
Goriely S.,  Mart{\'\i}nez~Pinedo G.,  2015, \mn@doi [Nuclear Physics A]
  {10.1016/j.nuclphysa.2015.07.020}, 944, 158

\bibitem[\protect\citeauthoryear{Goriely, Hilaire  \& Koning}{Goriely
  et~al.}{2008}]{goriely2008}
Goriely S.,  Hilaire S.,   Koning A.~J.,  2008, \mn@doi [Astronomy \&
  Astrophysics] {10.1051/0004-6361:20078825}, 487, 767

\bibitem[\protect\citeauthoryear{Goriely, Hilaire, Girod  \& P\'eru}{Goriely
  et~al.}{2009}]{Goriely2009b}
Goriely S.,  Hilaire S.,  Girod M.,   P\'eru S.,  2009, \mn@doi [Phys. Rev.
  Lett.] {10.1103/PhysRevLett.102.242501}, 102, 242501

\bibitem[\protect\citeauthoryear{Goriely, Chame  \& Pearson}{Goriely
  et~al.}{2010}]{goriely2010}
Goriely S.,  Chame N.,   Pearson J.,  2010, Physical Review C, 82, 035804

\bibitem[\protect\citeauthoryear{Goriely, Bauswein  \& Janka}{Goriely
  et~al.}{2011}]{goriely2011}
Goriely S.,  Bauswein A.,   Janka H.~T.,  2011, \mn@doi [Astrophysical Journal
  Letters] {10.1088/2041-8205/738/2/L32}, 738

\bibitem[\protect\citeauthoryear{Goriely, Sida, Lema{\^\i}tre, Panebianco,
  Dubray, Hilaire, Bauswein  \& Janka}{Goriely et~al.}{2013}]{goriely2013}
Goriely S.,  Sida J.~L.,  Lema{\^\i}tre J.~F.,  Panebianco S.,  Dubray N.,
  Hilaire S.,  Bauswein A.,   Janka H.~T.,  2013, \mn@doi [Physical Review
  Letters] {10.1103/PhysRevLett.111.242502}, 111, 1

\bibitem[\protect\citeauthoryear{Goriely, Bauswein, Just, Pllumbi  \&
  Janka}{Goriely et~al.}{2015}]{goriely2015a}
Goriely S.,  Bauswein A.,  Just O.,  Pllumbi E.,   Janka H.~T.,  2015, \mn@doi
  [Monthly Notices of the Royal Astronomical Society] {10.1093/mnras/stv1526},
  452, 3894

\bibitem[\protect\citeauthoryear{Goriely, Chamel  \& Pearson}{Goriely
  et~al.}{2016a}]{Goriely2016c}
Goriely S.,  Chamel N.,   Pearson J.~M.,  2016a, Phys. Rev. C, 93, 034337

\bibitem[\protect\citeauthoryear{Goriely, Bauswein, Janka, Panebianco, Sida,
  Lemaitre, Hialire  \& Dubray}{Goriely et~al.}{2016b}]{goriely2016b}
Goriely S.,  Bauswein A.,  Janka H.,  Panebianco S.,  Sida J.-L.,  Lemaitre J.,
   Hialire S.,   Dubray N.,  2016b, Journal of Physics: Conference Series, 665,
  012052

\bibitem[\protect\citeauthoryear{Goriely, Hilaire, P{\'e}ru  \& Sieja}{Goriely
  et~al.}{2018}]{Goriely18a}
Goriely S.,  Hilaire S.,  P{\'e}ru S.,   Sieja K.,  2018, Phys. Rev. C, 98,
  014327

\bibitem[\protect\citeauthoryear{{Grimmett}, {M{\"u}ller}, {Heger}, {Banerjee}
  \& {Obergaulinger}}{{Grimmett} et~al.}{2020}]{Grimmett2020}
{Grimmett} J.~J.,  {M{\"u}ller} B.,  {Heger} A.,  {Banerjee} P.,
  {Obergaulinger} M.,  2020, arXiv, 2010.06766

\bibitem[\protect\citeauthoryear{Hayashi, Fujibayashi, Kiuchi, Kyutoku,
  Sekiguchi  \& Shibata}{Hayashi et~al.}{2021}]{Hayashi2021}
Hayashi K.,  Fujibayashi S.,  Kiuchi K.,  Kyutoku K.,  Sekiguchi Y.,   Shibata
  M.,  2021, \mn@doi [arXiv] {10.48550/ARXIV.2111.04621}

\bibitem[\protect\citeauthoryear{Hilaire, Goriely, P{\'e}ru, Dubray, Dupuis  \&
  Bauge}{Hilaire et~al.}{2016}]{Hilaire16}
Hilaire S.,  Goriely S.,  P{\'e}ru S.,  Dubray N.,  Dupuis M.,   Bauge E.,
  2016, Eur. Phys. J. A, 52, 336

\bibitem[\protect\citeauthoryear{{Hill, V.}, {Christlieb, N.}, {Beers, T. C.},
  {Barklem, P. S.}, {Kratz, K.-L.}, {Nordstr\"om, B.}, {Pfeiffer, B.}  \&
  {Farouqi, K.}}{{Hill, V.} et~al.}{2017}]{Hill2017}
{Hill, V.} {Christlieb, N.} {Beers, T. C.} {Barklem, P. S.} {Kratz, K.-L.}
  {Nordstr\"om, B.} {Pfeiffer, B.}  {Farouqi, K.} 2017, \mn@doi [A\&A]
  {10.1051/0004-6361/201629092}, 607, A91

\bibitem[\protect\citeauthoryear{Hoffman, Woosley  \& Qian}{Hoffman
  et~al.}{1997}]{hoffman1997}
Hoffman R.~D.,  Woosley S.~E.,   Qian Y.-Z.,  1997, \mn@doi [The Astrophysical
  Journal] {10.1086/304181}, 482, 951

\bibitem[\protect\citeauthoryear{Holmbeck et~al.,}{Holmbeck
  et~al.}{2018}]{holmbeck2018}
Holmbeck E.~M.,  et~al., 2018, \mn@doi [The Astrophysical Journal]
  {10.3847/2041-8213/aac722}, 859, L24

\bibitem[\protect\citeauthoryear{Holmbeck, Frebel, McLaughlin, Mumpower,
  Sprouse  \& Surman}{Holmbeck et~al.}{2019}]{Holmbeck19}
Holmbeck E.,  Frebel A.,  McLaughlin G.,  Mumpower M.,  Sprouse T.~M.,   Surman
  R.,  2019, Astrophys. J, 881, 5

\bibitem[\protect\citeauthoryear{{Hossein Nouri} et~al.,}{{Hossein Nouri}
  et~al.}{2018}]{HosseinNouri2018}
{Hossein Nouri} F.,  et~al., 2018, \mn@doi [\prd] {10.1103/PhysRevD.97.083014},
  \href {https://ui.adsabs.harvard.edu/abs/2018PhRvD..97h3014H} {97, 083014}

\bibitem[\protect\citeauthoryear{Hotokezaka, Kiuchi, Kyutoku, Okawa, Sekiguchi,
  Shibata  \& Taniguchi}{Hotokezaka et~al.}{2013}]{hotokezaka2013}
Hotokezaka K.,  Kiuchi K.,  Kyutoku K.,  Okawa H.,  Sekiguchi Y.-i.,  Shibata
  M.,   Taniguchi K.,  2013, \mn@doi [Physical Review D]
  {10.1103/PhysRevD.87.024001}, 87, 024001

\bibitem[\protect\citeauthoryear{Hotokezaka, Beniamini  \& Piran}{Hotokezaka
  et~al.}{2018}]{Hotokezaka2018}
Hotokezaka K.,  Beniamini P.,   Piran T.,  2018, \mn@doi [International Journal
  of Modern Physics D] {10.1142/S0218271818420051}, 27, 1842005

\bibitem[\protect\citeauthoryear{{H{\"u}depohl}, {M{\"u}ller}, {Janka}, {Marek}
   \& {Raffelt}}{{H{\"u}depohl} et~al.}{2010}]{Hudepohl2010}
{H{\"u}depohl} L.,  {M{\"u}ller} B.,  {Janka} H.~T.,  {Marek} A.,   {Raffelt}
  G.~G.,  2010, \mn@doi [\prl] {10.1103/PhysRevLett.104.251101}, \href
  {https://ui.adsabs.harvard.edu/abs/2010PhRvL.104y1101H} {104, 251101}

\bibitem[\protect\citeauthoryear{{Janka}}{{Janka}}{2012}]{janka2012}
{Janka} H.-T.,  2012, \mn@doi [Ann. Rev. of Nucl. Part. Sci.]
  {10.1146/annurev-nucl-102711-094901}, 62, 407

\bibitem[\protect\citeauthoryear{{Janka} \& {Ruffert}}{{Janka} \&
  {Ruffert}}{2002}]{Janka2002}
{Janka} H.~T.,  {Ruffert} M.,  2002, in {Shara} M.~M.,  ed.,  Astronomical
  Society of the Pacific Conference Series Vol. 263, Stellar Collisions,
  Mergers and their Consequences. p.~333 (\mn@eprint {arXiv}
  {astro-ph/0101357})

\bibitem[\protect\citeauthoryear{{Janka}, {Eberl}, {Ruffert}  \&
  {Fryer}}{{Janka} et~al.}{1999}]{Janka1999}
{Janka} H.~T.,  {Eberl} T.,  {Ruffert} M.,   {Fryer} C.~L.,  1999, \mn@doi
  [\apjl] {10.1086/312397}, \href
  {https://ui.adsabs.harvard.edu/abs/1999ApJ...527L..39J} {527, L39}

\bibitem[\protect\citeauthoryear{{Ji}, {Frebel}, {Simon}  \& {Chiti}}{{Ji}
  et~al.}{2016}]{Ji2016}
{Ji} A.~P.,  {Frebel} A.,  {Simon} J.~D.,   {Chiti} A.,  2016, \mn@doi [\apj]
  {10.3847/0004-637X/830/2/93}, \href
  {https://ui.adsabs.harvard.edu/abs/2016ApJ...830...93J} {830, 93}

\bibitem[\protect\citeauthoryear{Just, Bauswein, Ardevol~Pulpillo, Goriely  \&
  Janka}{Just et~al.}{2015a}]{Just2015}
Just O.,  Bauswein A.,  Ardevol~Pulpillo R.,  Goriely S.,   Janka H.~T.,
  2015a, \mn@doi [Monthly Notices of the Royal Astronomical Society]
  {10.1093/mnras/stv009}, 448, 541

\bibitem[\protect\citeauthoryear{Just, Obergaulinger  \& Janka}{Just
  et~al.}{2015b}]{Just2015b}
Just O.,  Obergaulinger M.,   Janka H.-T.,  2015b, \mn@doi [Monthly Notices of
  the Royal Astronomical Society] {10.1093/mnras/stv1892}, 453, 3386

\bibitem[\protect\citeauthoryear{Just, Kullmann, Goriely, Bauswein, Janka  \&
  Collins}{Just et~al.}{2021a}]{Just2021}
Just O.,  Kullmann I.,  Goriely S.,  Bauswein A.,  Janka H.-T.,   Collins
  C.~E.,  2021a, \mn@doi [Monthly Notices of the Royal Astronomical Society]
  {10.1093/mnras/stab3327}

\bibitem[\protect\citeauthoryear{Just, Goriely, Janka, Nagataki  \&
  Bauswein}{Just et~al.}{2021b}]{Just2021c}
Just O.,  Goriely S.,  Janka H.-T.,  Nagataki S.,   Bauswein A.,  2021b,
  \mn@doi [Monthly Notices of the Royal Astronomical Society]
  {10.1093/mnras/stab2861}, 509, 1377

\bibitem[\protect\citeauthoryear{{Just}, {Aloy}, {Obergaulinger}  \&
  {Nagataki}}{{Just} et~al.}{2022}]{Just2022}
{Just} O.,  {Aloy} M.~A.,  {Obergaulinger} M.,   {Nagataki} S.,  2022, arXiv
  e-prints, \href {https://ui.adsabs.harvard.edu/abs/2022arXiv220514158J} {p.
  arXiv:2205.14158}

\bibitem[\protect\citeauthoryear{Kasen, Badnell  \& Barnes}{Kasen
  et~al.}{2013}]{Kasen2013}
Kasen D.,  Badnell N.,   Barnes J.,  2013, Astrophys. J, 774, 25

\bibitem[\protect\citeauthoryear{Kasen, Metzger, Barnes, Quataert  \&
  {Ramirez-Ruiz}}{Kasen et~al.}{2017}]{kasen2017}
Kasen D.,  Metzger B.,  Barnes J.,  Quataert E.,   {Ramirez-Ruiz} E.,  2017,
  \mn@doi [Nature] {10.1038/nature24453}, 551, 80

\bibitem[\protect\citeauthoryear{Kawaguchi, Fujibayashi, Shibata, Tanaka  \&
  Wanajo}{Kawaguchi et~al.}{2021}]{Kawaguchi2021}
Kawaguchi K.,  Fujibayashi S.,  Shibata M.,  Tanaka M.,   Wanajo S.,  2021,
  \mn@doi [The Astrophysical Journal] {10.3847/1538-4357/abf3bc}, 913, 100

\bibitem[\protect\citeauthoryear{{Kelic}, {Ricciardi}  \& {Schmidt}}{{Kelic}
  et~al.}{2008}]{Kelic2008}
{Kelic} A.,  {Ricciardi} M.~V.,   {Schmidt} K.~H.,  2008, in Dynamical Aspects
  of Nuclear Fission. pp 203--215, \mn@doi{10.1142/9789812837530\_0016}

\bibitem[\protect\citeauthoryear{Kelic, Ricciardi  \& Schmidt}{Kelic
  et~al.}{2009}]{Kelic2009}
Kelic A.,  Ricciardi M.~V.,   Schmidt K.-H.,  2009, \mn@doi [arXiv e-prints]
  {10.48550/ARXIV.0906.4193}, p. arXiv:0906.4193

\bibitem[\protect\citeauthoryear{{Kilpatrick} et~al.,}{{Kilpatrick}
  et~al.}{2017}]{Kilpatrick2017}
{Kilpatrick} C.~D.,  et~al., 2017, \mn@doi [Science] {10.1126/science.aaq0073},
  \href {https://ui.adsabs.harvard.edu/abs/2017Sci...358.1583K} {358, 1583}

\bibitem[\protect\citeauthoryear{{Kiuchi}, {Sekiguchi}, {Kyutoku}, {Shibata},
  {Taniguchi}  \& {Wada}}{{Kiuchi} et~al.}{2015}]{Kiuchi2015}
{Kiuchi} K.,  {Sekiguchi} Y.,  {Kyutoku} K.,  {Shibata} M.,  {Taniguchi} K.,
  {Wada} T.,  2015, \mn@doi [\prd] {10.1103/PhysRevD.92.064034}, \href
  {https://ui.adsabs.harvard.edu/abs/2015PhRvD..92f4034K} {92, 064034}

\bibitem[\protect\citeauthoryear{{Kiuchi}, {Kyutoku}, {Sekiguchi}  \&
  {Shibata}}{{Kiuchi} et~al.}{2018}]{Kiuchi2018}
{Kiuchi} K.,  {Kyutoku} K.,  {Sekiguchi} Y.,   {Shibata} M.,  2018, \mn@doi
  [\prd] {10.1103/PhysRevD.97.124039}, \href
  {https://ui.adsabs.harvard.edu/abs/2018PhRvD..97l4039K} {97, 124039}

\bibitem[\protect\citeauthoryear{Klapdor, Metzinger  \& Oda}{Klapdor
  et~al.}{1984}]{klapdor1984}
Klapdor H.,  Metzinger J.,   Oda T.,  1984, \mn@doi [Atomic Data and Nuclear
  Data Tables] {https://doi.org/10.1016/0092-640X(84)90017-2}, 31, 81

\bibitem[\protect\citeauthoryear{{Kodama} \& {Takahashi}}{{Kodama} \&
  {Takahashi}}{1975}]{Kodama1975}
{Kodama} T.,  {Takahashi} K.,  1975, \mn@doi [\nphysa]
  {10.1016/0375-9474(75)90381-4}, \href
  {https://ui.adsabs.harvard.edu/abs/1975NuPhA.239..489K} {239, 489}

\bibitem[\protect\citeauthoryear{Kondev, Wang, Huang, Naimi  \& Audi}{Kondev
  et~al.}{2021}]{kondev2021}
Kondev F.,  Wang M.,  Huang W.,  Naimi S.,   Audi G.,  2021, \mn@doi [Chinese
  Physics C] {10.1088/1674-1137/abddae}, 45, 030001

\bibitem[\protect\citeauthoryear{Koning \& Delaroche}{Koning \&
  Delaroche}{2003}]{koning2003}
Koning A.,  Delaroche J.,  2003, \mn@doi [Nuclear Physics A]
  {https://doi.org/10.1016/S0375-9474(02)01321-0}, 713, 231

\bibitem[\protect\citeauthoryear{Koning \& Rochman}{Koning \&
  Rochman}{2012}]{koning2012}
Koning A.~J.,  Rochman D.,  2012, \mn@doi [Nuclear Data Sheets]
  {http://dx.doi.org/10.1016/j.nds.2012.11.002}, 113, 2841

\bibitem[\protect\citeauthoryear{Koura, Tachibana  \& Yoshida}{Koura
  et~al.}{2002}]{Koura2002}
Koura H.,  Tachibana T.,   Yoshida T.,  2002, \mn@doi [Journal of Nuclear
  Science and Technology] {10.1080/00223131.2002.10875212}, 39, 774

\bibitem[\protect\citeauthoryear{{Kr{\"u}ger} \& {Foucart}}{{Kr{\"u}ger} \&
  {Foucart}}{2020}]{Kruger2020}
{Kr{\"u}ger} C.~J.,  {Foucart} F.,  2020, \mn@doi [\prd]
  {10.1103/PhysRevD.101.103002}, \href
  {https://ui.adsabs.harvard.edu/abs/2020PhRvD.101j3002K} {101, 103002}

\bibitem[\protect\citeauthoryear{{Kulkarni}}{{Kulkarni}}{2005}]{Kulkarni2005}
{Kulkarni} S.~R.,  2005, arXiv, 0510256

\bibitem[\protect\citeauthoryear{Kullmann, Goriely, Just, Ardevol-Pulpillo,
  Bauswein  \& Janka}{Kullmann et~al.}{2021}]{kullmann2021}
Kullmann I.,  Goriely S.,  Just O.,  Ardevol-Pulpillo R.,  Bauswein A.,   Janka
  H.-T.,  2021, \mn@doi [Monthly Notices of the Royal Astronomical Society]
  {10.1093/mnras/stab3393}

\bibitem[\protect\citeauthoryear{{Kyutoku}, {Ioka}  \& {Shibata}}{{Kyutoku}
  et~al.}{2013}]{Kyutoku2013}
{Kyutoku} K.,  {Ioka} K.,   {Shibata} M.,  2013, \mn@doi [\prd]
  {10.1103/PhysRevD.88.041503}, \href
  {https://ui.adsabs.harvard.edu/abs/2013PhRvD..88d1503K} {88, 041503}

\bibitem[\protect\citeauthoryear{Lattimer \& Schramm}{Lattimer \&
  Schramm}{1974}]{lattimer1974}
Lattimer J.~M.,  Schramm D.~N.,  1974, Astrophysical Journal, 192, 145

\bibitem[\protect\citeauthoryear{Lema{\^\i}tre, Goriely, Hilaire  \&
  Sida}{Lema{\^\i}tre et~al.}{2019}]{Lemaitre2019}
Lema{\^\i}tre J.-F.,  Goriely S.,  Hilaire S.,   Sida J.-L.,  2019, \mn@doi
  [Physical Review C] {10.1103/PhysRevC.99.034612}, 99, 034612

\bibitem[\protect\citeauthoryear{Lema{\^\i}tre, Goriely, Bauswein  \&
  Janka}{Lema{\^\i}tre et~al.}{2021}]{Lemaitre2021}
Lema{\^\i}tre J.-F.,  Goriely S.,  Bauswein A.,   Janka H.-T.,  2021, Phys.
  Rev. C, 103, 025806

\bibitem[\protect\citeauthoryear{Li \& Paczy{\'n}ski}{Li \&
  Paczy{\'n}ski}{1998}]{li1998}
Li L.-X.,  Paczy{\'n}ski B.,  1998, \mn@doi [The Astrophysical Journal]
  {10.1086/311680}, 507, L59

\bibitem[\protect\citeauthoryear{Li et~al.,}{Li et~al.}{2022}]{Li2022}
Li H.~F.,  et~al., 2022, \mn@doi [Phys. Rev. Lett.]
  {10.1103/PhysRevLett.128.152701}, 128, 152701

\bibitem[\protect\citeauthoryear{Liddick et~al.,}{Liddick
  et~al.}{2016}]{Liddick2016}
Liddick S.~N.,  et~al., 2016, \mn@doi [Phys. Rev. Lett.]
  {10.1103/PhysRevLett.116.242502}, 116, 242502

\bibitem[\protect\citeauthoryear{Liu, Wang, Deng  \& Wu}{Liu
  et~al.}{2011}]{Liu11}
Liu M.,  Wang N.,  Deng Y.,   Wu X.,  2011, Phys. Rev. C, 84, 014333

\bibitem[\protect\citeauthoryear{{Lund}, {Engel}, {McLaughlin}, {Mumpower},
  {Ney}  \& {Surman}}{{Lund} et~al.}{2022}]{Lund2022}
{Lund} K.~A.,  {Engel} J.,  {McLaughlin} G.~C.,  {Mumpower} M.~R.,  {Ney}
  E.~M.,   {Surman} R.,  2022, arXiv e-prints, \href
  {https://ui.adsabs.harvard.edu/abs/2022arXiv220806373L} {p. arXiv:2208.06373}

\bibitem[\protect\citeauthoryear{Lunney, Pearson  \& Thibault}{Lunney
  et~al.}{2003}]{Lunney03}
Lunney D.,  Pearson J.,   Thibault C.,  2003, Rev. Mod. Phys., 75, 1021

\bibitem[\protect\citeauthoryear{MacFadyen \& Woosley}{MacFadyen \&
  Woosley}{1999}]{macfadyen1999}
MacFadyen A.~I.,  Woosley S.~E.,  1999, \mn@doi [The Astrophysical Journal]
  {10.1086/307790}, 524, 262

\bibitem[\protect\citeauthoryear{Marketin, Huther  \&
  {Martinez-Pinedo}}{Marketin et~al.}{2016}]{Marketin2016}
Marketin T.,  Huther L.,   {Martinez-Pinedo} G.,  2016, Physical Review C, 93,
  025805

\bibitem[\protect\citeauthoryear{Martin, Arcones, Nazarewicz  \& Olsen}{Martin
  et~al.}{2016}]{Martin2016}
Martin D.,  Arcones A.,  Nazarewicz W.,   Olsen E.,  2016, \mn@doi [Phys. Rev.
  Lett.] {10.1103/PhysRevLett.116.121101}, 116, 121101

\bibitem[\protect\citeauthoryear{Martin, Perego, Kastaun  \& Arcones}{Martin
  et~al.}{2018}]{martin2018}
Martin D.,  Perego A.,  Kastaun W.,   Arcones A.,  2018, \mn@doi [Classical and
  Quantum Gravity] {10.1088/1361-6382/aa9f5a}, 35, 034001

\bibitem[\protect\citeauthoryear{McKay, Denissenkov, Herwig, Perdikakis  \&
  Schatz}{McKay et~al.}{2019}]{McKay2020}
McKay J.~E.,  Denissenkov P.~A.,  Herwig F.,  Perdikakis G.,   Schatz H.,
  2019, \mn@doi [Monthly Notices of the Royal Astronomical Society]
  {10.1093/mnras/stz3322}, 491, 5179

\bibitem[\protect\citeauthoryear{{Mendoza-Temis}, Wu, Langanke,
  {Mart{\'\i}nez-Pinedo}, Bauswein  \& Janka}{{Mendoza-Temis}
  et~al.}{2015}]{Mendoza-Temis2015}
{Mendoza-Temis} J. D.~J.,  Wu M.~R.,  Langanke K.,  {Mart{\'\i}nez-Pinedo} G.,
  Bauswein A.,   Janka H.~T.,  2015, \mn@doi [Physical Review C - Nuclear
  Physics] {10.1103/PhysRevC.92.055805}, 92, 1

\bibitem[\protect\citeauthoryear{Metzger \& Fern{\'a}ndez}{Metzger \&
  Fern{\'a}ndez}{2014}]{Metzger2014}
Metzger B.~D.,  Fern{\'a}ndez R.,  2014, \mn@doi [Monthly Notices of the Royal
  Astronomical Society] {10.1093/mnras/stu802}, 441, 3444

\bibitem[\protect\citeauthoryear{Metzger et~al.,}{Metzger
  et~al.}{2010}]{metzger2010}
Metzger B.~D.,  et~al., 2010, MNRAS, p.~13

\bibitem[\protect\citeauthoryear{Meyer}{Meyer}{1989}]{meyer1989}
Meyer B.~S.,  1989, Astrophysical Journal, 343, 254

\bibitem[\protect\citeauthoryear{Miller et~al.,}{Miller
  et~al.}{2019}]{miller2019}
Miller J.~M.,  et~al., 2019, \mn@doi [Physical Review D]
  {10.1103/PhysRevD.100.023008}, 100, 023008

\bibitem[\protect\citeauthoryear{{Mirizzi}, {Tamborra}, {Janka}, {Saviano},
  {Scholberg}, {Bollig}, {H{\"u}depohl}  \& {Chakraborty}}{{Mirizzi}
  et~al.}{2016}]{Mirizzi2016}
{Mirizzi} A.,  {Tamborra} I.,  {Janka} H.~T.,  {Saviano} N.,  {Scholberg} K.,
  {Bollig} R.,  {H{\"u}depohl} L.,   {Chakraborty} S.,  2016, \mn@doi [Nuovo
  Cimento Rivista Serie] {10.1393/ncr/i2016-10120-8}, \href
  {https://ui.adsabs.harvard.edu/abs/2016NCimR..39....1M} {39, 1}

\bibitem[\protect\citeauthoryear{M\"oller, Pfeiffer  \& Kratz}{M\"oller
  et~al.}{2003}]{moller2003}
M\"oller P.,  Pfeiffer B.,   Kratz K.-L.,  2003, \mn@doi [Phys. Rev. C]
  {10.1103/PhysRevC.67.055802}, 67, 055802

\bibitem[\protect\citeauthoryear{M{\"o}ller, Sierk, Ichikawa  \&
  Sagawa}{M{\"o}ller et~al.}{2016}]{Moller2016}
M{\"o}ller P.,  Sierk A.,  Ichikawa T.,   Sagawa H.,  2016, \mn@doi [Atomic
  Data and Nuclear Data Tables] {https://doi.org/10.1016/j.adt.2015.10.002},
  109-110, 1

\bibitem[\protect\citeauthoryear{{M{\"o}sta}, {Roberts}, {Halevi}, {Ott},
  {Lippuner}, {Haas}  \& {Schnetter}}{{M{\"o}sta} et~al.}{2018}]{Mosta2018}
{M{\"o}sta} P.,  {Roberts} L.~F.,  {Halevi} G.,  {Ott} C.~D.,  {Lippuner} J.,
  {Haas} R.,   {Schnetter} E.,  2018, \apj, 864, 171

\bibitem[\protect\citeauthoryear{Mumpower, Surman, McLaughlin  \&
  Aprahamian}{Mumpower et~al.}{2016}]{Mumpower2016}
Mumpower M.~R.,  Surman R.,  McLaughlin G.~C.,   Aprahamian A.,  2016, \mn@doi
  [Progress in Particle and Nuclear Physics] {10.1016/j.ppnp.2015.09.001}, 86,
  86

\bibitem[\protect\citeauthoryear{Myers \& \ifmmode \acute{S}\else
  \'{S}\fi{}wia\ifmmode~\mbox{\c{}}\else \c{}\fi{}tecki}{Myers \& \ifmmode
  \acute{S}\else \'{S}\fi{}wia\ifmmode~\mbox{\c{}}\else
  \c{}\fi{}tecki}{1999}]{Myers1999}
Myers W.~D.,  \ifmmode \acute{S}\else \'{S}\fi{}wia\ifmmode~\mbox{\c{}}\else
  \c{}\fi{}tecki W.~J.,  1999, \mn@doi [Phys. Rev. C]
  {10.1103/PhysRevC.60.014606}, 60, 014606

\bibitem[\protect\citeauthoryear{{Ney}, {Engel}, {Li}  \& {Schunck}}{{Ney}
  et~al.}{2020}]{Ney2020}
{Ney} E.~M.,  {Engel} J.,  {Li} T.,   {Schunck} N.,  2020, \mn@doi [\prc]
  {10.1103/PhysRevC.102.034326}, \href
  {https://ui.adsabs.harvard.edu/abs/2020PhRvC.102c4326N} {102, 034326}

\bibitem[\protect\citeauthoryear{Nikas, Perdikakis, Beard, Surman, Mumpower  \&
  Tsintari}{Nikas et~al.}{2020}]{Nikas2020}
Nikas S.,  Perdikakis G.,  Beard M.,  Surman R.,  Mumpower M.~R.,   Tsintari
  P.,  2020, arXiv:2010.01698 [astro-ph, physics:nucl-th]

\bibitem[\protect\citeauthoryear{{Nishimura}, {Takiwaki}  \&
  {Thielemann}}{{Nishimura} et~al.}{2015}]{Nishimura2015}
{Nishimura} N.,  {Takiwaki} T.,   {Thielemann} F.-K.,  2015, \apj, 810, 109

\bibitem[\protect\citeauthoryear{Nishimura, Podoly{\'a}k, Fang  \&
  Suzuki}{Nishimura et~al.}{2016}]{Nishimura2016}
Nishimura N.,  Podoly{\'a}k Z.,  Fang D.-L.,   Suzuki T.,  2016, \mn@doi
  [Physics Letters B] {https://doi.org/10.1016/j.physletb.2016.03.025}, 756,
  273

\bibitem[\protect\citeauthoryear{Oechslin, Janka  \& Marek}{Oechslin
  et~al.}{2007}]{Oechslin2007}
Oechslin R.,  Janka H.-T.,   Marek A.,  2007, \mn@doi [Astronomy and
  Astrophysics] {10.1051/0004-6361:20066682}, 467, 395

\bibitem[\protect\citeauthoryear{{Otsuki}, {Tagoshi}, {Kajino}  \&
  {Wanajo}}{{Otsuki} et~al.}{2000}]{Otsuki2000}
{Otsuki} K.,  {Tagoshi} H.,  {Kajino} T.,   {Wanajo} S.-y.,  2000, \mn@doi
  [\apj] {10.1086/308632}, \href
  {https://ui.adsabs.harvard.edu/abs/2000ApJ...533..424O} {533, 424}

\bibitem[\protect\citeauthoryear{Palenzuela, Liebling  \& Mi\~nano}{Palenzuela
  et~al.}{2022}]{Palenzuela2022}
Palenzuela C.,  Liebling S.,   Mi\~nano B.,  2022, \mn@doi [Phys. Rev. D]
  {10.1103/PhysRevD.105.103020}, 105, 103020

\bibitem[\protect\citeauthoryear{{Panov}, {Freiburghaus}  \&
  {Thielemann}}{{Panov} et~al.}{2001}]{Panov2001}
{Panov} I.~V.,  {Freiburghaus} C.,   {Thielemann} F.~K.,  2001, \mn@doi
  [\nphysa] {10.1016/S0375-9474(01)00797-7}, \href
  {https://ui.adsabs.harvard.edu/abs/2001NuPhA.688..587P} {688, 587}

\bibitem[\protect\citeauthoryear{{Panov}, {Korneev}  \& {Thielemann}}{{Panov}
  et~al.}{2008}]{Panov2008}
{Panov} I.~V.,  {Korneev} I.~Y.,   {Thielemann} F.~K.,  2008, \mn@doi
  [Astronomy Letters] {10.1007/s11443-008-3006-1}, \href
  {https://ui.adsabs.harvard.edu/abs/2008AstL...34..189P} {34, 189}

\bibitem[\protect\citeauthoryear{{Pearson}}{{Pearson}}{2001}]{Pearson00}
{Pearson} J.~M.,  2001, Hyp. Int., 132, 59

\bibitem[\protect\citeauthoryear{Pearson, Chamel, Potekhin, Fantina, Ducoin,
  Dutta  \& Goriely}{Pearson et~al.}{2018}]{pearson2018}
Pearson J.~M.,  Chamel N.,  Potekhin A.~Y.,  Fantina A.~F.,  Ducoin C.,  Dutta
  A.~K.,   Goriely S.,  2018, \mn@doi [Monthly Notices of the Royal
  Astronomical Society] {10.1093/mnras/sty2413}, 481, 2994

\bibitem[\protect\citeauthoryear{Perego, Rosswog, Cabez{\'o}n, Korobkin,
  K{\"a}ppeli, Arcones  \& Liebend{\"o}rfer}{Perego et~al.}{2014}]{perego2014}
Perego A.,  Rosswog S.,  Cabez{\'o}n R.~M.,  Korobkin O.,  K{\"a}ppeli R.,
  Arcones A.,   Liebend{\"o}rfer M.,  2014, \mn@doi [Monthly Notices of the
  Royal Astronomical Society] {10.1093/mnras/stu1352}, 443, 3134

\bibitem[\protect\citeauthoryear{Perego, Radice  \& Bernuzzi}{Perego
  et~al.}{2017}]{Perego17}
Perego A.,  Radice D.,   Bernuzzi S.,  2017, Astrophys. J. Lett, 850, L37

\bibitem[\protect\citeauthoryear{{Perego}, {Thielemann}  \&
  {Cescutti}}{{Perego} et~al.}{2021}]{Perego2021}
{Perego} A.,  {Thielemann} F.~K.,   {Cescutti} G.,  2021, in Bambi C.,
  Katsanevas S.,   Kokkotas K.,  eds, , Handbook of Gravitational Wave
  Astronomy.
Springer, Singapore, p.~13, \mn@doi{10.1007/978-981-15-4702-7_13-1}

\bibitem[\protect\citeauthoryear{{Placco} et~al.,}{{Placco}
  et~al.}{2017}]{Placco2017}
{Placco} V.~M.,  et~al., 2017, \mn@doi [\apj] {10.3847/1538-4357/aa78ef}, \href
  {https://ui.adsabs.harvard.edu/abs/2017ApJ...844...18P} {844, 18}

\bibitem[\protect\citeauthoryear{Plompen et~al.}{Plompen
  et~al.}{2017}]{Goriely17a}
Plompen A.,  et~al., eds, 2017, Towards more accurate and reliable predictions
  for nuclear applications  EPJ Web of Conferences Vol. 146.
EDP Sciences, France

\bibitem[\protect\citeauthoryear{Qian \& Woosley}{Qian \&
  Woosley}{1996}]{qian1996}
Qian Y.~Z.,  Woosley S.~E.,  1996, \mn@doi [The Astrophysical Journal]
  {10.1086/177973}, 1, 331

\bibitem[\protect\citeauthoryear{Radice, Galeazzi, Lippuner, Roberts, Ott  \&
  Rezzolla}{Radice et~al.}{2016}]{radice2016}
Radice D.,  Galeazzi F.,  Lippuner J.,  Roberts L.~F.,  Ott C.~D.,   Rezzolla
  L.,  2016, \mn@doi [Monthly Notices of the Royal Astronomical Society]
  {10.1093/mnras/stw1227}, 460, 3255

\bibitem[\protect\citeauthoryear{Radice, Perego, Hotokezaka, Fromm, Bernuzzi
  \& Roberts}{Radice et~al.}{2018a}]{radice2018a}
Radice D.,  Perego A.,  Hotokezaka K.,  Fromm S.~A.,  Bernuzzi S.,   Roberts
  L.~F.,  2018a, \mn@doi [The Astrophysical Journal]
  {10.3847/1538-4357/aaf054}, 869, 130

\bibitem[\protect\citeauthoryear{Radice, Perego, Hotokezaka, Bernuzzi, Fromm
  \& Roberts}{Radice et~al.}{2018b}]{radice2018b}
Radice D.,  Perego A.,  Hotokezaka K.,  Bernuzzi S.,  Fromm S.~A.,   Roberts
  L.~F.,  2018b, \mn@doi [The Astrophysical Journal]
  {10.3847/2041-8213/aaf053}, 869, L35

\bibitem[\protect\citeauthoryear{{Radice}, {Bernuzzi}  \& {Perego}}{{Radice}
  et~al.}{2020}]{Radice2020b}
{Radice} D.,  {Bernuzzi} S.,   {Perego} A.,  2020, \mn@doi [Annual Review of
  Nuclear and Particle Science] {10.1146/annurev-nucl-013120-114541}, \href
  {https://ui.adsabs.harvard.edu/abs/2020ARNPS..70...95R} {70, 95}

\bibitem[\protect\citeauthoryear{Rauscher}{Rauscher}{2020}]{Rauscher2020}
Rauscher T.,  2020, \mn@doi [Journal of Physics: Conference Series]
  {10.1088/1742-6596/1643/1/012062}, 1643, 012062

\bibitem[\protect\citeauthoryear{{Reichert}, {Obergaulinger}, {Eichler}, {Aloy}
   \& {Arcones}}{{Reichert} et~al.}{2021}]{Reichert2021}
{Reichert} M.,  {Obergaulinger} M.,  {Eichler} M.,  {Aloy} M.~{\'A}.,
  {Arcones} A.,  2021, \mnras, 501, 5733

\bibitem[\protect\citeauthoryear{{Reichert}, {Obergaulinger}, {Aloy}, {Gabler},
  {Arcones}  \& {Thielemann}}{{Reichert} et~al.}{2022}]{Reichert2022}
{Reichert} M.,  {Obergaulinger} M.,  {Aloy} M.-A.,  {Gabler} M.,  {Arcones} A.,
    {Thielemann} F.-K.,  2022, arXiv e-prints, \href
  {https://ui.adsabs.harvard.edu/abs/2022arXiv220611914R} {p. arXiv:2206.11914}

\bibitem[\protect\citeauthoryear{{Roberts}, {Woosley}  \& {Hoffman}}{{Roberts}
  et~al.}{2010}]{roberts2010}
{Roberts} L.~F.,  {Woosley} S.~E.,   {Hoffman} R.~D.,  2010, \mn@doi [\apj]
  {10.1088/0004-637X/722/1/954}, \href
  {https://ui.adsabs.harvard.edu/abs/2010ApJ...722..954R} {722, 954}

\bibitem[\protect\citeauthoryear{Roberts, Kasen, Lee  \&
  {Ramirez-Ruiz}}{Roberts et~al.}{2011}]{roberts2011}
Roberts L.~F.,  Kasen D.,  Lee W.~H.,   {Ramirez-Ruiz} E.,  2011, \mn@doi [The
  Astrophysical Journal] {10.1088/2041-8205/736/1/L21}, 736, L21

\bibitem[\protect\citeauthoryear{{Roberts} et~al.,}{{Roberts}
  et~al.}{2017}]{Roberts2017}
{Roberts} L.~F.,  et~al., 2017, \mn@doi [\mnras] {10.1093/mnras/stw2622}, \href
  {https://ui.adsabs.harvard.edu/abs/2017MNRAS.464.3907R} {464, 3907}

\bibitem[\protect\citeauthoryear{{Roederer} et~al.,}{{Roederer}
  et~al.}{2016}]{Roederer2016}
{Roederer} I.~U.,  et~al., 2016, \mn@doi [\aj] {10.3847/0004-6256/151/3/82},
  \href {https://ui.adsabs.harvard.edu/abs/2016AJ....151...82R} {151, 82}

\bibitem[\protect\citeauthoryear{Rosswog, Liebend{\"o}rfer, Thielemann, Davies,
  Benz  \& Piran}{Rosswog et~al.}{1999}]{rosswog1999}
Rosswog S.,  Liebend{\"o}rfer M.,  Thielemann F.~K.,  Davies M.~B.,  Benz W.,
  Piran T.,  1999, Astron. Astrophys., \href
  {https://ui.adsabs.harvard.edu/abs/1999A\&A...341..499R} {341, 499}

\bibitem[\protect\citeauthoryear{{Rosswog}, {Feindt}, {Korobkin}, {Wu},
  {Sollerman}, {Goobar}  \& {Martinez-Pinedo}}{{Rosswog}
  et~al.}{2017}]{Rosswog2017}
{Rosswog} S.,  {Feindt} U.,  {Korobkin} O.,  {Wu} M.~R.,  {Sollerman} J.,
  {Goobar} A.,   {Martinez-Pinedo} G.,  2017, \mn@doi [Classical and Quantum
  Gravity] {10.1088/1361-6382/aa68a9}, \href
  {https://ui.adsabs.harvard.edu/abs/2017CQGra..34j4001R} {34, 104001}

\bibitem[\protect\citeauthoryear{{Ruffert} \& {Janka}}{{Ruffert} \&
  {Janka}}{1999}]{Ruffert1999}
{Ruffert} M.,  {Janka} H.~T.,  1999, \aap, \href
  {https://ui.adsabs.harvard.edu/abs/1999A&A...344..573R} {344, 573}

\bibitem[\protect\citeauthoryear{{Ruffert} \& {Janka}}{{Ruffert} \&
  {Janka}}{2001}]{Ruffert2001}
{Ruffert} M.,  {Janka} H.~T.,  2001, \mn@doi [\aap]
  {10.1051/0004-6361:20011453}, \href
  {https://ui.adsabs.harvard.edu/abs/2001A&A...380..544R} {380, 544}

\bibitem[\protect\citeauthoryear{{Ruffert}, {Janka}, {Takahashi}  \&
  {Schaefer}}{{Ruffert} et~al.}{1997}]{Ruffert1997}
{Ruffert} M.,  {Janka} H.~T.,  {Takahashi} K.,   {Schaefer} G.,  1997, \aap,
  \href {https://ui.adsabs.harvard.edu/abs/1997A&A...319..122R} {319, 122}

\bibitem[\protect\citeauthoryear{Ryssens, Scamps, Goriely  \& Bender}{Ryssens
  et~al.}{2022}]{Ryssens2022}
Ryssens W.,  Scamps G.,  Goriely S.,   Bender M.,  2022, Eur. Phys. J. A, 58,
  246

\bibitem[\protect\citeauthoryear{{Schatz}, {Toenjes}, {Pfeiffer}, {Beers},
  {Cowan}, {Hill}  \& {Kratz}}{{Schatz} et~al.}{2002}]{Schatz2002}
{Schatz} H.,  {Toenjes} R.,  {Pfeiffer} B.,  {Beers} T.~C.,  {Cowan} J.~J.,
  {Hill} V.,   {Kratz} K.-L.,  2002, \mn@doi [\apj] {10.1086/342939}, \href
  {https://ui.adsabs.harvard.edu/abs/2002ApJ...579..626S} {579, 626}

\bibitem[\protect\citeauthoryear{Schmidt, Jurado, Amouroux  \& Schmitt}{Schmidt
  et~al.}{2016}]{Schmidt2016}
Schmidt K.-H.,  Jurado B.,  Amouroux C.,   Schmitt C.,  2016, \mn@doi [Nuclear
  Data Sheets] {https://doi.org/10.1016/j.nds.2015.12.009}, 131, 107

\bibitem[\protect\citeauthoryear{Sekiguchi, Kiuchi, Kyutoku  \&
  Shibata}{Sekiguchi et~al.}{2015}]{sekiguchi2015}
Sekiguchi Y.,  Kiuchi K.,  Kyutoku K.,   Shibata M.,  2015, \mn@doi [Physical
  Review D] {10.1103/PhysRevD.91.064059}, 91, 064059

\bibitem[\protect\citeauthoryear{Shelley \& Pastore}{Shelley \&
  Pastore}{2021}]{Shelley2021}
Shelley M.,  Pastore A.,  2021, \mn@doi [Universe] {10.3390/universe7050131}, 7

\bibitem[\protect\citeauthoryear{Shen, Cooke, Ramirez-Ruiz, Madau, Mayer  \&
  Guedes}{Shen et~al.}{2015}]{Shen2015}
Shen S.,  Cooke R.,  Ramirez-Ruiz E.,  Madau P.,  Mayer L.,   Guedes J.,  2015,
  Astrophys. J, 807, 115

\bibitem[\protect\citeauthoryear{Shibata \& Hotokezaka}{Shibata \&
  Hotokezaka}{2019a}]{Shibata19}
Shibata M.,  Hotokezaka K.,  2019a, Ann. Rev. Nucl. Part. Sc.{\S}, 69, 41

\bibitem[\protect\citeauthoryear{{Shibata} \& {Hotokezaka}}{{Shibata} \&
  {Hotokezaka}}{2019b}]{Shibata2019}
{Shibata} M.,  {Hotokezaka} K.,  2019b, \mn@doi [Annual Review of Nuclear and
  Particle Science] {10.1146/annurev-nucl-101918-023625}, \href
  {https://ui.adsabs.harvard.edu/abs/2019ARNPS..69...41S} {69, 41}

\bibitem[\protect\citeauthoryear{Shibata, Fujibayashi  \& Sekiguchi}{Shibata
  et~al.}{2021}]{Shibata2021b}
Shibata M.,  Fujibayashi S.,   Sekiguchi Y.,  2021, \mn@doi [Phys. Rev. D]
  {10.1103/PhysRevD.104.063026}, 104, 063026

\bibitem[\protect\citeauthoryear{{Siegel}}{{Siegel}}{2022}]{Siegel2022}
{Siegel} D.~M.,  2022, \mn@doi [Nature Reviews Physics]
  {10.1038/s42254-022-00439-1}, \href
  {https://ui.adsabs.harvard.edu/abs/2022NatRP...4..306S} {4, 306}

\bibitem[\protect\citeauthoryear{Siegel \& Metzger}{Siegel \&
  Metzger}{2018}]{Siegel2018}
Siegel D.~M.,  Metzger B.~D.,  2018, \mn@doi [The Astrophysical Journal]
  {10.3847/1538-4357/aabaec}, 858, 52

\bibitem[\protect\citeauthoryear{Siegel, Barnes  \& Metzger}{Siegel
  et~al.}{2019}]{siegel2019}
Siegel D.~M.,  Barnes J.,   Metzger B.~D.,  2019, \mn@doi [Nature]
  {10.1038/s41586-019-1136-0}, 569

\bibitem[\protect\citeauthoryear{{Sieja} \& {Goriely}}{{Sieja} \&
  {Goriely}}{2021}]{sieja2021}
{Sieja} K.,  {Goriely} S.,  2021, \mn@doi [European Physical Journal A]
  {10.1140/epja/s10050-021-00439-2}, \href
  {https://ui.adsabs.harvard.edu/abs/2021EPJA...57..110S} {57, 110}

\bibitem[\protect\citeauthoryear{{Siqueira Mello} et~al.,}{{Siqueira Mello}
  et~al.}{2013}]{Mello2013}
{Siqueira Mello} C.,  et~al., 2013, \mn@doi [\aap]
  {10.1051/0004-6361/201219949}, \href
  {https://ui.adsabs.harvard.edu/abs/2013A&A...550A.122S} {550, A122}

\bibitem[\protect\citeauthoryear{Sneden, McWilliam, Preston, Cowan, Burris  \&
  Armosky}{Sneden et~al.}{1996}]{Sneden96}
Sneden C.,  McWilliam A.,  Preston G.,  Cowan J.,  Burris D.,   Armosky B.,
  1996, Astrophys. J, 467, 819

\bibitem[\protect\citeauthoryear{Sneden et~al.,}{Sneden
  et~al.}{2003}]{sneden2003c}
Sneden C.,  et~al., 2003, \mn@doi [The Astrophysical Journal] {10.1086/375491},
  591, 936

\bibitem[\protect\citeauthoryear{Sprouse, Navarro~Perez, Surman, Mumpower,
  McLaughlin  \& Schunck}{Sprouse et~al.}{2020}]{sprouse2020a}
Sprouse T.~M.,  Navarro~Perez R.,  Surman R.,  Mumpower M.~R.,  McLaughlin
  G.~C.,   Schunck N.,  2020, \mn@doi [Physical Review C]
  {10.1103/PhysRevC.101.055803}, 101, 055803

\bibitem[\protect\citeauthoryear{Steiner, Hempel  \& Fischer}{Steiner
  et~al.}{2013}]{steiner2013}
Steiner A.~W.,  Hempel M.,   Fischer T.,  2013, \mn@doi [The Astrophysical
  Journal] {10.1088/0004-637X/774/1/17}, 774, 17

\bibitem[\protect\citeauthoryear{Stoitsov, Dobaczewski, Nazarewicz, Pittel  \&
  Dean}{Stoitsov et~al.}{2003}]{Stoitsov2003}
Stoitsov M.~V.,  Dobaczewski J.,  Nazarewicz W.,  Pittel S.,   Dean D.~J.,
  2003, \mn@doi [Phys. Rev. C] {10.1103/PhysRevC.68.054312}, 68, 054312

\bibitem[\protect\citeauthoryear{Surman \& Mumpower}{Surman \&
  Mumpower}{2018}]{Surman18}
Surman R.,  Mumpower M.,  2018, EPJ Web of Conferences, 178, 04002

\bibitem[\protect\citeauthoryear{Surman, Mumpower, Sinclair, Jones, Hix  \&
  McLaughlin}{Surman et~al.}{2014}]{surman2014}
Surman R.,  Mumpower M.,  Sinclair R.,  Jones K.~L.,  Hix W.~R.,   McLaughlin
  G.~C.,  2014, \mn@doi [AIP Advances] {10.1063/1.4867191}, 4

\bibitem[\protect\citeauthoryear{Tachibana, Yamada  \& Yoshida}{Tachibana
  et~al.}{1990}]{tachibana1990}
Tachibana T.,  Yamada M.,   Yoshida Y.,  1990, \mn@doi [Progress of Theoretical
  Physics] {10.1143/ptp/84.4.641}, 84, 641

\bibitem[\protect\citeauthoryear{Takahashi, Witti  \& Janka}{Takahashi
  et~al.}{1994}]{takahashi1994}
Takahashi K.,  Witti J.,   Janka H.~T.,  1994, Astron. Astrophys., \href
  {https://ui.adsabs.harvard.edu/abs/1994A\&A...286..857T} {286, 857}

\bibitem[\protect\citeauthoryear{Tanaka \& Hotokezaka}{Tanaka \&
  Hotokezaka}{2013}]{Tanaka2013}
Tanaka M.,  Hotokezaka K.,  2013, \mn@doi [The Astrophysical Journal]
  {10.1088/0004-637X/775/2/113}, 775, 113

\bibitem[\protect\citeauthoryear{{Vassh} et~al.,}{{Vassh}
  et~al.}{2019}]{Vassh2019}
{Vassh} N.,  et~al., 2019, \mn@doi [Journal of Physics G Nuclear Physics]
  {10.1088/1361-6471/ab0bea}, \href
  {https://ui.adsabs.harvard.edu/abs/2019JPhG...46f5202V} {46, 065202}

\bibitem[\protect\citeauthoryear{Vautherin \& Brink}{Vautherin \&
  Brink}{1972}]{Vautherin72}
Vautherin D.,  Brink D.~M.,  1972, Phys. Rev. C, 5, 626

\bibitem[\protect\citeauthoryear{{Villar} et~al.,}{{Villar}
  et~al.}{2017}]{Villar2017}
{Villar} V.~A.,  et~al., 2017, \mn@doi [\apjl] {10.3847/2041-8213/aa9c84},
  \href {https://ui.adsabs.harvard.edu/abs/2017ApJ...851L..21V} {851, L21}

\bibitem[\protect\citeauthoryear{Walker, Litvinov  \& Geissel}{Walker
  et~al.}{2013}]{Walker2013}
Walker P.,  Litvinov Y.~A.,   Geissel H.,  2013, \mn@doi [International Journal
  of Mass Spectrometry] {https://doi.org/10.1016/j.ijms.2013.04.007}, 349-350,
  247

\bibitem[\protect\citeauthoryear{Wanajo}{Wanajo}{2018}]{wanajo2018}
Wanajo S.,  2018, \mn@doi [Astrophys. J] {10.3847/1538-4357/aae0f2}, 868, 65

\bibitem[\protect\citeauthoryear{Wanajo, Janka  \& M{\"u}ller}{Wanajo
  et~al.}{2011}]{wanajo2011}
Wanajo S.,  Janka H.~T.,   M{\"u}ller B.,  2011, \mn@doi [Astrophysical Journal
  Letters] {10.1088/2041-8205/726/2/L15}, 726, 2

\bibitem[\protect\citeauthoryear{Wanajo, Sekiguchi, Nishimura, Kiuchi, Kyutoku
  \& Shibata}{Wanajo et~al.}{2014}]{wanajo2014}
Wanajo S.,  Sekiguchi Y.,  Nishimura N.,  Kiuchi K.,  Kyutoku K.,   Shibata M.,
   2014, \mn@doi [Astrophysical Journal Letters] {10.1088/2041-8205/789/2/L39},
  789

\bibitem[\protect\citeauthoryear{Wanajo, M\"uller, Janka  \& Heger}{Wanajo
  et~al.}{2018}]{wanajo2018c}
Wanajo S.,  M\"uller B.,  Janka H.-T.,   Heger A.,  2018, Astrophys. J, 852, 40

\bibitem[\protect\citeauthoryear{Wang, Liu  \& Wu}{Wang et~al.}{2010}]{Wang10}
Wang N.,  Liu M.,   Wu X.,  2010, Phys. Rev. C, 81, 044322

\bibitem[\protect\citeauthoryear{Wang, Liu, Wu  \& Meng}{Wang
  et~al.}{2014}]{Wang2014}
Wang N.,  Liu M.,  Wu X.,   Meng J.,  2014, \mn@doi [Physics Letters B]
  {https://doi.org/10.1016/j.physletb.2014.05.049}, 734, 215

\bibitem[\protect\citeauthoryear{Wang, Huang, Kondev, Audi  \& Naimi}{Wang
  et~al.}{2021}]{Wang2021}
Wang M.,  Huang W.,  Kondev F.,  Audi G.,   Naimi S.,  2021, \mn@doi [Chinese
  Physics C] {10.1088/1674-1137/abddaf}, 45, 030003

\bibitem[\protect\citeauthoryear{{Winteler}, {K{\"a}ppeli}, {Perego},
  {Arcones}, {Vasset}, {Nishimura}, {Liebend{\"o}rfer}  \&
  {Thielemann}}{{Winteler} et~al.}{2012}]{winteler2012}
{Winteler} C.,  {K{\"a}ppeli} R.,  {Perego} A.,  {Arcones} A.,  {Vasset} N.,
  {Nishimura} N.,  {Liebend{\"o}rfer} M.,   {Thielemann} F.-K.,  2012, \mn@doi
  [Astrophys. J] {10.1088/2041-8205/750/1/L22}, 750, L22

\bibitem[\protect\citeauthoryear{{Witti}, {Janka}  \& {Takahashi}}{{Witti}
  et~al.}{1994}]{witti1994}
{Witti} J.,  {Janka} H.~T.,   {Takahashi} K.,  1994, \aap, \href
  {https://ui.adsabs.harvard.edu/abs/1994A&A...286..841W} {286, 841}

\bibitem[\protect\citeauthoryear{Wu, Fern{\'a}ndez, {Mart{\'\i}nez-Pinedo}  \&
  Metzger}{Wu et~al.}{2016}]{Wu2016}
Wu M.-R.,  Fern{\'a}ndez R.,  {Mart{\'\i}nez-Pinedo} G.,   Metzger B.~D.,
  2016, \mn@doi [Monthly Notices of the Royal Astronomical Society]
  {10.1093/mnras/stw2156}, 463, 2323

\bibitem[\protect\citeauthoryear{Wu, Barnes, {Mart{\'\i}nez-Pinedo}  \&
  Metzger}{Wu et~al.}{2019}]{wu2019}
Wu M.-R.,  Barnes J.,  {Mart{\'\i}nez-Pinedo} G.,   Metzger B.~D.,  2019,
  \mn@doi [Physical Review Letters] {10.1103/PhysRevLett.122.062701}, 122,
  062701

\bibitem[\protect\citeauthoryear{Xu \& Goriely}{Xu \& Goriely}{2012}]{Xu2012}
Xu Y.,  Goriely S.,  2012, \mn@doi [Phys. Rev. C] {10.1103/PhysRevC.86.045801},
  86, 045801

\bibitem[\protect\citeauthoryear{Xu, Goriely, Jorissen, Chen  \& Arnould}{Xu
  et~al.}{2013}]{xu2013}
Xu Y.,  Goriely S.,  Jorissen A.,  Chen G.,   Arnould M.,  2013, Astron.
  Astrophys., 549, 10

\bibitem[\protect\citeauthoryear{Xu, Goriely, Koning  \& Hilaire}{Xu
  et~al.}{2014}]{Xu2014}
Xu Y.,  Goriely S.,  Koning A.~J.,   Hilaire S.,  2014, \mn@doi [Phys. Rev. C]
  {10.1103/PhysRevC.90.024604}, 90, 024604

\bibitem[\protect\citeauthoryear{{Zhu} et~al.,}{{Zhu} et~al.}{2018}]{zhu2018}
{Zhu} Y.,  et~al., 2018, \mn@doi [\apjl] {10.3847/2041-8213/aad5de}, \href
  {https://ui.adsabs.harvard.edu/abs/2018ApJ...863L..23Z} {863, L23}

\bibitem[\protect\citeauthoryear{Zhu, Lund, Barnes, Sprouse, Vassh, McLaughlin,
  Mumpower  \& Surman}{Zhu et~al.}{2021}]{Zhu2021}
Zhu Y.~L.,  Lund K.~A.,  Barnes J.,  Sprouse T.~M.,  Vassh N.,  McLaughlin
  G.~C.,  Mumpower M.~R.,   Surman R.,  2021, \mn@doi [The Astrophysical
  Journal] {10.3847/1538-4357/abc69e}, 906, 94

\bibitem[\protect\citeauthoryear{{van de Voort}, {Pakmor}, {Grand}, {Springel},
  {G{\'o}mez}  \& {Marinacci}}{{van de Voort} et~al.}{2020}]{Voort2020}
{van de Voort} F.,  {Pakmor} R.,  {Grand} R. J.~J.,  {Springel} V.,
  {G{\'o}mez} F.~A.,   {Marinacci} F.,  2020, \mn@doi [\mnras]
  {10.1093/mnras/staa754}, \href
  {https://ui.adsabs.harvard.edu/abs/2020MNRAS.494.4867V} {494, 4867}

\bibitem[\protect\citeauthoryear{{van de Voort}, {Pakmor}, {Bieri}  \&
  {Grand}}{{van de Voort} et~al.}{2022}]{Voort2022}
{van de Voort} F.,  {Pakmor} R.,  {Bieri} R.,   {Grand} R. J.~J.,  2022,
  \mn@doi [\mnras] {10.1093/mnras/stac710}, \href
  {https://ui.adsabs.harvard.edu/abs/2022MNRAS.512.5258V} {512, 5258}

\bibitem[\protect\citeauthoryear{{von Weizs\"acker}}{{von
  Weizs\"acker}}{1935}]{VonWeizsacker1935}
{von Weizs\"acker} C.~F.,  1935, {Z. Phys.}, 96, 431

\makeatother
\end{thebibliography}

\end{document}